\documentclass[letterpaper, 10 pt, conference]{ieeeconf}  

\IEEEoverridecommandlockouts                              

\usepackage{amsmath} 
\usepackage{amssymb}  
\usepackage{bm}
\usepackage{units}
\usepackage{hyphsubst}
\usepackage{nicefrac}
\usepackage{textgreek}

\usepackage{graphics} 
\usepackage[pdftex]{graphicx}
\usepackage[pdftex]{xcolor}
\usepackage{subcaption}
\usepackage{caption}
\usepackage{tikz,pgfplots}
\usetikzlibrary{external,backgrounds,fit,calc,positioning,matrix,decorations.markings,arrows,arrows.meta,fadings,decorations.pathmorphing,shapes.geometric}
\usepgfplotslibrary{colormaps}

\usepackage{makecell}

\graphicspath{{Bilder/}}

\title{\LARGE \bf
Highway traffic data: macroscopic, microscopic and criticality analysis for capturing relevant traffic scenarios and\\traffic modeling based on the highD data set
}

\author{Friedrich Kruber$^{1}$, Jonas Wurst$^{1}$, Samarjit Chakraborty$^{2}$, Michael Botsch$^{1}$
\thanks{$^{1}$Technische Hochschule Ingolstadt, Research Center CARISSMA, Esplanade 10, 85049 Ingolstadt, Germany,
\{firstname.lastname\}@thi.de\
$^{2}$Technical University of Munich, Real-Time Computer Systems, Arcisstra{\ss}e 21, 80333 Munich, Germany, \{firstname\}@tum.de}
}

\begin{document}

\maketitle
\thispagestyle{empty}
\pagestyle{empty}

\begin{abstract}
This work provides a comprehensive analysis on naturalistic driving behavior for highways based on the highD data set. Two thematic fields are considered. First, some macroscopic and microscopic traffic statistics are provided. These include the traffic flow rate and the traffic density, as well as velocity, acceleration and distance distributions. Additionally, the dependencies to each other are examined and compared to related work.
The second part investigates the distributions of criticality measures. The Time-To-Collision, Time-Headway and a third measure, which couples both, are analyzed. These measures are also combined with other indicators. Scenarios, in which these measures reach a critical level, are separately discussed. The results are compared to related work as well. The two main contributions of this work can be stated as follows. First, the analysis on the criticality measures can be used to find suitable thresholds for rare traffic scenarios. Second, the statistics provided in this work can also be utilized for traffic modeling, for example in simulation environments.
\end{abstract}


\section{Introduction}
\label{intrduction_section}
An overview of real world traffic patterns is valuable for everyone working on vehicle or traffic related research and applications. The presented work provides several insights into the highD data set, which was published in November 2018 \cite{highDdataset}. Traffic on German motorways was recorded from a drone perspective. It contains data of approximately $\unit[45\,000]{km}$ of driving, recorded at a frame rate of $\unit[25]{Hz}$.

Large data sets on naturalistic driving behavior are rarely publicly available. Hence, publications which analyze the distribution for criticality measures are rare as well. All statistics in this work are derived from one data source. The microscopic view on the data can be interpreted with regard to the macroscopic basis of the data set.  

The macroscopic view is depicted with the fundamental diagrams of traffic flow. Additionally, a normalized load per lane, depending on the motorway capacity limits and the flow rates is provided. The number of lane changes depending on the flow rate and density is depicted as well. 
 
The microscopic analysis includes velocity distributions, longitudinal and lateral acceleration distributions and Distance Headway (DHW) distributions.
Regarding the criticality of traffic scenarios, the distributions of the Time-To-Collision (TTC) and THW are provided. One more criticality measure is analyzed, which is composed of both, the TTC and THW. Each of the two components provides different aspects of a traffic scenario as discussed in Section \ref{relatedwork_section}. The occurrence of criticality measure values are also combined with other indicators like the longitudinal acceleration.

The paper is organized as follows. Section \ref{relatedwork_section} depicts a selection of related work. Section \ref{variables_section} briefly describes the variables of interest. Section \ref{data_section} gives an overview on the highD data set and shows some issues concerning the data quality. Section \ref{macroscopic_section} addresses the macroscopic, Section \ref{microscopic_section} the microscopic analysis. In Section \ref{results_section} the analysis on the appearance of critical scenarios are presented. Finally, this paper is concluded in Section \ref{conclusions_section}.

\section{Related Work}
\label{relatedwork_section}
This work is related to two research topics: first macroscopic and microscopic traffic research and second, criticality measures for vehicle safety applications.
\subsection{Traffic Analysis}
Traffic flow theory has been intensively researched for decades, starting with Greenshields publication and the first fundamental diagram in 1933 \cite{Greenshields.1933}. The selection of publications mentioned here, will be taken up again in the following sections. In \cite{Filzek.2002}, field tests on German motorways were performed to compare adaptive cruise control systems to human driving. The results concerning THW and TTC are compared to the highD data set. The publication \cite{Busch.1984} analyzes the lane change behavior on motorways and provides macroscopic statistics from the observations. Some of the statistics are compared to this data set. In \cite{Neubert.2000}, road traffic was mainly analyzed by induction loop data in order to simulate traffic flow. The work of \cite{Hall.96} provides a detailed description and literature research of traffic stream characteristics and compares several approaches to obtain the necessary data.

\subsection{Criticality measures}
Besides the widely used TTC and THW measures, several other measures are proposed in various research work. For example in \cite{Kitajima.2009, Junietz.2017, Maurer.2016b} several measures are compared. Publication \cite{Kondoh.2008} analyzes THW and TTC and states with reference to other publications, that the THW determines the car-following strategies of drivers and the magnitude of the THW influences the drivers stress level. The inverse of TTC, TTC\textsuperscript{-1}, corresponds to the expansion rate of the visual angle of the vehicle ahead when closing in on it, which represents the drivers input. Additionally, in \cite{Kondoh.2008} a combined measure named Risk Perception (RP) is proposed, which is analyzed with the highD data set in this work.

In \cite{Chen.2016}, the data of the 100-cars-study \cite{nhtsa.2006} is analyzed by means of the TTC and the Enhanced TTC (ETTC) for all braking events during the car following mode. The ETTC is defined in the ISO 22839 norm and assumes a constant relative acceleration. The publication \cite{Chen.2016} concludes that both, TTC and ETTC increase with higher velocities. The median was higher for the TTC values compared to the ETTC values. Additionally, the variance across all drivers was higher for the TTC.

The publication \cite{Benmimoun.2012} proposes three risk level thresholds for THW and TTC. Level 1 describes scenarios, which exceed ``normal'' driving. Level 2 scenarios show a higher collision risk and level 3 scenarios indicate an imminent collision risk. The THW is combined with a relative velocity threshold. The TTC threshold is combined with a brakelight indication, see Table \ref{table:ATZ_risk_profiles} in Section \ref{ttcthw_section}. Longitudinal and lateral accelerations can increase or decrease the level according to their defined tresholds, see Section \ref{ttcthw_section}. This paper provides occurrences of scenarios in the highD data set according to that risk definition. The authors in \cite{Benmimoun.2012} state, that the false positive rate for detecting critical scenarios was reduced considering the driver intention, which can be achieved for example by considering the yaw-rate, turn or brake signals, accessible from the bus system. Another aspect stated in \cite{Benmimoun.2012} is, that drivers usually see the presence or absence of a vehicle in front of their leader vehicle and adjust their driving strategy to that. 

The publication \cite{Chen.2013} combines TTC with a safety braking distance, which considers vehicle and environmental variables. This approach assumes an existing estimation of the environmental conditions.  

A study performed on a test track revealed that only 25\% of 64 drivers, who were initially distracted, were able to avoid a collision when alerting them at $TTC = \unit[2.1]{s}$ at $\mathrm{DHW} \approx \unit[33]{m}$ and $v \approx \unitfrac[55]{km}{h}$ \cite{nhtsa.2011}. The drivers were allowed to brake and to perform a lateral avoidance maneuver. Automatic brake assistant systems are therefore required to warn at latest from $TTC \geq \unit[2.0]{s}$ up to $TTC \geq \unit[2.4]{s}$ in the United States, depending on the scenario \cite{nhtsa.2013}. 

In the 100-cars-study \cite{nhtsa.2006}, several triggers as the TTC and accelerations are used in order to find relevant events. More details are given in Table \ref{table:100_cars_study_triggers} in Section \ref{ttcthw_section}. The occurrences of such events for the highD data set are provided in this work as well.

A survey on motion prediction and risk assessment considers various aspects on motion prediction and the risk definition. The authors highlight, that the risk definition is influenced by the choice of the motion model \cite{Lefevre.2014}. Besides deterministic risk indicators like TTC, which assume a collision, probabilistic measures are discussed as well. The probabilistic collision predictions take the uncertainty of the predicted trajectories of the human driven ego vehicle and the surrounding objects into account.

Generally, the TTC and THW, as defined in Section \ref{variables_section}, are suitable indicators for highway data. Nevertheless, they should be coupled with a motion model and additional features to reduce false positives, as shown in this work. 

These two indicators fail for crossing scenarios. In such cases the indicators have to be adapted or replaced as discussed for example in \cite{Maurer.2016b}.

\section{Variables of interest}
\label{variables_section}
The variables of interest are described briefly in this section. First, the macroscopic variables and second, the variables concerning the criticality indicators are described.

Throughout this work, the longitudinal velocity is abbreviated with velocity.

\subsection{Macroscopic variables}
The three macroscopic variables considered in this work are the traffic flow rate, the traffic density and the average velocity of the traffic stream. These variables are used to plot the fundamental diagrams. The road occupancy, as an alternative to the density, is not considered.

The flow rate is defined as
\begin{equation}
q = \unitfrac[]{vehicles}{time}.
\end{equation}
The flow rate is measured at a certain point over time. Usually stationary induction loops are installed on the road. Today, flow rates on an hourly basis are available for many metering points \cite{Bast.2019}.

The traffic density is defined as
\begin{equation}
\rho = \unitfrac[\,]{vehicles}{distance}.
\end{equation}
The traffic density $\rho$ does not consider the vehicle length. 
To include the vehicle length, the equation has to be adapted to
\begin{equation}
\rho_a = l_r \bigg( \frac{1}{N} \sum_{n=1}^{N} (l_n + \mathrm{DHW}_n) \bigg)^{-1},
\end{equation}
where $l_r$ denotes the length of the observed road segment, $N$ the number of vehicles, $l_n$ the length of the $n$-th vehicle and $\mathrm{DHW}_n$ the distance between two vehicles. 
In order to measure the traffic density, a road segment of at least several hundred meters should be observed, for example with surveillance cameras \cite{Hall.96}. This data is usually not publicy available. Alternatively, the density can be estimated by summarizing the road occupancy over time or if one has the flow rate and average velocities available \cite{Neubert.2000}.

Flow and density refer to different measurement frameworks: The flow rate is measured over time at a specific point in space. The density is measured over space at a specific point in time \cite{Hall.96}.
 
The third and last variable, the average velocity, can be estimated as ``time mean velocity'' or ``space mean velocity'' \cite{Hall.96}. The difference of both is minor for stable traffic flow and therefore neglected in the following.

\subsection{Criticality measures}
The following three criticality measures are analyzed in this work: the TTC, THW and RP.

The TTC is defined as
\begin{equation}
  TTC = \frac{x_l - x_f}{v_f-v_l} = \frac{x_r}{v_r},
\end{equation}  
where $x$ denotes the position and $v$ the velocity. The subscript $l$ denotes the leader vehicle, the subscript $f$ the follower vehicle and subscript $r$ the relative position or velocity, respectively. The TTC equation assumes constant velocities for the prediction horizon. In the highD data set the TTC values appear with positive and negative sign. A positive TTC indicates, that the relative distance $x_r$ decreases over time and a collision is going to happen at the future time instance TTC, given the assumption of constant velocities. A negative time before collision states that $\mathrm{TTC} \rightarrow\infty$.
Alternatively, one can use the ETTC, which assumes a constant relative acceleration.

The THW is defined as
\begin{equation}
  THW = \frac{x_l - x_f}{v_f}= \frac{x_r}{v_f}.
  \end{equation} 
Contrary to the TTC, the THW only considers the follower vehicle velocity. 

The TTC delivers valuable information during a closing scenario, while the THW delivers information for tailgate driving scenarios and can be thought of as a latent risk potential.  

The RP is a combination of the TTC and THW proposed in \cite{Kondoh.2008}. The RP is defined as
\begin{equation}
  RP = \frac{A}{THW} + \frac{B}{TTC},
  \end{equation} 
where $A$ and $B$ are two weight factors. In \cite{Kondoh.2008}, the weight factors are examined by means of a naturalistic driving behavior data set. The authors propose a set of $A=1$ and $B=4$. Given these values, all drivers performed a braking maneuver, if $RP \geq \ 2.0$. The authors added the restriction, that a follower vehicle drove for at least four seconds before and after the initial brake application behind its leader vehicle. Since this set of variables is used as a benchmark in Section \ref{rp_section}, it is denoted as $\mathrm{S}_{\mathrm{RP}_{\mathrm{214}}}$ in the following. 

The RP analysis in this work considers scenarios with positive signed TTC values, in order to compare the results with \cite{Kondoh.2008}, where only closing scenarios were considered. As far as the RP serves as a risk perception indicator, one might include negative TTC values as well.  
In order to use both, positive and negative TTC values, the RP equation must be adapted for the following reason. A large negative TTC value indicates, that the relative velocity is small, i\,e. the follower vehicle velocity is only slightly below the leader vehicle velocity, so that the relative distance will decrease only slowly over time in order to reduce the criticality level. Therefore, large negative TTC values should punish the criticality measure stronger, which is not the case with the RP equation as stated above.

\section{highD data set}
\label{data_section}
\begin{figure}[h]
\centering
	\includegraphics[width=\linewidth]{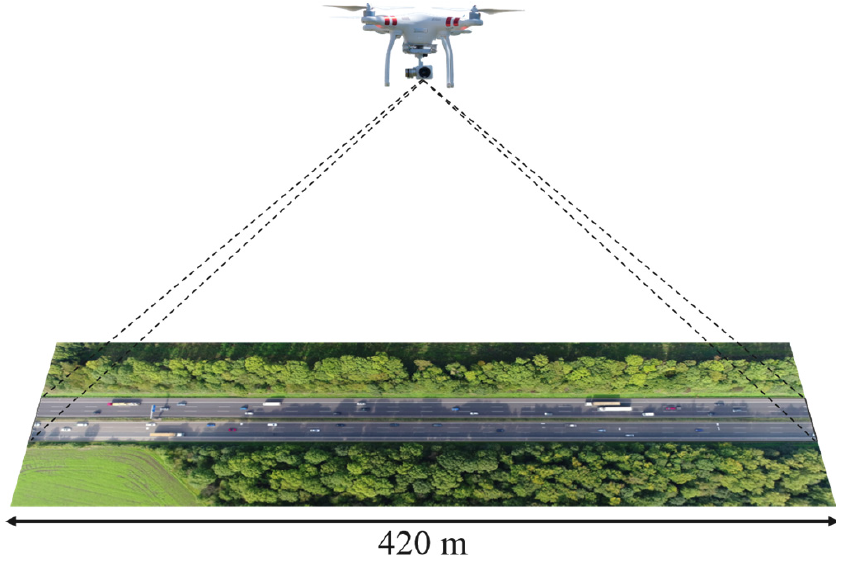}
	\vspace{-0.5cm}
	\caption{Illustration of the drone based video recordings in highD. The image is taken from the publication \cite{highDdataset}.}
	\label{fig:highD}
	\vspace{-0.2cm}
\end{figure}

The highD data set is available on request and was published alongside with the publication \cite{highDdataset}. Videos of German motorways were recorded from a drone perspective as depicted in Fig. \ref{fig:highD}.
The following enumeration, adopted from \cite{highDdataset}, depicts some key facts about the data set:
\begin{itemize}
\item around 110\,500 vehicles were recorded,
\item the total driven distance is \unit[45\,000]{km},
\item the total driven time is \unit[447]{h},
\item 11\,000 lane changes, from which 5\,600 were completely performed in the observed area,
\item 850 cut-in maneuvers: THW from around \unit[0.1 - 4]{s} with a distribution peak at \unit[1]{s},
\item proportion of 77\% passenger cars, 23\% trucks,
\item 6 locations around Cologne (2 or 3 lanes per direction),
\item 60 videos recorded at \unit[25]{fps} and 4K resolution,
\item the average video length is \unit[17]{min} (total \unit[16.5]{h}),
\item the pixel resolution is $\unit[1]{px} \widehat{=} \unit[10]{cm}$, 	
\item each vehicle is visible for a median of \unit[14]{s},
\item the recording time was between 08 a.m. and 5 p.m.,
\item sunny weather and low wind conditions during the recordings.
\end{itemize}

The videos were stabilized and rotated by openCV algorithms, additionally to the drones gimbal based stabilization system. The images are rotated, that lane markings are oriented horizontally. 
The semantic segmentation was performed using a modified U-Net neural network \cite{Ronneberger.2015}. To adapt the neural network to the highD data set, around 3000 image patches were manually labeled. To generate more training examples, the labeled data was flipped, Gaussian noise was added, the patches were re-sized and contrast was changed.
The trajectories were smoothed and validated in a post-processing step. 

The data is provided in csv files. Phyton and Matlab scripts are provided on GitHub in order to read in the files and to plot the tracks. Already preprocessed variables are for example the velocities, accelerations, THW, TTC and vehicle indexes for neighboring lanes for each recorded time frame. For some variables the minimum, maximum and mean values per track are provided as well in the track meta data. 

The vehicles are divided into two classes: cars and trucks. The ``trucks'' class includes different types of vehicle load vehicles, as velocities of $v \geq \unitfrac[80]{km}{h}$ are recorded for this class (see Fig. \ref{fig:vel_histogram_cars_trucks}). In this work the term ``trucks'' is used as a substitute for heavy load vehicles as well. 

\subsection{Data quality}
The data is checked manually for tracks with $\mathrm{TTC}_{\mathrm{min}} \leq \unit[0.8]{s}$ and $\mathrm{THW}_{\mathrm{min}} < \unit[0]{s}$.\footnote{highD data set version 1.0}
These thresholds were chosen because they are very rare when computed properly ($\mathrm{TTC}_{\mathrm{min}} \leq \unit[0.8]{s}$) or should not appear at all ($\mathrm{THW}_{\mathrm{min}} < \unit[0]{s}$). 

In case of the TTC samples, 36 tracks are validated. From those five are discarded due to object recognition issues. In two cases a vehicle from the neighboring lane occupied the lane of the ego vehicle partly. All other samples occur during lane changes. The TTC value in the data set is computed for a specific lane until the vehicle bounding box crosses the lane markings completely. Often the trajectories of the vehicles involved do not overlap anymore at $\mathrm{TTC}_{\mathrm{min}}$. As shown in Section \ref{ttcthw_section}, this applies also for many cases with larger TTC values.  

In case of the THW samples, 64 tracks are validated. From these two are discarded due to object recognition issues. All other samples occurred during a stop-and-go situation. A vehicle in standstill was recognized to be moving backwards with velocities below $v = \unitfrac[1]{km}{h}$, which caused negative THW values. All of these THW samples are discarded for the analysis. Six additional tracks are removed because of implausible acceleration, TTC values or driving direction.

In total 75 tracks are discarded from the data set for this work. Generally, the last recorded time step has corrupted values and is removed for each variable and track. This has already been done for the track meta data provided in the data set (e.\,g. the $\mathrm{TTC}_{\mathrm{min}}$ values for each track). Removing the corrupted vehicle tracks, in total $M=110\,381$ vehicle tracks remain. From these, in total 1521 vehicles had no leader vehicle in the range of view. 

\section{Macroscopic analysis}
\label{macroscopic_section}
This section depicts the macroscopic analysis for the variables stated in Section \ref{variables_section}. Additionally the load per road and lane is analyzed. 

\subsection{Road and lane load}
This subsection analyzes the road and lane load in dependence of the recorded video and macroscopic variables as well. In the highD data set, both driving directions are recorded simultaneously. Vehicles driving from the right to the left side are denoted with the variable value of $drivingDirection=1$, which corresponds to the upper road. The lane ID numbering starts at $\mathrm{ID_L}=1$, which is the emergency lane on the upper road. No vehicle was recorded on that lane. The outer right lane on the upper road continues with $\mathrm{ID_L}=2$, followed by the middle lane with $\mathrm{ID_L}=3$ and the left lane with $\mathrm{ID_L}=4$. The structural separation between the upper and lower road has its own ID. In case of motorways with 3 lanes, the ID for the structural separation is $\mathrm{ID_L}=5$. The index count is continued with the outer left lane of the lower road ($\mathrm{ID_L}=6$ in case of a 3 lane motorway). The emergency lane on the lower road has no ID. 

This indexing principle is not consistent for all locations. Location number 6 has an acceleration lane on the upper road. This acceleration lane ends in the middle of the visible range and is continued as an emergency lane, i.\,e. this lane is the most upper lane. Still, the lane ID in this case is $\mathrm{ID_L}=2$. 

All corresponding lane IDs for each location are depicted in Table \ref{tb:laneIDs}. Additionally the velocity limit for each location is depicted in the table.

\begin{table}[h]
\centering
\begin{tabular}{|l||c|c|c|c|c|c|}
\hline 
\rule[0ex]{0pt}{2.5ex} \textbf{Location ID} $\longrightarrow$ & 1 & 2 & 3 & 4 & 5 & 6 \\ 
\hline 
\rule[0ex]{0pt}{2.5ex} velocity Limit [$\unitfrac[]{km}{h}$] & 120 & -- & 130 & -- & -- & -- \\ 
\hline
\hline  
\rule[0ex]{0pt}{2.5ex} \textbf{Upper Road} & \multicolumn{6}{c|}{LANE ID} \\ 
\hline
\hline  
\rule[0ex]{0pt}{2.5ex} Lane 0 (acc. lane) & -- & -- & -- & -- & -- & 2 \\ 
\hline 
\rule[0ex]{0pt}{2.5ex} Lane 1 (right) & 2 & 2 & 2 & 2 & 2 & 3 \\ 
\hline 
\rule[0ex]{0pt}{2.5ex} Lane 2 & 3 & 3 & 3 & 3 & 3 & 4 \\ 
\hline 
\rule[0ex]{0pt}{2.5ex} Lane 3 & 4 & -- & 4 & 4 & -- & 5 \\ 
\hline
\hline 
\rule[0ex]{0pt}{2.5ex} \textbf{Lower Road} & \multicolumn{6}{c|}{LANE ID} \\ 
\hline
\hline  
\rule[0ex]{0pt}{2.5ex} Lane 1 (right) & 8 & 6 & 8 & 8 & 6 & 9 \\ 
\hline 
\rule[0ex]{0pt}{2.5ex} Lane 2  & 7 & 5 & 7 & 7 & 5 & 8 \\ 
\hline 
\rule[0ex]{0pt}{2.5ex} Lane 3 & 6 & -- & 6 & 6 & -- & 7 \\ 
\hline 
\end{tabular} 
\caption{Lane IDs for the upper and lower road. Location 2 and 5 have only two lanes. Location 6 has an acceleration lane on the upper road.}
\label{tb:laneIDs}
\end{table}

\begin{figure}[h]
\centering
	\includegraphics[width=\linewidth]{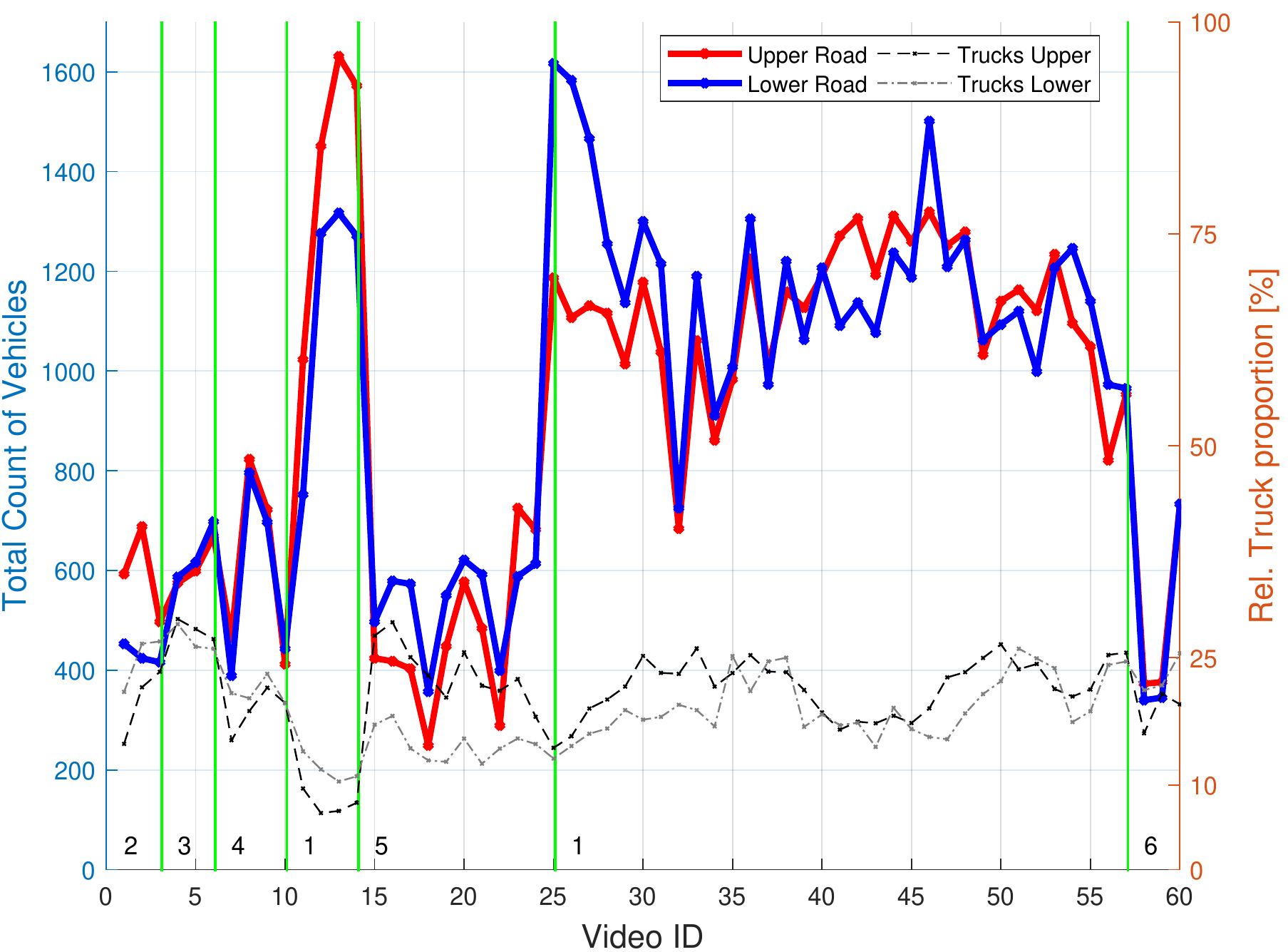}
	\vspace{-0.5cm}
	\caption{Total count of vehicles per road and video (left y-axis) and relative proportion of trucks for both roads (right y-axis). The vertical green lines and the numbers at the bottom indicate the location ID.}
	\label{fig:VehPerRoad_plot}
	\vspace{-0.2cm}
\end{figure}

Fig. \ref{fig:VehPerRoad_plot} depicts the total number of recorded vehicles per video. The vertical green lines and numbers at the bottom indicate the location ID. Additionally, the relative proportion of trucks per road is given.  

\begin{figure}[h]
\centering
	\includegraphics[width=\linewidth]{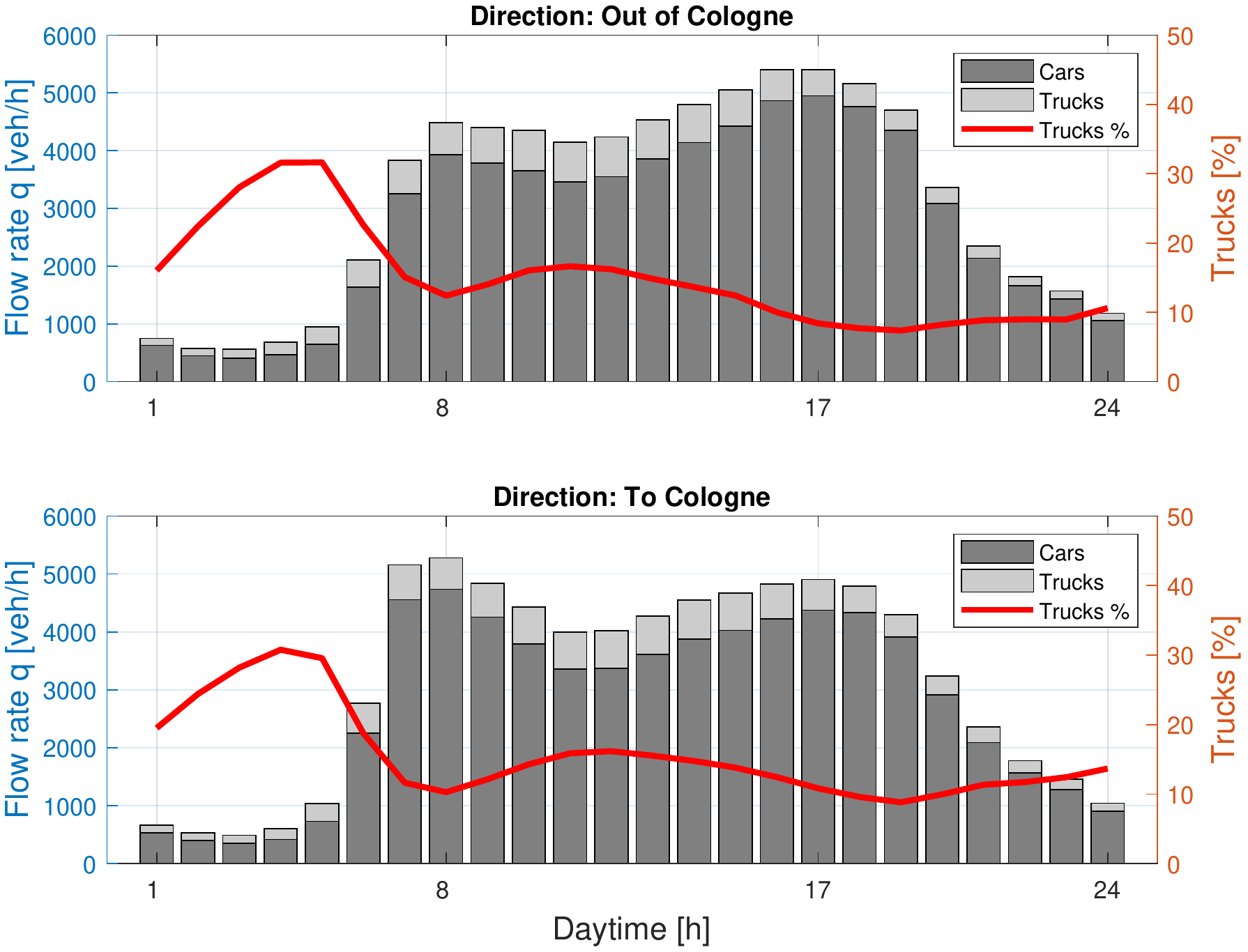}
	\vspace{-0.5cm}
	\caption{Traffic flow rate $q$ for working days of a metering point near Cologne\textsuperscript{2} in 2017 (left y-axis). Relative proportion of trucks (right y-axis).}
	\label{fig:flowrate_cologne}
	\vspace{-0.2cm}
\end{figure}

In order to evaluate the truck proportion in the data set, the proportion is compared to induction loop data. Fig. \ref{fig:flowrate_cologne} depicts the flow rate for the year 2017 from a metering point\footnote{Metering Point: Autobahn A4, K{\"o}ln-Klettenberg, BASt-Nr. 5053}, 
which is closely located to one of the highD locations depicted in \cite{highDdataset}. The induction loop data is available at \cite{Bast.2019}. It shows the averaged traffic flow rates for working days. Days on public holidays are excluded. One can observe the effect of the commuters in direction to the city in the morning and out of the city in the evening by comparing the upper road with the lower road in Fig. \ref{fig:flowrate_cologne}. The overall heavy load vehicle proportion for this metering point in 2017 is 13.3\%.

By comparing Fig. \ref{fig:VehPerRoad_plot} with Fig. \ref{fig:flowrate_cologne}, one finds that the proportion of trucks is relatively large in the highD data set. This can be partly explained by the daytime, in which the recordings were done. According to \cite{Busch.1984}, the proportion of heavy load vehicles increases with lower traffic flow rates, which can be confirmed by Fig. \ref{fig:flowrate_cologne}. Most of the videos of the highD data set were recorded between 8.30 a.m. and 1.00 pm, where the proportions of trucks is higher than in the afternoon rush hours. Only the videos 11-14 were recorded in the afternoon, which explains the lower proportion of trucks as depicted in Fig. \ref{fig:VehPerRoad_plot}. Still, the absolute values for the truck proportions in the highD data set are higher than the average over the year for that specific metering point. This holds also for the month of September, where the recording took place. The yearly averaged proportions are checked for several metering points in that region. Only one available metering point showed a heavy vehicle load proportion, which is comparable to the highD data. The overall heavy load proportion for this metering point in 2017 is 21.4\%.\footnote{Metering Point: Autobahn A61, Kerpen (S), BASt-Nr. 5621} However, this metering point is separated by at least one motorway intersection, compared to the depicted locations in the highD publication.  
  
Concluding this aspect, the high proportion of heavy load vehicles in the highD data set can only be partly explained by the daytime. Other reasons are unknown. The bias of the truck proportion has to be considered in terms of traffic modeling, as the distributions of the variables are affected due to the high truck rate.

\begin{figure}[h]
\centering
	\includegraphics[width=\linewidth]{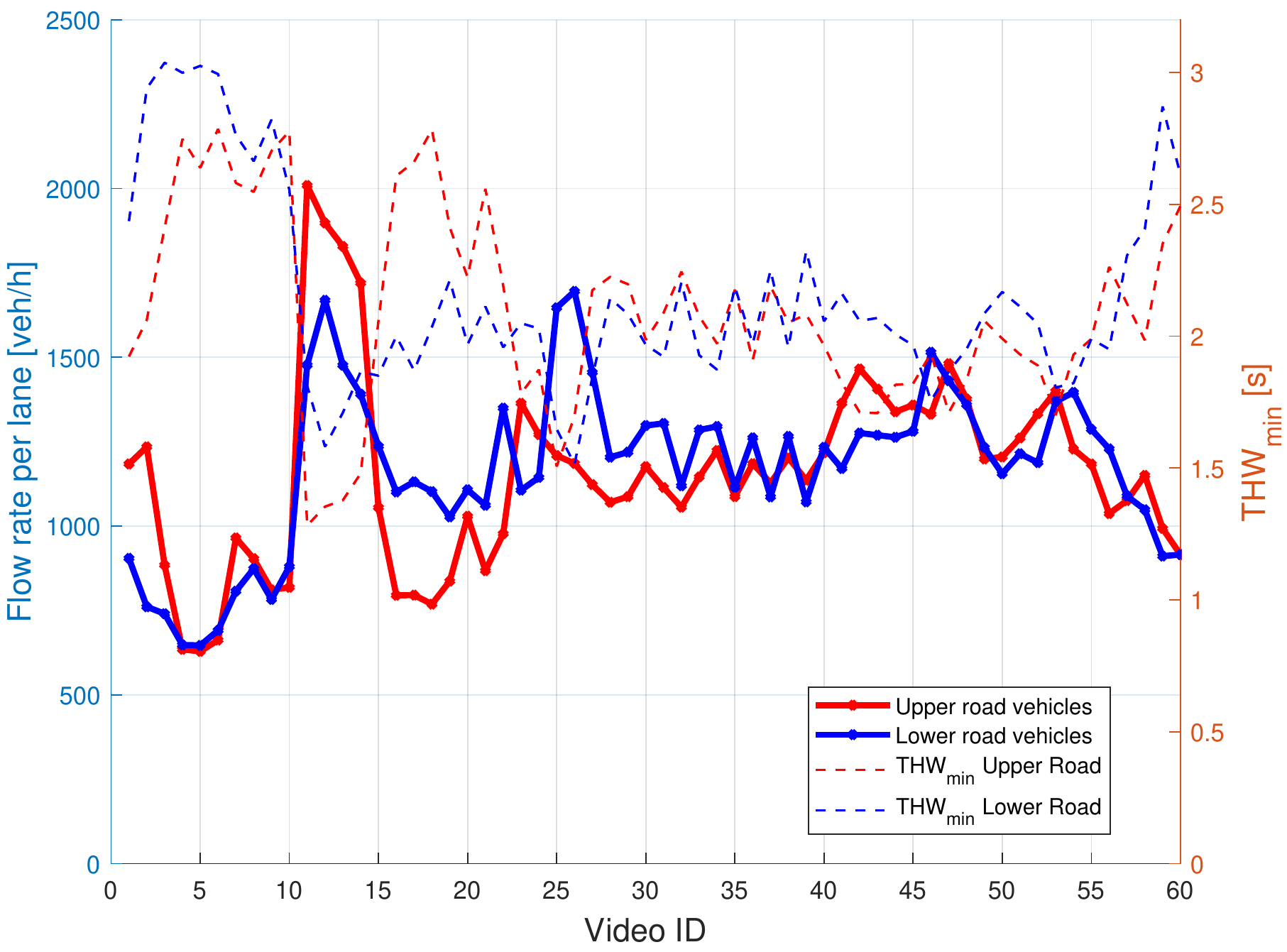}
	\vspace{-0.5cm}
\caption{Flow rates per lane are depicted with the solid lines. The dashed lines indicate the average $\mathrm{THW}_{\mathrm{min}}$. The negative correlation between the flow rate and $\mathrm{THW}_{\mathrm{min}}$ can be seen immediately ($R = -0.94$).}
	\label{fig:VehPerLaneAndHour_plot}
	\vspace{-0.2cm}
\end{figure}

Fig. \ref{fig:VehPerLaneAndHour_plot} depicts the average traffic load per lane and hour. Additionally, the average $\mathrm{THW}_{\mathrm{min}}$ is depicted. The negative correlation between the flow rate and the average $\mathrm{THW}_{\mathrm{min}}$ can be seen immediately, which is in accordance to the field test in \cite{Filzek.2002}. The correlation coefficient is $R = -0.94$.

\begin{figure}[h]
\centering
	\includegraphics[width=\linewidth]{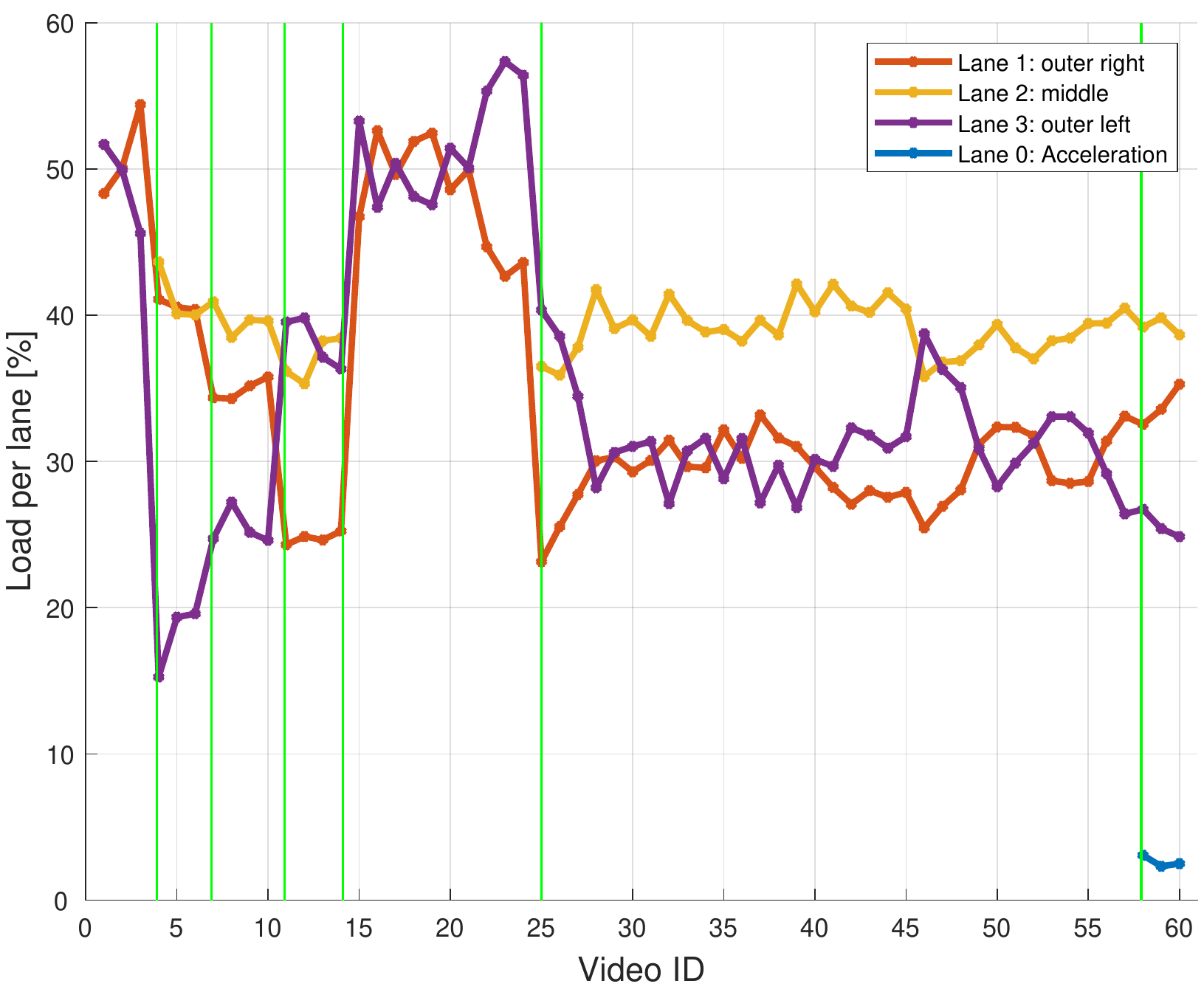}
	\vspace{-0.5cm}
\caption{Relative lane load per video. Higher flow rates increase the proportion of the outer left lane. Lower flow rates increase the proportion to the outer right lane. In case of a 3 lane motorway, in most cases the middle lane has the highest proportion in this data set.}
	\label{fig:laneUsagePerVideo}
	\vspace{-0.2cm}
\end{figure}

Fig. \ref{fig:laneUsagePerVideo} depicts the relative proportion of the load per lane. For 3 lane motorways in most cases the middle lane has the highest proportion in this data set. This indicates a lower to medium traffic flow according to \cite{Busch.1984}. On the right bottom part of Fig. \ref{fig:laneUsagePerVideo} the proportion of the acceleration lane at location number 6 is depicted. The lane load proportions in Fig. \ref{fig:laneUsagePerVideo} are computed by taking into account every time frame recorded. As the acceleration lane is only visible across half of the recorded road segment, the proportion depicted in the figure is lower than the actual one. No extrapolation is performed, since drivers do not tend to make full use of the acceleration lane if the free space on the neighboring lane is sufficient. 

\begin{figure}[h]
\centering
	\includegraphics[width=\linewidth]{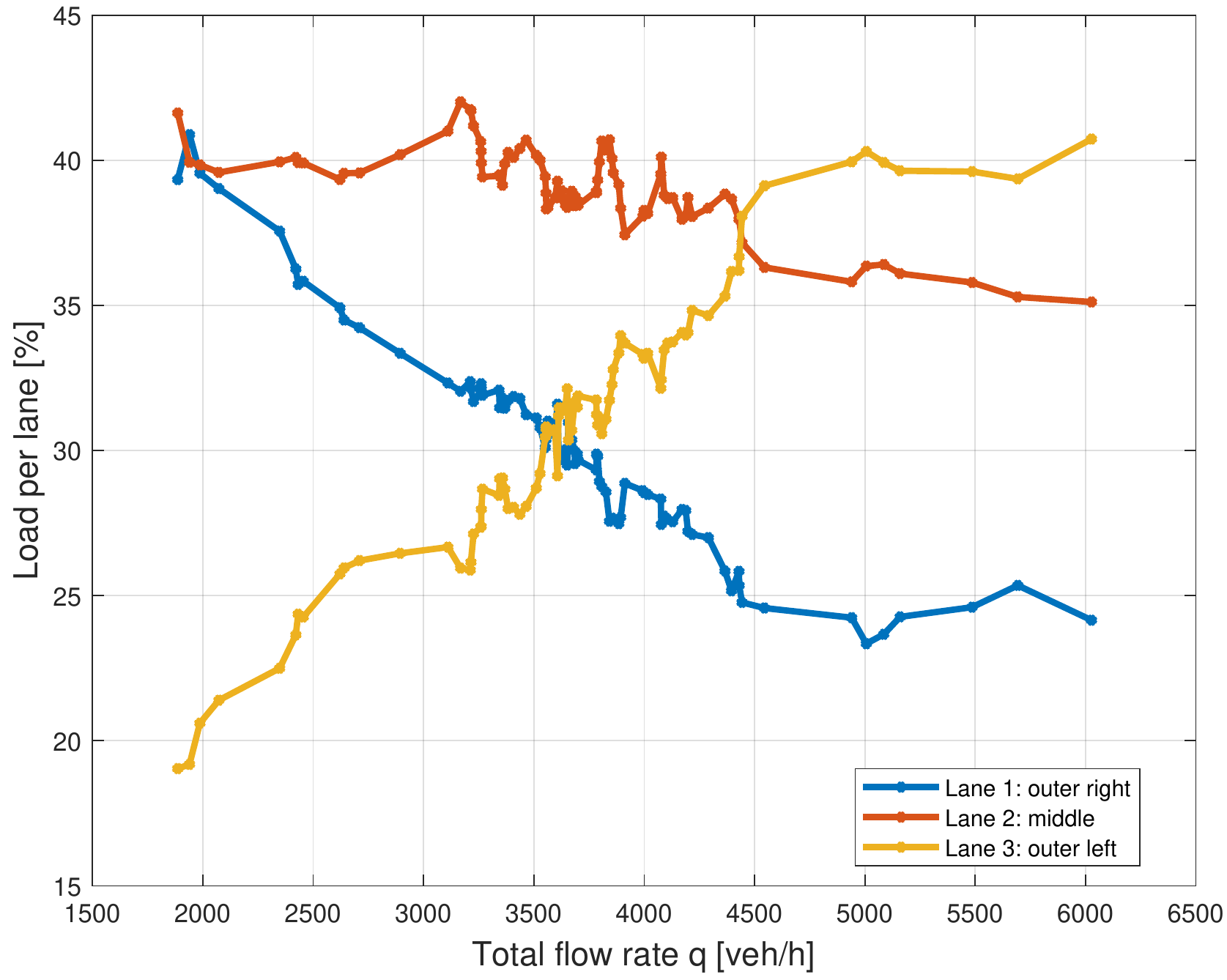}
	\vspace{-0.5cm}
\caption{The relative lane load depending on the averaged traffic flow rate. The curves are processed with a smoothing filter. For this plot only roads with 3 lanes are considered.}
	\label{fig:flowrate_vs_laneusage}
	\vspace{-0.2cm}
\end{figure}

Fig. \ref{fig:flowrate_vs_laneusage} depicts the relative lane load depending on the traffic flow rate of the road. Higher traffic flow rates increase the proportion of vehicles on the outer left lane. Smaller flow rates increase the proportion to the outer right lane. Drivers are legally obliged to drive on the right lane, if not driving at a higher velocity than the leader vehicle, which is more likely to occur during lower flow rates. In order to reach velocities above the truck velocity limits at higher flow rates, drivers switch to the middle and and with further increasing traffic flow more often to the left lane. Only recordings of 3 lane motorways are selected (location number 1, 3, 4). The observations from \cite{Busch.1984} coincide by means of the tendency, that higher flow rates increase the proportion on the left lane, decrease the proportion on the middle lane slightly and a decrease the proportion on the right lane heavily. The actual values for the relative lane load dependent on the flow rate differ from \cite{Busch.1984}. 

\begin{figure}[h]
\centering
	\includegraphics[width=\linewidth]{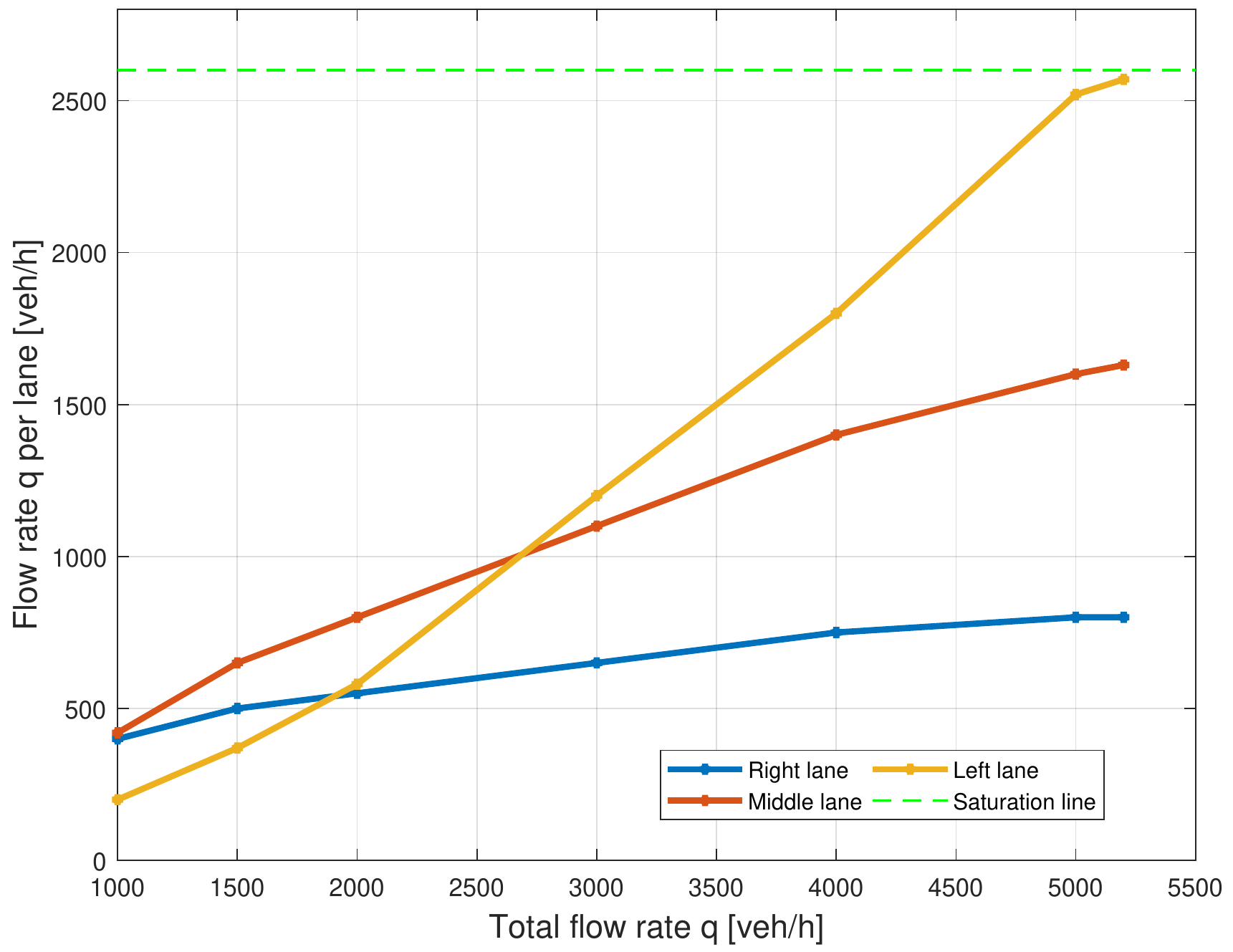}
	\vspace{-0.5cm}
\caption{A schematic comparison of the average load for a 3 lane motorway from \cite{Busch.1984}. The right lane saturates around $q = \unitfrac[800]{veh}{h}$, the left lane around $q = \unitfrac[2600]{veh}{h}$ and short term peaks above.}
	\label{fig:laneload_over_flowrate}
	\vspace{-0.2cm}
\end{figure}

The publication \cite{Busch.1984} proposes flow rates per lane as depicted in Fig. \ref{fig:laneload_over_flowrate}.
The right lane saturates around $q = \unitfrac[800]{veh}{h}$, the left lane around $q = \unitfrac[2600]{veh}{h}$ \cite{Busch.1984, Krau.1998} and short term peaks above.
Note, that Fig. \ref{fig:flowrate_vs_laneusage} depicts the relative load in the highD data set, while  Fig. \ref{fig:laneload_over_flowrate} depicts the absolute flow rates per lane as proposed in \cite{Busch.1984}. 

\subsection{Lane changes}
The number of lane changes is an indicator for the regional criticality on a road segment, since collisions due to lane changes are ranked third in the collision statistics for German motorways, behind high velocities (40\% total, 26\% with personal injury) and short distances (28\% total, 31\% with personal injury) \cite{ACE.2010, BMVI.2018}.
Note that the values stem from two studies and are not directly comparable. They serve only for orientation. The collision values for lane changes are not provided in the referenced literature.
 
\begin{figure}[h]
\centering
	\includegraphics[width=\linewidth]{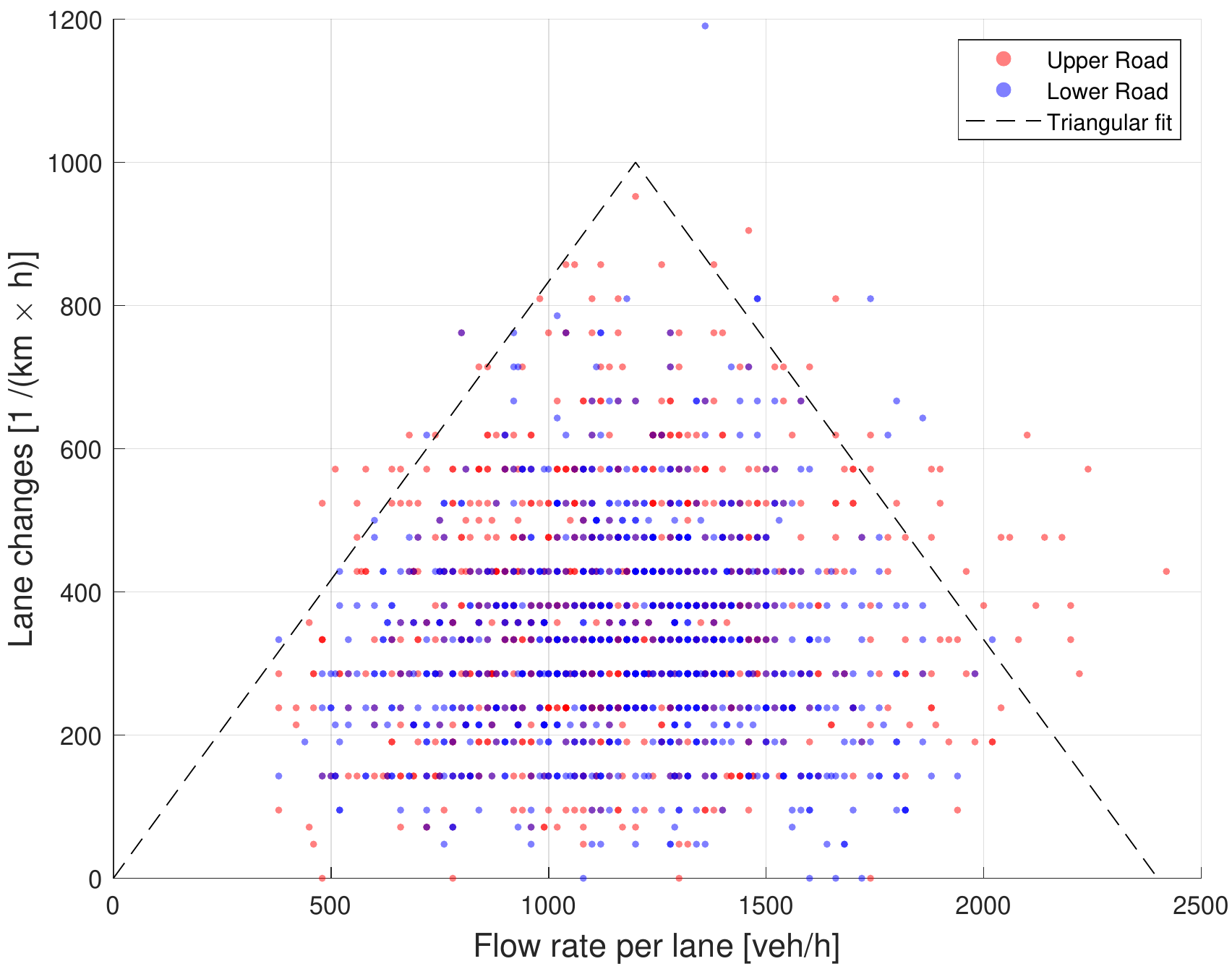}
	\vspace{-0.5cm}
\caption{The number of lane changes per lane, hour and kilometer depending on the averaged traffic flow rate per lane. The lane change rate can be well fitted by a triangular fit (dashed line).}
	\label{fig:flowrate_vs_lnchange}
	\vspace{-0.2cm}
\end{figure}

\begin{figure}[h]
\centering
	\includegraphics[width=\linewidth]{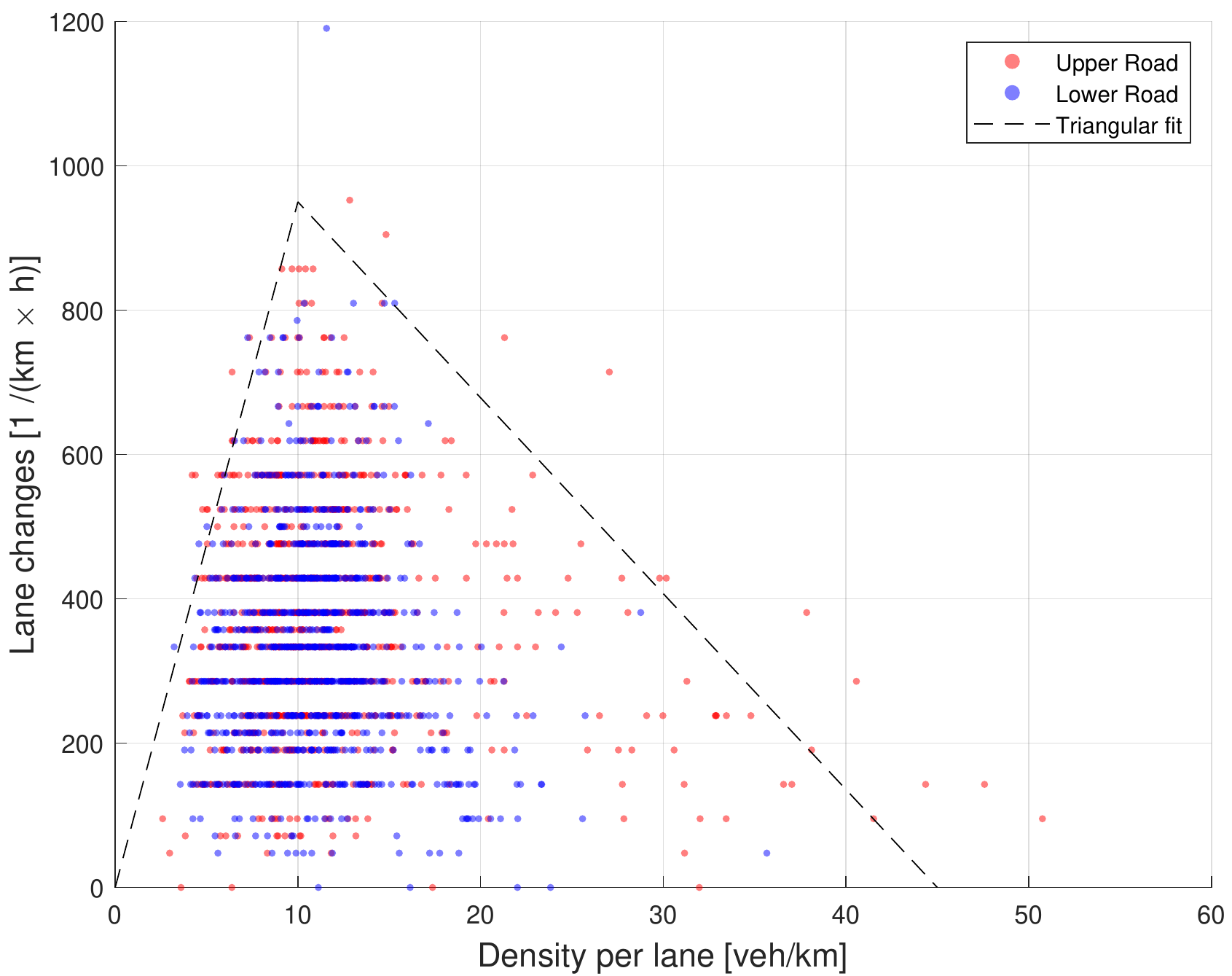}
	\vspace{-0.5cm}
\caption{The number of lane changes per lane, hour and kilometer depending on the traffic density. The dashed line depicts a triangular fit.}
	\label{fig:density_vs_lnchange}
	\vspace{-0.2cm}
\end{figure}

Fig. \ref{fig:flowrate_vs_lnchange} and Fig. \ref{fig:density_vs_lnchange} depict the normalized number of lane changes over the flow rate and traffic density. 
The data points depicted in these two figures are aggregated values of one minute slices.

Publication \cite{Busch.1984} proposes a maximum lane change rate at around 3500 vehicles per hour for a 3 lane motorway. In \cite{Busch.1984} the lane change rate is described by step functions which form a triangular shape. Similar to \cite{Busch.1984}, a triangular fit for the highD data set is depicted in \ref{fig:flowrate_vs_lnchange}. It encloses at least 97\% of all depicted one minute slices for the upper and lower road. The results on the lane change rate coincide with the observation in \cite{Busch.1984}. Regarding the dependency on the traffic density, the lane change rate increases up to around 10-12 vehicles per kilometer and lane \cite{Busch.1984}. Again, a triangular fit is depicted in the figure, which encloses at least 97\% of all one minute slices.

The lane change rate in the highD data set might be biased to some extend, as the overall longitudinal acceleration is shifted slightly to the positive direction. The reason for the acceleration shift might be caused by the recording location, e.\,g. if an intersection is located closely in front of the visible range, see Section \ref{acceleration_subsection} for further details.

Additionally to the overall lane change rate, the publication \cite{Busch.1984} is analyzing the shift of lane changes depending on the flow rate and the relative position of a lane (right, middle, left).
An increase of lane changes between the left and middle lane was observed, if the traffic flow rate increased. A significant change of performed lane changes between the right and middle lane was not observed in \cite{Busch.1984}.

\subsection{Fundamental diagrams}
This subsection depicts the fundamental diagrams on a video based representation with two y axes and additionally in their typical representation with one y axis.

\begin{figure}[h]
\centering
	\includegraphics[width=\linewidth]{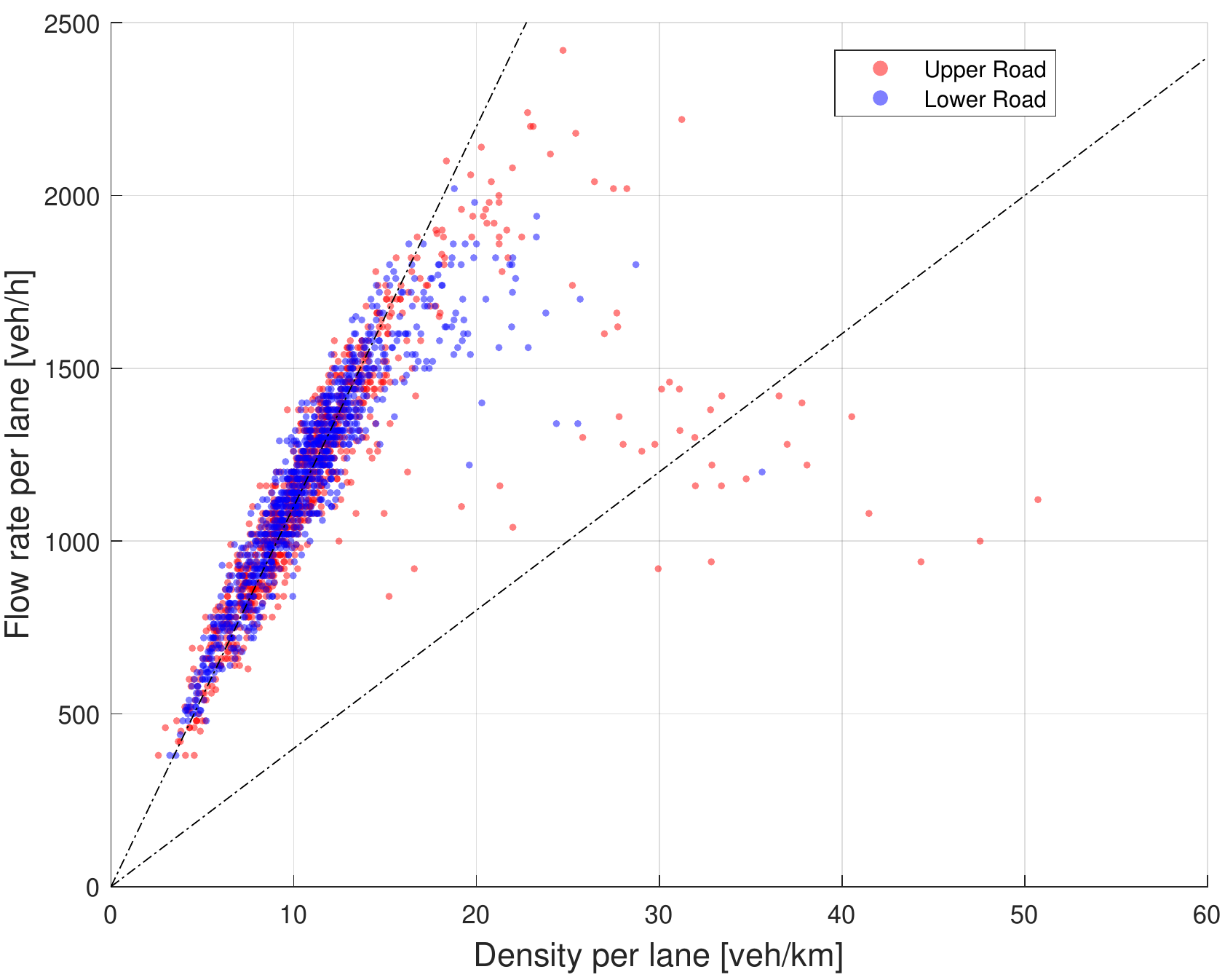}
	\vspace{-0.5cm}
\caption{Traffic density versus flow rate. Up to a certain density the curve can be fitted linearly, which represents the transition from free flow (left side) to bounded flow. The second dashed line roughly depicts a transition from bounded to stop-and-go traffic. Jammed traffic was not recorded within one minute periods.}
	\label{fig:density_vs_flowrate}
	\vspace{-0.2cm}
\end{figure}

First, the terms free flow, bounded flow and congested/jammed traffic are briefly discussed based on \cite{Neubert.2000} and referenced literature. The three types are roughly depicted by the dashed lines in the $\rho - q$ diagram in Fig. \ref{fig:density_vs_flowrate}. 

During free flow traffic, the dependency of the flow rate to the density can be fitted linearly, as the fluctuations are minor. The positive signed slope of the curve represents the average velocity for that traffic type. 

Exceeding a certain density threshold, the average velocity drops significantly. The critical density threshold depends on the drivers and environmental parameters, including the road infrastructure. This type is called bounded or synchronized traffic. Beyond the critical density threshold, the characteristics change. The relative velocities decrease and therefore the rate of lane changes decrease as well. The bounded traffic type can be divided into three classes. First, the homogeneous flow, where density and velocities stay rather constant. This state can be depicted as a point in the fundamental diagram. In the second case, the velocity is homogeneous, but the flow rate and density vary over time and space. This state can be depicted as a line with negative gradient in the fundamental diagram. In the third and most frequently appearing class for bounded traffic, all three parameters vary, while vehicles still keep a certain velocity. Considering time series data, the consecutive datapoints of this inhomogeneous traffic stream class jump stochastically in the fundamental diagram.
Obviously, a further increase of the traffic density yields a stop-and-go situation followed by the jammed traffic with zero velocity.

\begin{figure}[h]
\centering
	\includegraphics[width=\linewidth]{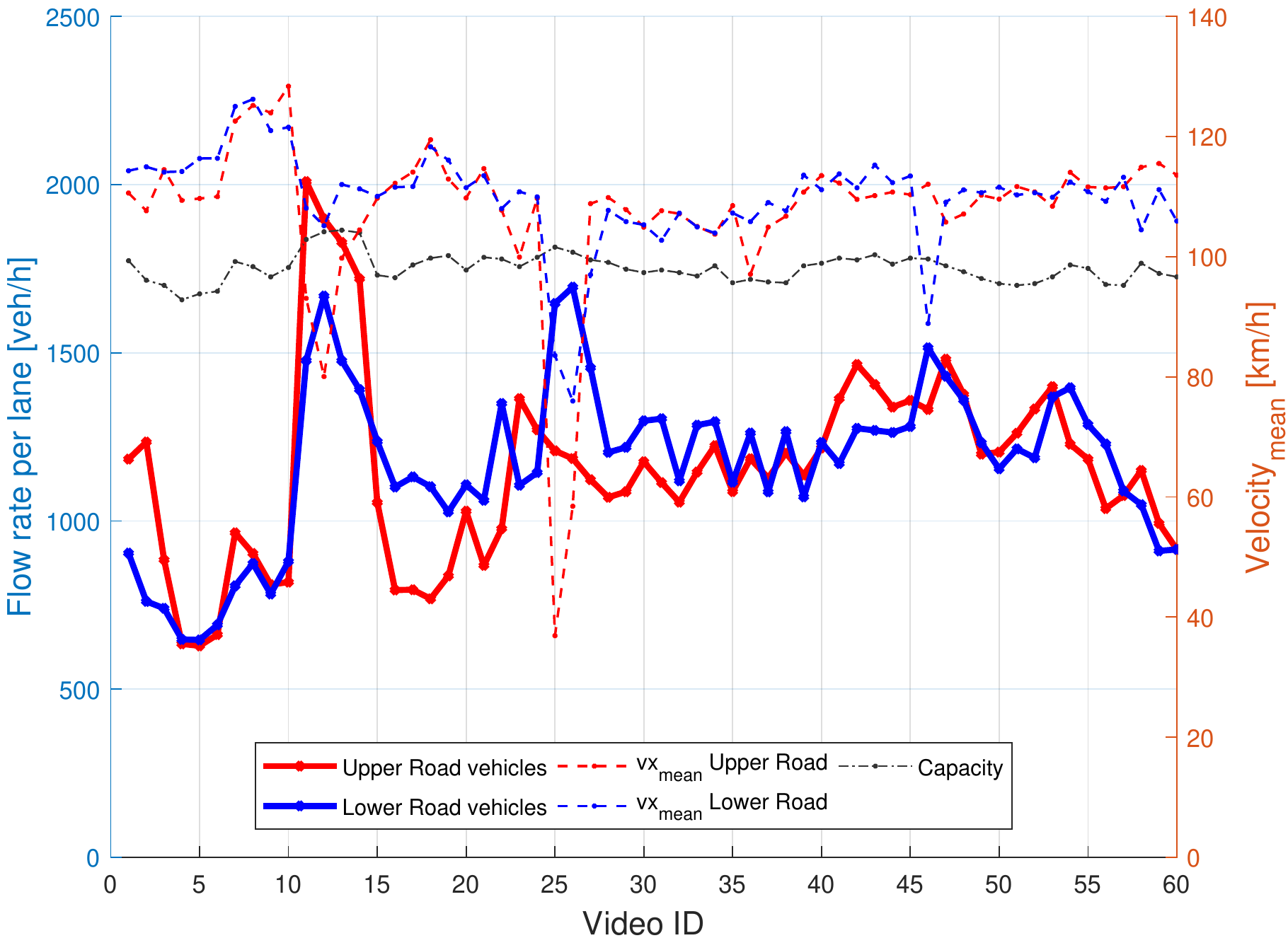}
	\vspace{-0.5cm}
\caption{A comparison of the average velocity (dashed line) versus the flow rate per lane (solid line). In six recordings the flow rate reached a value around the capacity limit according to \cite{FGSV.2015}, resulting in a velocity drop.}
	\label{fig:velocity_vs_vehPerLaneAndHour}
	\vspace{-0.2cm}
\end{figure}

Now, the video based diagrams are discussed.
Fig. \ref{fig:velocity_vs_vehPerLaneAndHour} compares the average velocity versus the flow rate $q$ per lane for each video. Note, that the variables are averaged here per video recording. In six recordings the number of vehicles reached a value close or above to the average capacity limit per lane, according to the definition of \cite{FGSV.2015} (gray line). The capacity limit includes the proportion of heavy load vehicles in each video recording. 

A lane wise consideration (right, middle, left) is neglected in the guideline \cite{FGSV.2015} and here as well. The values are interpolated from the definition of \cite{FGSV.2015} and adapted for each video. As long as the traffic flow is stable with approximetily $v \geq \unitfrac[40]{km}{h}$ in this data set, the relationship between the capacity threshold and average velocity is given, for example in the videos 12 and 13. If the traffic flow is unstable, as in video 25 (upper road), where $v \leq \unitfrac[40]{km}{h}$, the flow rate is far below the capacity line. The guideline \cite{FGSV.2015} proposes a velocity drop to $v \approx \unitfrac[100]{km}{h}$ with approximately 1500 vehicles per hour and lane and $v \approx \unitfrac[80]{km}{h}$ with approximately 1700 vehicles per hour an lane. These numbers assume an average heavy load vehicle proportion of 20\% and a road inclination of less than 2\%.

\begin{figure}[h]
\centering
	\includegraphics[width=\linewidth]{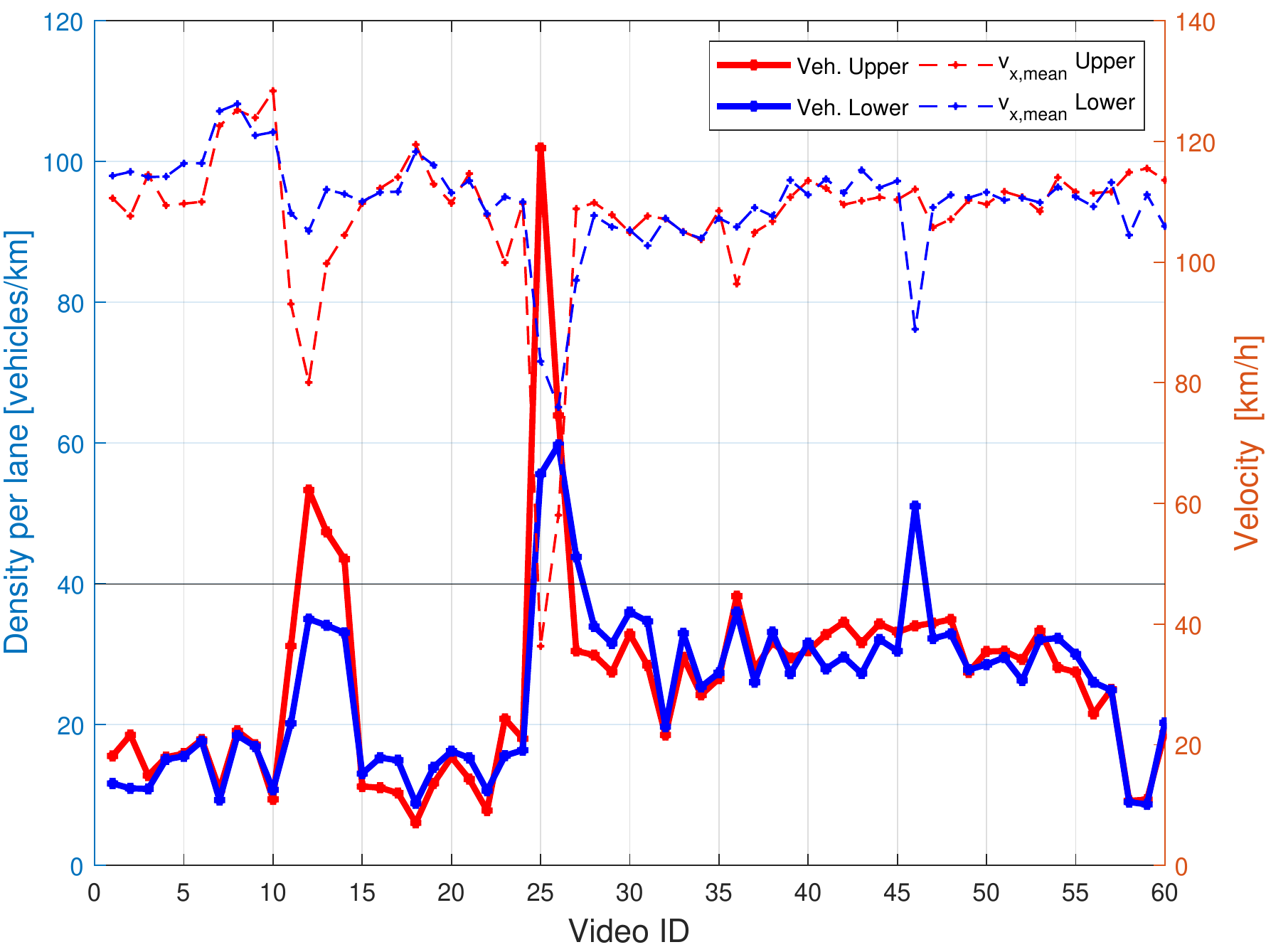}
	\vspace{-0.5cm}
\caption{A comparison of the average velocity (dashed line) versus the traffic density per lane (solid line). A traffic density of around 40 or more vehicles per lane and kilometer, averaged per video, corresponds to a significant drop in the average velocity (black line).}
	\label{fig:VehiclePerKMandLane}
	\vspace{-0.2cm}
\end{figure}

Fig. \ref{fig:VehiclePerKMandLane} depicts the velocity versus the density for each video. A traffic density of around 40 or more vehicles per lane and kilometer corresponds to a significant drop in the average velocity. The correlation coefficient for the upper road is $R = -0.84$, for the lower road $R = -0.74$.

The following part of this subsection depicts the classical fundamental diagrams based on the highD data set.
The relationship between the density and flow rate is depicted in Fig. \ref{fig:density_vs_flowrate}. For free flow traffic up to around $\rho = \unitfrac[10-15]{veh}{km}$, the curve can be fitted linearly. The correlation coefficient for $\rho < \unitfrac[15]{veh}{km}$ is $R = 0.94$. The zone for bounded traffic up to stop-and-go traffic is depicted within the two dashed lines. The transition from stop-and-go to jammed traffic arises with a density of approximately $\rho = \unitfrac[30-40]{veh}{km}$, while saturating at around $\rho = \unitfrac[45-50]{veh}{km}$ per lane \cite{Busch.1984, Neubert.2000}. The plot of the highD data set coincides with the results of the empirical studies in \cite{Busch.1984, Neubert.2000}. In \cite{Neubert.2000}, the curve shapes are provided for each lane separately as well. As mentioned above, the outer left lane has a higher capacity, yielding a different curve shape. 

\begin{figure}[h]
\centering
	\includegraphics[width=\linewidth]{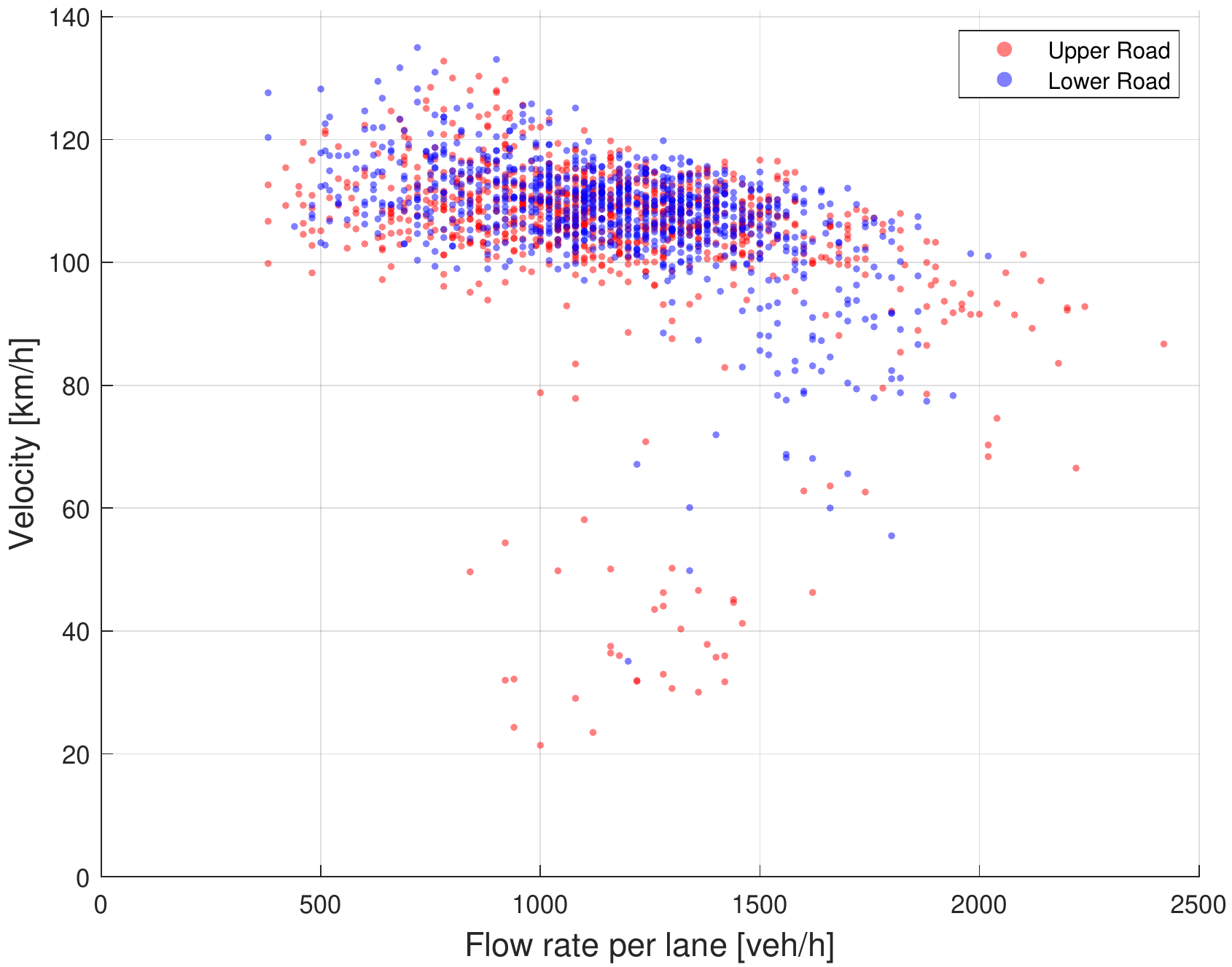}
	\vspace{-0.5cm}
\caption{Flow rate versus velocity. Two different velocity ranges can be found at the same flow rate, which leads to the classification of stable and unstable traffic flow.}
	\label{fig:flowrate_vs_velocity}
	\vspace{-0.2cm}
\end{figure}

Fig. \ref{fig:flowrate_vs_velocity} depicts the average flow rate over the velocity. Two different velocity ranges can be found at the same flow rate, which leads to the classification of stable and unstable traffic flow.

\begin{figure}[h]
\centering
	\includegraphics[width=\linewidth]{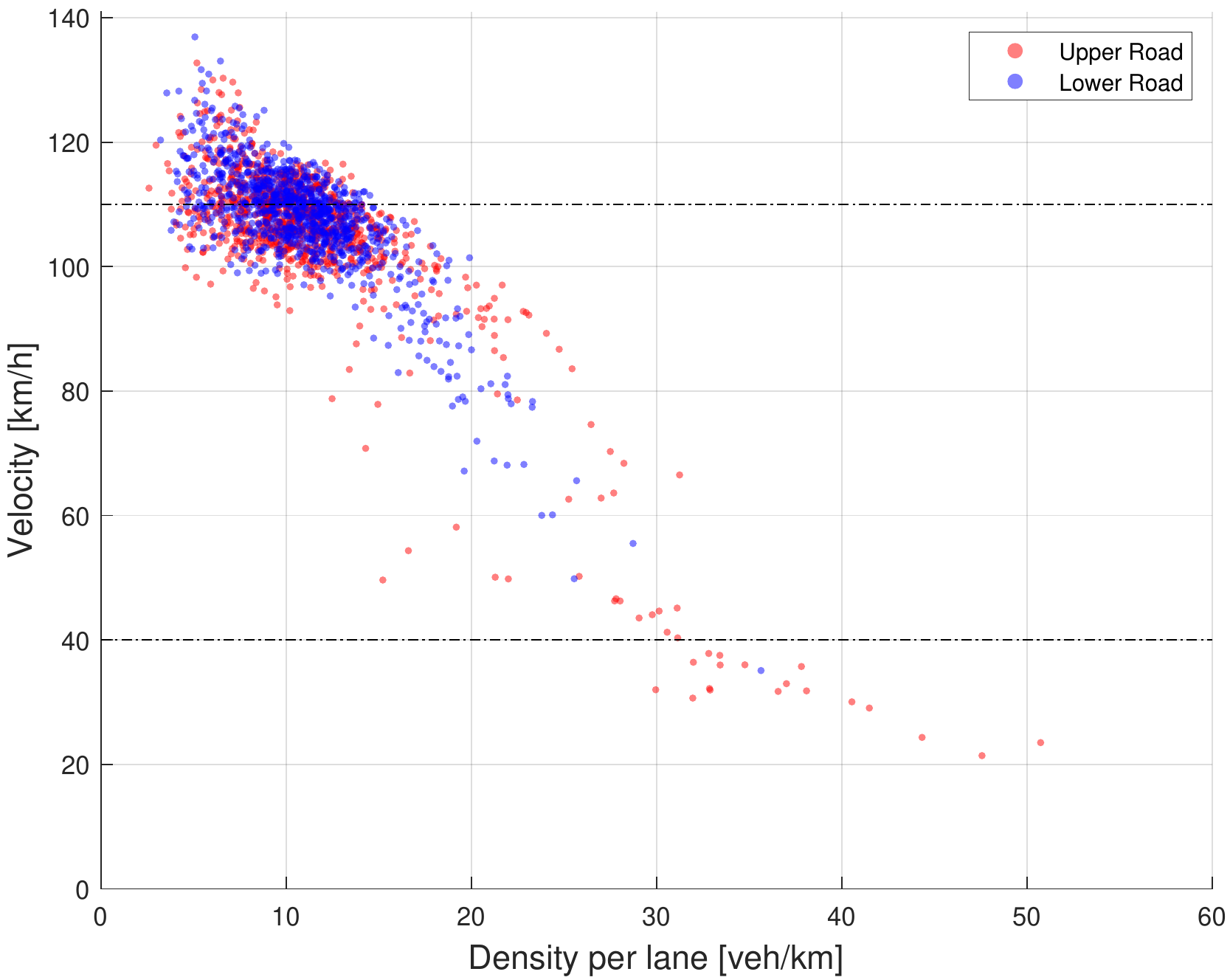}
	\vspace{-0.5cm}
\caption{Density versus velocity. The two dashed line roughly indicate the transitions from free-flow to bounded flow and finally to stop-and-go traffic.}
	\label{fig:density_vs_velocity}
	\vspace{-0.2cm}
\end{figure}

Fig. \ref{fig:density_vs_velocity} depicts the velocity over the traffic density. The graph is approximately divided into the three traffic types (dashed lines). The average velocity decreases slightly up to a certain density threshold. Above the threshold the velocity drops significantly. Most of the recorded data is located in the transition zone of free flow to bounded traffic, while still retaining a velocity of $v \geq \unitfrac[100]{km}{h}$. 

\begin{figure}[h]
\centering
	\includegraphics[width=\linewidth]{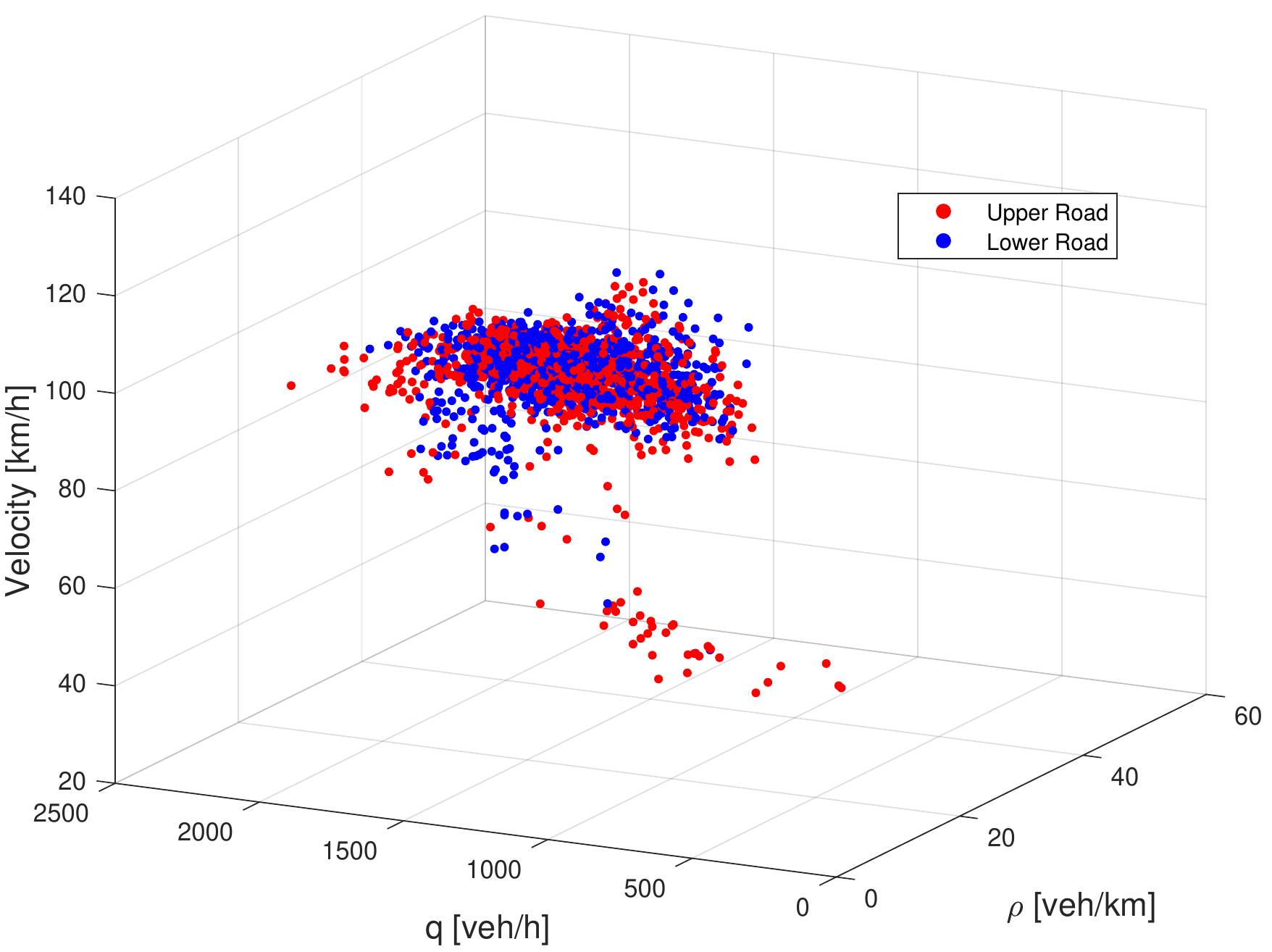}
	\vspace{-0.5cm}
\caption{All three fundamental variables in one graph. High densities lead to low velocities and flow rates. Most of the recorded traffic is in the region of free-flow or intermediate flow rates.}
	\label{fig:density_vs_flowrate_vs_velocity}
	\vspace{-0.2cm}
\end{figure}

Finally, Fig. \ref{fig:density_vs_flowrate_vs_velocity} depicts a three-dimensional plot, which combines all three variables. It depicts the traffic stream with velocities of $v \geq \unitfrac[100]{km}{h}$ for most of the recording time and increasing flow rates up to the critical density point. Exceeding the density threshold leads to lower velocities and flow rates.

\section{Microscopic analysis}
\label{microscopic_section}
Microscopic variables consider each vehicle separately and are then aggregated in this work.
\subsection{Velocity distributions}
\begin{figure}[h]
\centering
	\includegraphics[width=\linewidth]{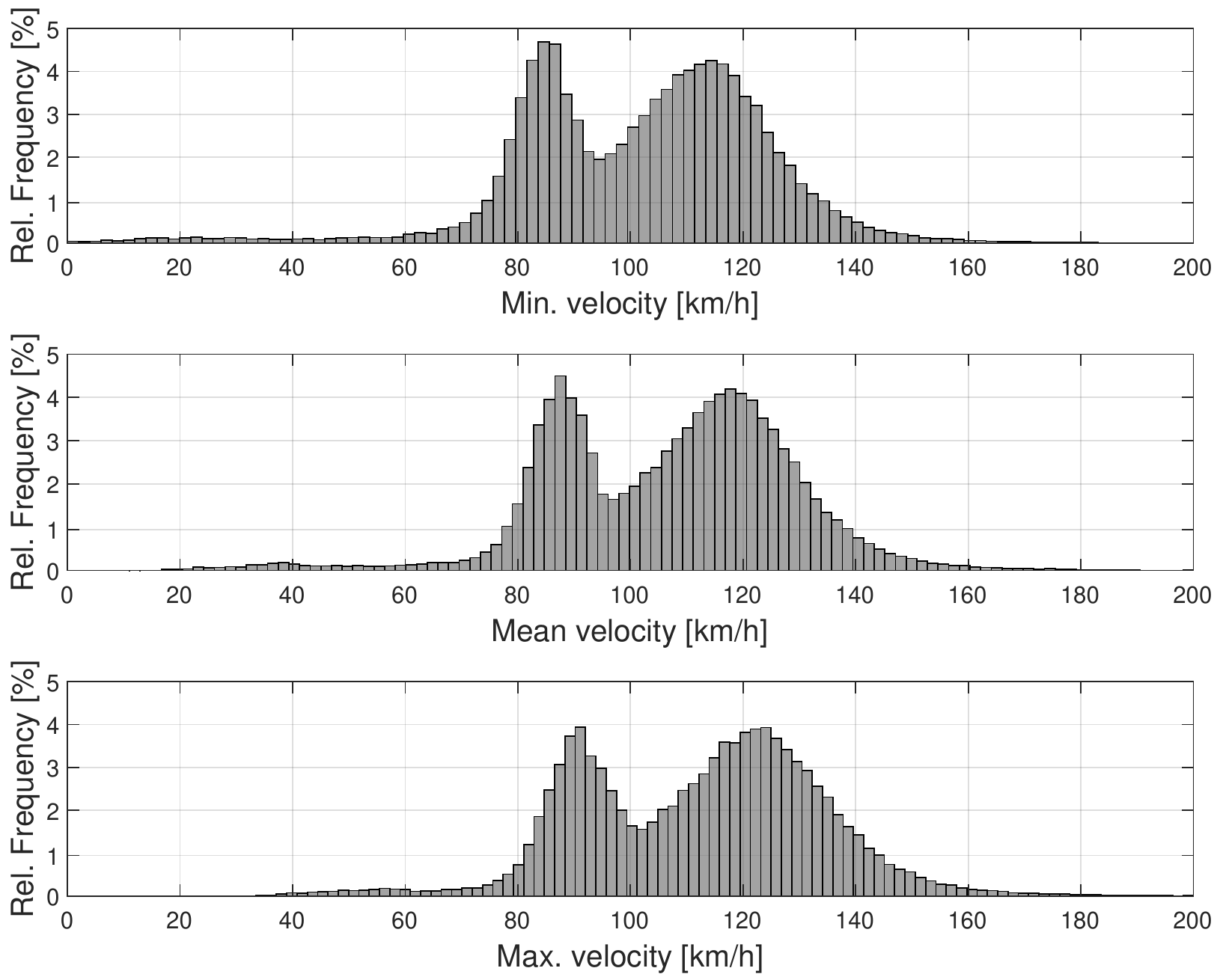}
	\vspace{-0.5cm}
	\caption{The distribution of the minimum, mean and maximum velocity over all tracks. The two prominent peaks are around $v \approx \unitfrac[90]{km}{h}$ and $v \approx \unitfrac[120]{km}{h}$. The left peak contains a contribution from cars and trucks.}
	\label{fig:vel_hist}
	\vspace{-0.2cm}
\end{figure}

The velocity distribution is depicted in Fig. \ref{fig:vel_hist}. The two prominent peaks are around $v \approx \unitfrac[90]{km}{h}$ and $v \approx \unitfrac[120]{km}{h}$. The left peak contains a contribution from cars and trucks. Slow average velocities of $v \leq \unitfrac[60]{km}{h}$ are recorded in 2.3\% of all tracks, and $v \leq \unitfrac[80]{km}{h}$ in 6.4\%. In total 22 vehicle reached a mean velocity of $v \geq\ \unitfrac[200]{km}{h}$, the highest registered velocity is $v = \unitfrac[246]{km}{h}$. 

\begin{figure}[h]
\centering
	\includegraphics[width=\linewidth]{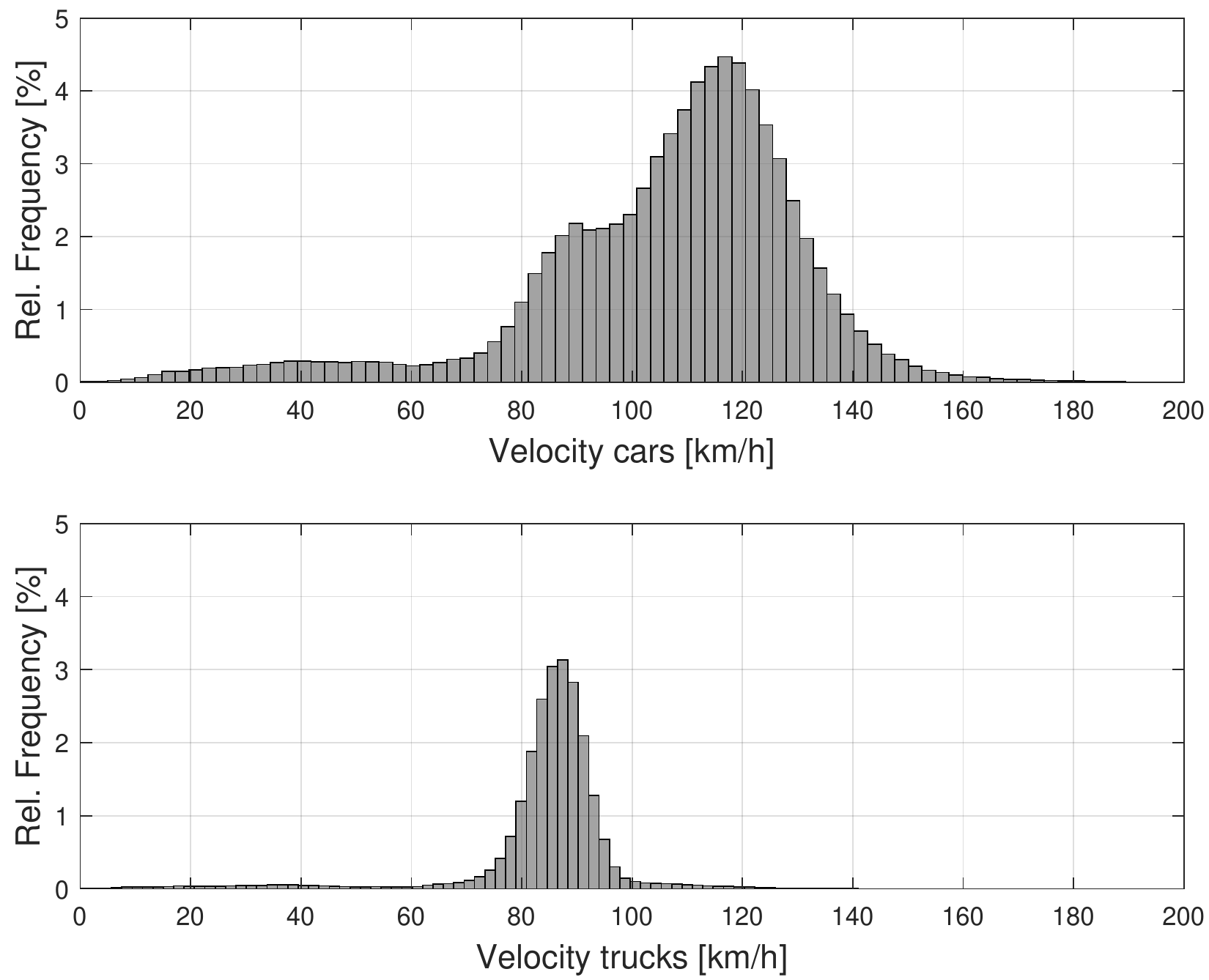}
	\vspace{-0.5cm}
	\caption{The distribution of the velocity for cars and trucks over all recorded time frames. The peak around $v \approx \unitfrac[90]{km}{h}$ depicted in Fig. \ref{fig:vel_hist} contains a contribution from both vehicle classes.}
	\label{fig:vel_histogram_cars_trucks}
	\vspace{-0.2cm}
\end{figure}

Fig. \ref{fig:vel_histogram_cars_trucks} depicts the velocity distribution over all time frames separated for cars and trucks. Here one can see, that the peak around $v \approx \unitfrac[90]{km}{h}$ in Fig. \ref{fig:vel_hist} contains a contribution from both vehicle classes.

\begin{figure}[h]
\centering
	\includegraphics[width=\linewidth]{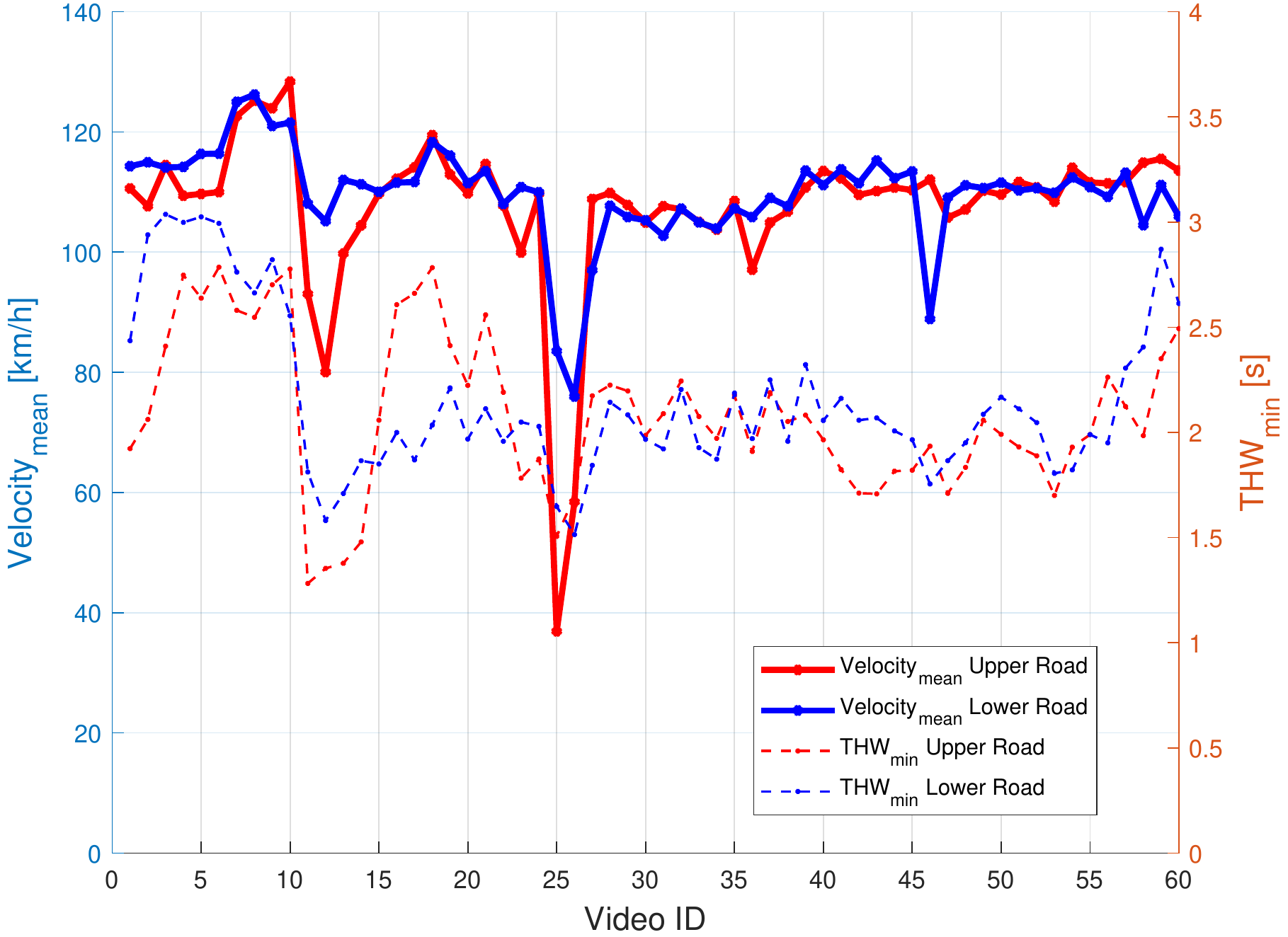}
	\vspace{-0.5cm}
\caption{The average velocity versus the average $\mathrm{THW}_{\mathrm{min}}$. A clear relationship is not recognizable without taking other variables into consideration.}
	\label{fig:velocity_vs_THW}
	\vspace{-0.2cm}
\end{figure}

\begin{figure}[h]
\centering
	\includegraphics[width=\linewidth]{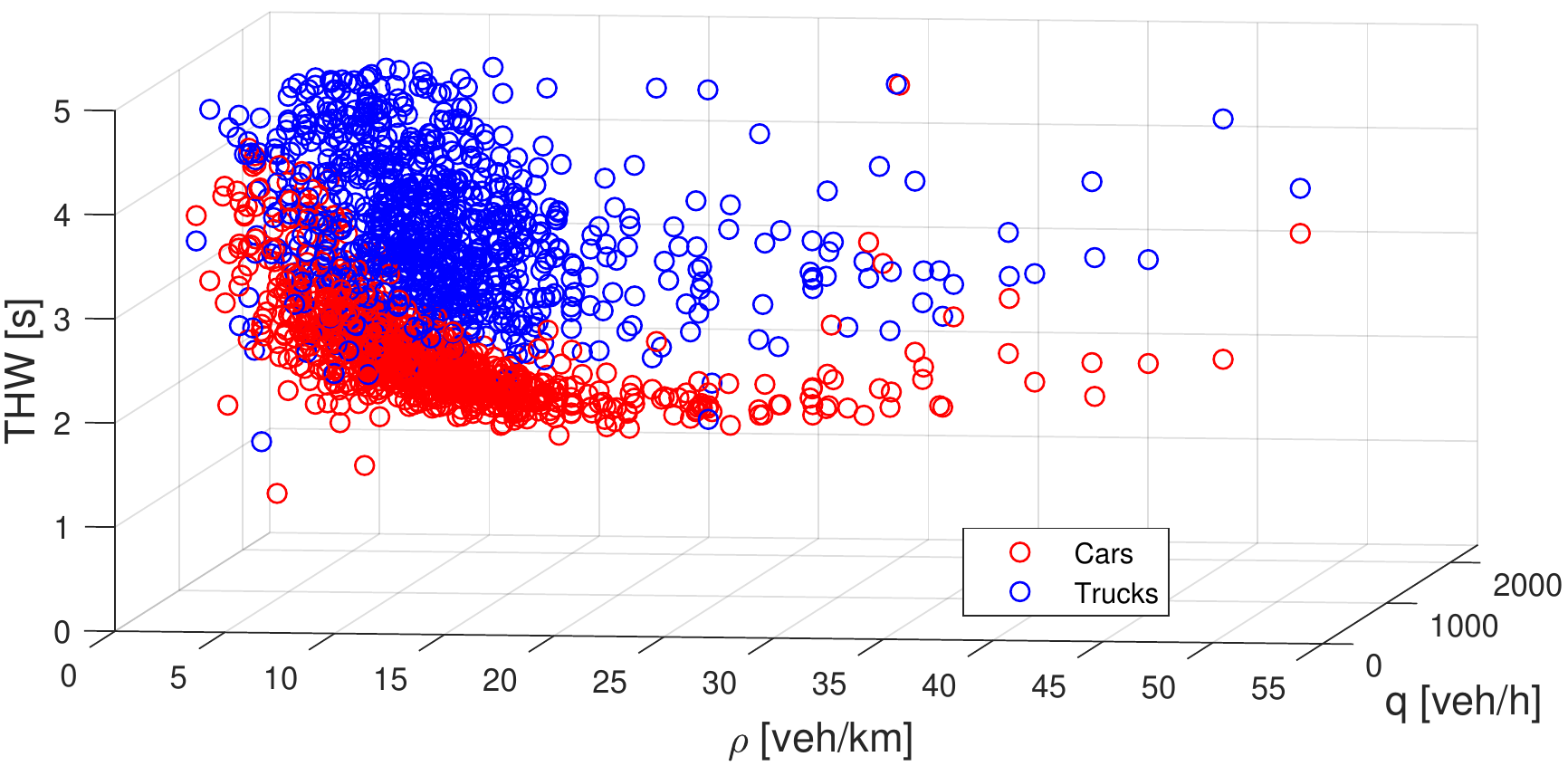}
	\vspace{-0.5cm}
\caption{The traffic density, flow rate and the average THW over all recorded time frames, aggregated to one minute slices.}
	\label{fig:density_vs_flowrate_vs_THW}
	\vspace{-0.2cm}
\end{figure}

Fig. \ref{fig:velocity_vs_THW} depicts the average velocity versus the average $\mathrm{THW}_{\mathrm{min}}$. A clear relationship is not recognizable. While in some cases a drop in the average velocity also decreases the $\mathrm{THW}_{\mathrm{min}}$, this can not be seen in all cases, e.\,g. compare video 13 and 26 of the lower road (blue). Other variables have to be additionally considered. Therefore, the average velocity is replaced by the density and flow rate in Fig. \ref{fig:density_vs_flowrate_vs_THW}. Additionally the truck rate is an important variable to be considered as depicted in Fig. \ref{fig:density_vs_flowrate_vs_THW}.

The average velocity depends on the flow rate and traffic density. The average THW depends to some extend on the proportion of trucks. Within the data set, the overall proportion of trucks is around 23\%. The proportion of trucks reaching a $\mathrm{THW}_{\mathrm{min}} \leq \unit[0.5]{s}$ is only 7.9\% compared to all THW values recorded with this threshold. The proportion of trucks increases with higher THW thresholds. For example $\mathrm{THW}_{\mathrm{min}} \leq \unit[0.9]{s}$ yields a rate of 9.6\%. This leads to the conclusion, that trucks generally keep larger THW than cars or perform overtake maneuvers less often.

Figure \ref{fig:density_vs_flowrate_vs_THW} depicts the relationship between the flow rate, density and THW from all available time frames, aggregated to one minute slices. 
Overall, the THW is considerable lower for cars than for trucks. The datapoints for cars show a clear tendency towards smaller THW with an increasing density and flow rate. On the other hand, the THW for trucks have a larger spread. The THW values reach a bottom of \unit[1.2]{s} around the transition zone between bounded traffic and stop-and-go traffic ($\rho \approx \unitfrac[30]{veh}{km}$).
From there on a tendency towards larger THW values can be recognized. Drivers keep in average a certain minimum distance of some meters to their leader vehicle during stop-and-go traffic. This yields larger averaged THW values due to the low velocity. Nevertheless, small $\mathrm{THW}_{\mathrm{min}}$ values are reached often as shown in Section \ref{results_section}. This is due to the oscillating distances and velocities during stop-and-go traffic.  

\subsection{Acceleration distribution}
\label{acceleration_subsection}
In the following, a considerable active braking maneuver is assumed to be at $a_x \leq \unitfrac[-1.5]{m}{s^2}$ for typical highway velocities. During stop-and-go situations, applying the brake pedal causes smaller negative accelerations, which can be plausibilized with the Video 25 for the upper road. These kind of casual braking maneuvers at low velocities are not considered to be relevant in this work.

\begin{figure}[h]
\centering
	\includegraphics[width=.99\linewidth]{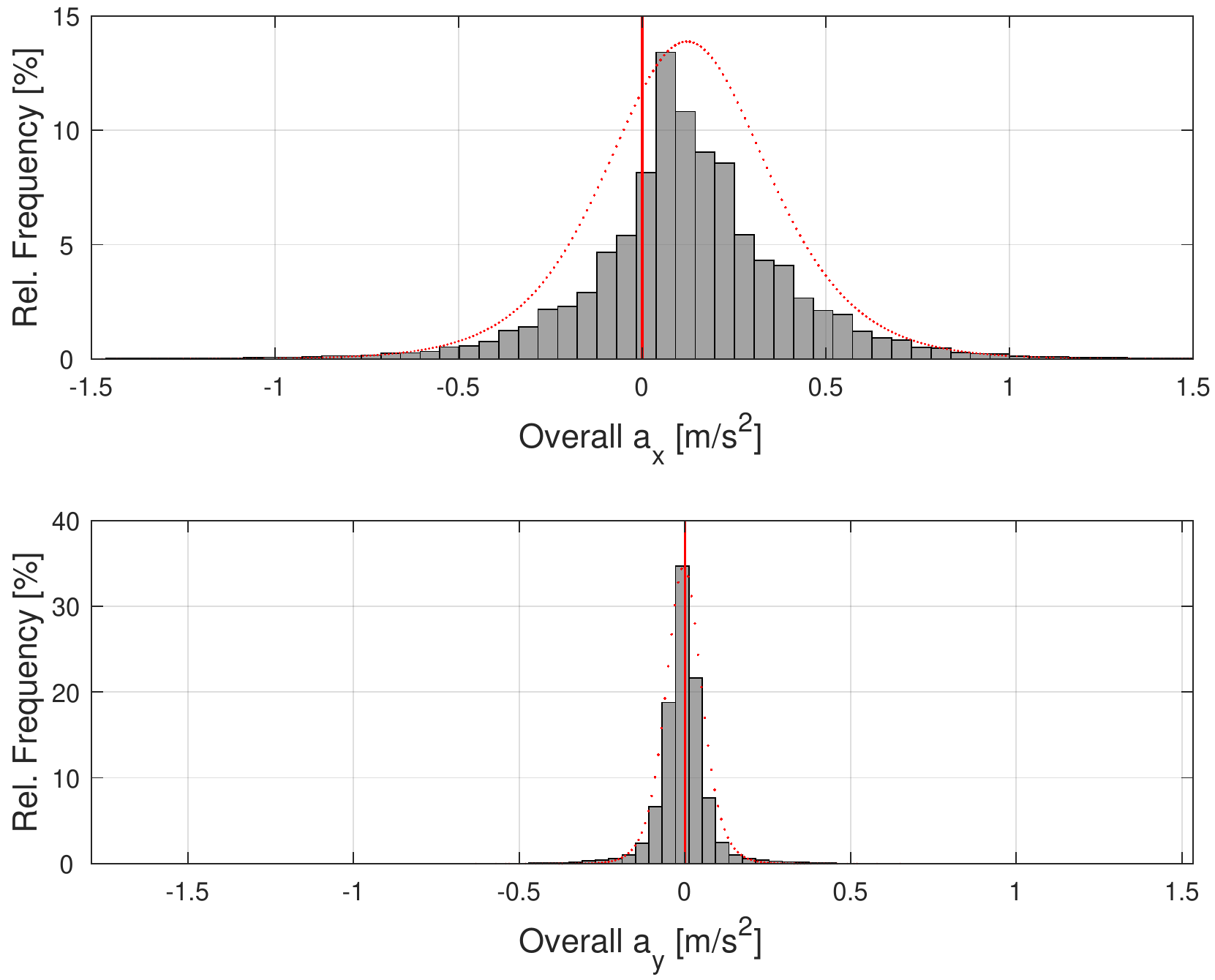}
	\vspace{-0.5cm}
	\caption{The overall distribution of the longitudinal and lateral acceleration over all recorded time frames.}
	\label{fig:ax_ay_all_histogram}
	\vspace{-0.2cm}
\end{figure}

Fig. \ref{fig:ax_ay_all_histogram} depicts the distribution of the longitudinal and lateral acceleration for all recorded time frames. The depicted logistic probability density functions (pdf) are $a_x \sim Logistic(0.122, 0.147)$ and $a_y \sim Logistic(-0.003, 0.035)$. 

The longitudinal acceleration has a mean offset of $a_{x,o} = \unitfrac[0.12]{m}{s^2}$ to the positive direction. The data from all provided videos has a positive offset in average. The strongest offset with $a_{x,o} \geq \unitfrac[0.27]{m}{s^2}$ can be found in Video 39-45. These videos were recorded on a Monday morning at location number 1. 

For a large data set, one would expect an average acceleration of zero. It is unclear, whether the offset appears due to the road infrastructure, e.\,g. ramps just in front of the visible region, or other reasons. 

\begin{figure}[h]
\centering
	\includegraphics[width=.99\linewidth]{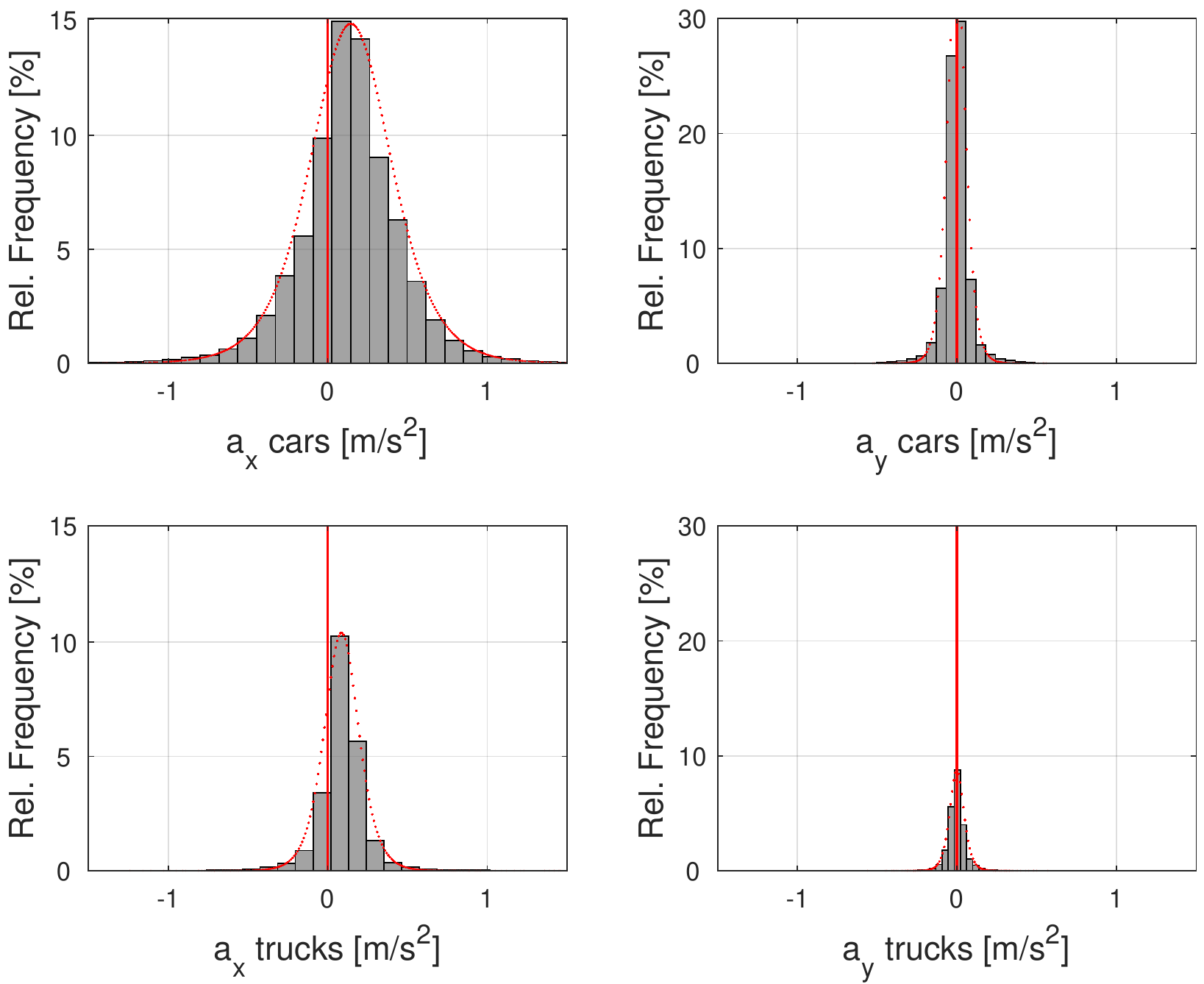}
	\vspace{-0.5cm}
	\caption{The overall distribution of the longitudinal and lateral acceleration for all recorded time frames separately for each vehicle class.}
	\label{fig:acc_x_y_histogram_cars_trucks}
	\vspace{-0.2cm}
\end{figure}

The acceleration distribution is additionally depicted for both vehicle classes over all available time frames in Fig. \ref{fig:acc_x_y_histogram_cars_trucks}.
The depicted logistic pdf are $a_{x_\mathrm{car}} \sim Logistic(0.137, 0.167)$, $a_{x_\mathrm{truck}} \sim Logistic(0.088, 0.073)$, $a_{y_\mathrm{car}} \sim Logistic(-0.003, 0.037)$ and $a_{y_\mathrm{truck}} \sim Logistic(-0.002, 0.029)$.  

\begin{figure}[h]
\centering
	\includegraphics[width=.99\linewidth]{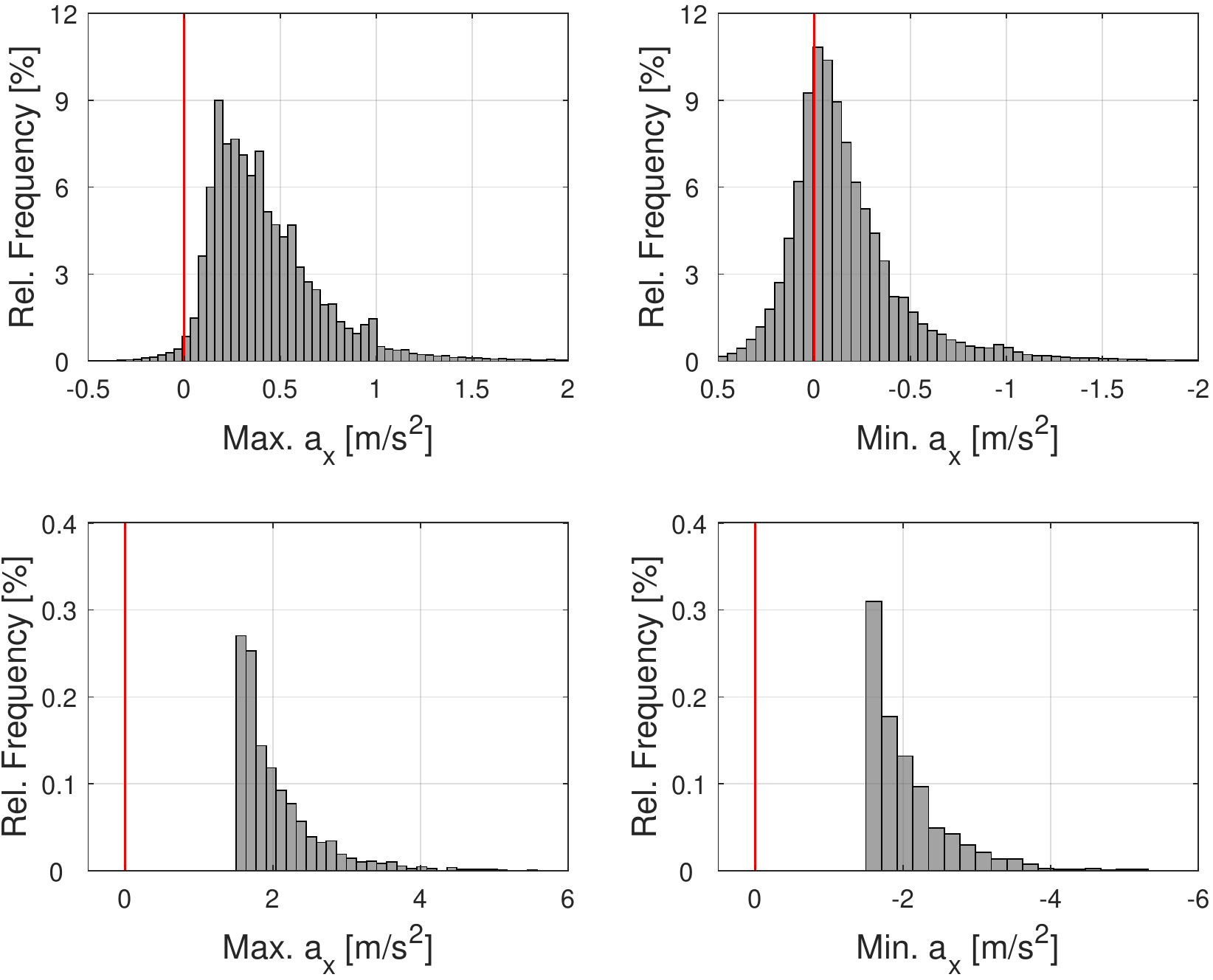}
	\vspace{-0.5cm}
	\caption{The distribution of the minimum and maximum longitudinal acceleration over all tracks. The bottom sub-figures depict higher accelerations of $a_x > \unitfrac[1.5]{m}{s^2}$ and stronger deceleration of $a_x < \unitfrac[-1.5]{m}{s^2}$.}
	\label{fig:accx_hist}
	\vspace{-0.2cm}
\end{figure}

Fig. \ref{fig:accx_hist} depicts the distribution of the minimum and maximum longitudinal acceleration over all tracks. The bottom sub-figures depict higher accelerations of $a_x > \unitfrac[1.5]{m}{s^2}$ and deceleration of $a_x < \unitfrac[-1.5]{m}{s^2}$. Around 29.5\% of all vehicles do not decelerate, while 1.7\% of all vehicle do not accelerate in the field of view. Only 0.9\% of all vehicles perform an active braking maneuver. Reducing the threshold to $a_x < \unitfrac[-1.0]{m}{s^2}$ yields a active braking proportion of 2.7\%.
The maximum positive longitudinal acceleration recorded is $a_{x_{\mathrm{max}}} = \unitfrac[5.6]{m}{s^2}$, the minimum is $a_{x_{\mathrm{min}}} = \unitfrac[-6.3]{m}{s^2}$. Strong deceleration of $a_x < \unitfrac[-5.0]{m}{s^2}$ appears in only 5 recorded tracks, a deceleration of $a_x < \unitfrac[-4.0]{m}{s^2}$ in 15 tracks. If one takes into account the fact that the weather was clear at the recordings, no vehicle delayed close to maximum. 

\begin{figure}[h]
\centering
	\includegraphics[width=.99\linewidth]{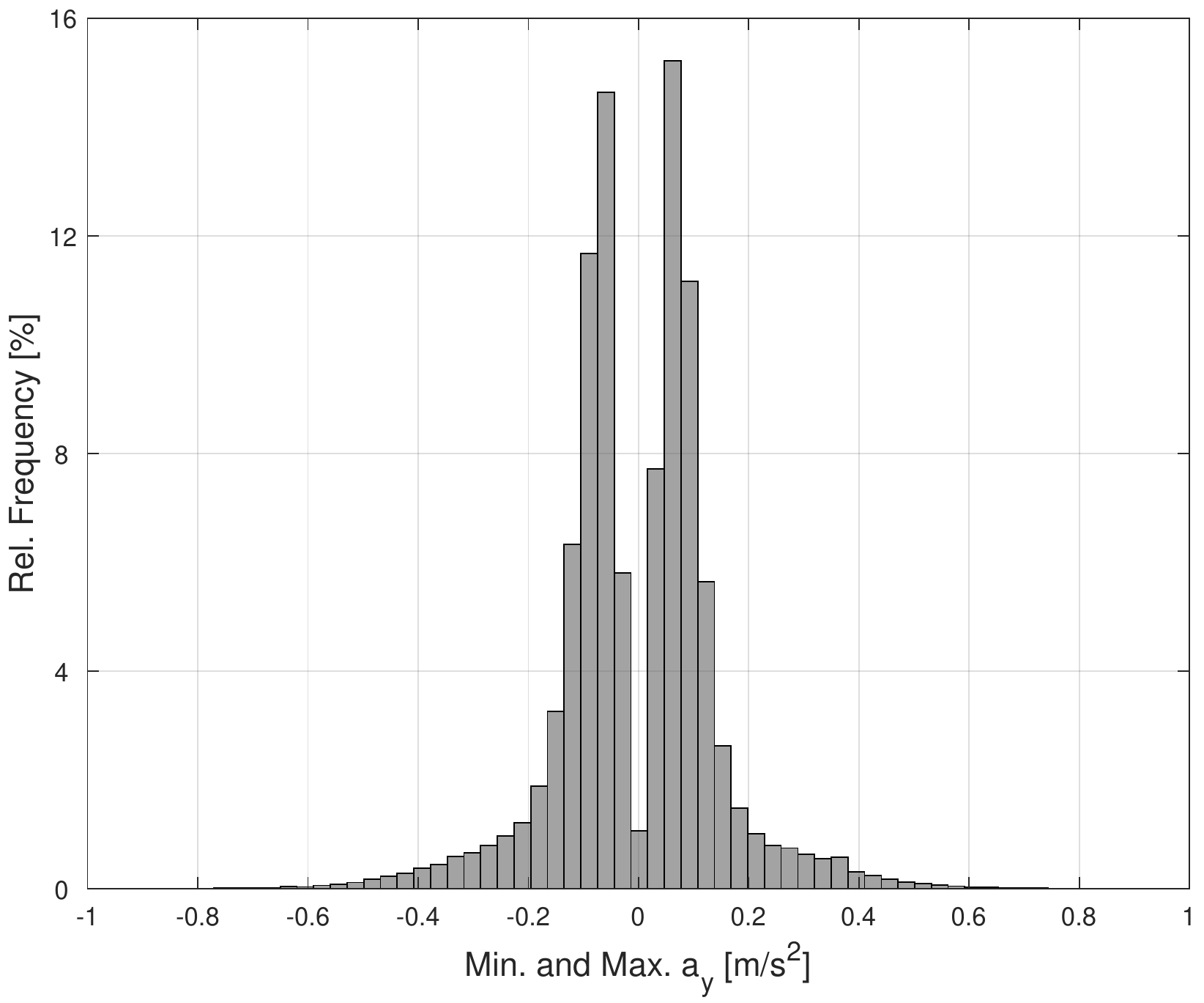}
	\vspace{-0.5cm}
	\caption{The distribution of the minimum and maximum lateral acceleration over all tracks}
	\label{fig:accy_hist}
	\vspace{-0.2cm}
\end{figure}

The distribution of the maximum and minimum lateral acceleration over all tracks is depicted in Fig. \ref{fig:accy_hist}. The maximum recorded lateral acceleration is $a_{y_{\mathrm{max}}} = \unitfrac[|1.6|]{m}{s^2}$.

\begin{figure}[h]
\centering
	\includegraphics[width=.99\linewidth]{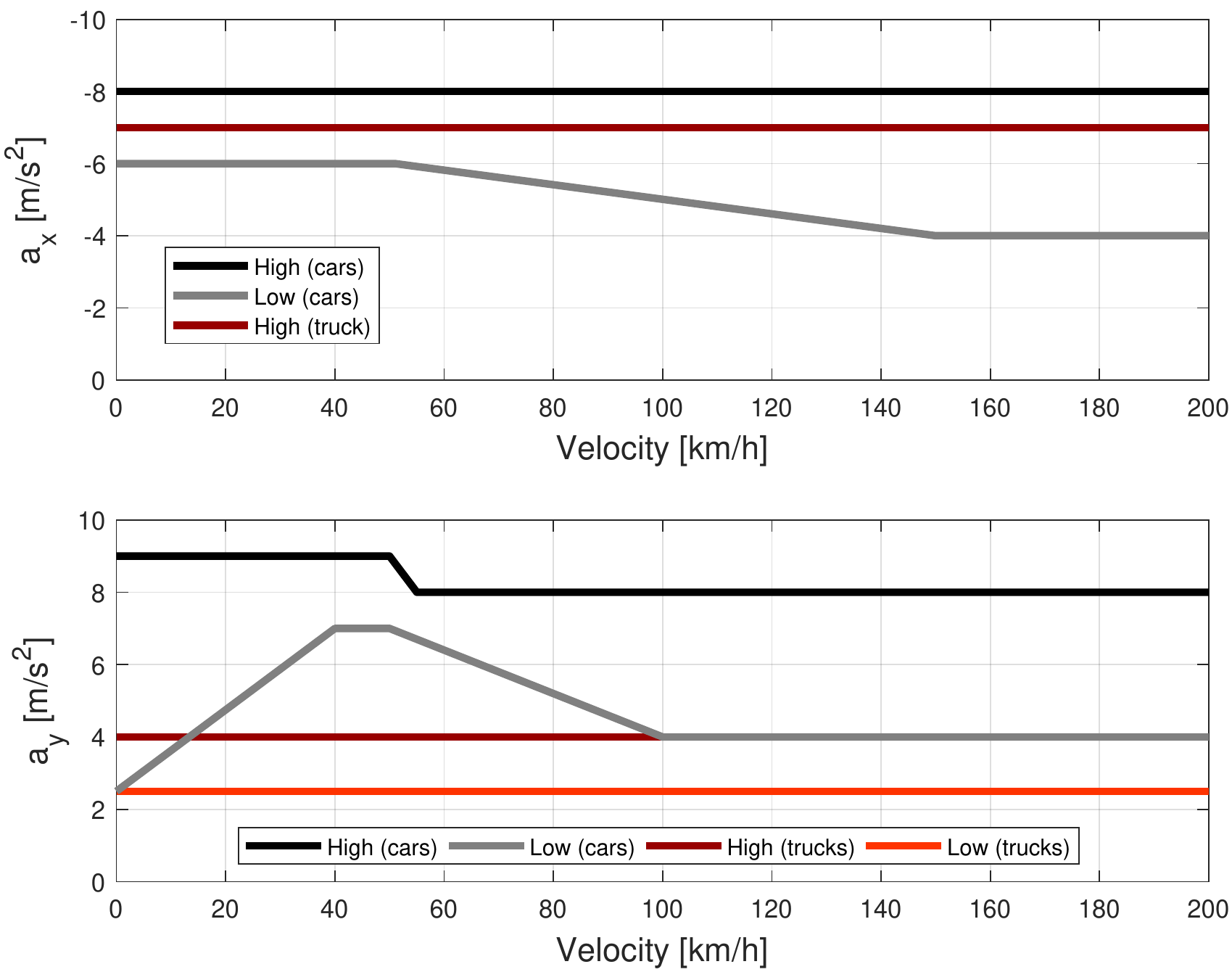}
	\vspace{-0.5cm}
	\caption{A two-step risk level definition (high and low) for longitudinal and lateral acceleration from \cite{Benmimoun.2012}.}
	\label{fig:ATZ_Beschleunigung_Threshold}
	\vspace{-0.2cm}
\end{figure}

The maximum and minimum acceleration values mentioned above can be compared to the two-step risk definition for accelerations from \cite{Benmimoun.2012} depicted in Fig. \ref{fig:ATZ_Beschleunigung_Threshold}. The low level threshold for longitudinal acceleration is reached in very few tracks. None of the tracks exceed a risk level for the lateral acceleration. According to this definition, no highly critical scenarios appeared due to accelerations.  

\subsection{TTC, THW and DHW distribution}
This section depicts the graphs for the distribution of $\mathrm{TTC}_{\mathrm{min}}$, $\mathrm{THW}_{\mathrm{min}}$ and $\mathrm{DHW}_{\mathrm{min}}$. Non-relevant data, e.\,g. very high TTC values of $\mathrm{TTC} \geq \unit[200]{s}$, is neglected in the figures for a better legibility. The share of neglected tracks is specified in the corresponding figure caption. A detailed analysis on the THW and TTC for capturing critical scenarios follows in Section \ref{results_section}.

\begin{figure}[h]
\centering
	\includegraphics[width=\linewidth]{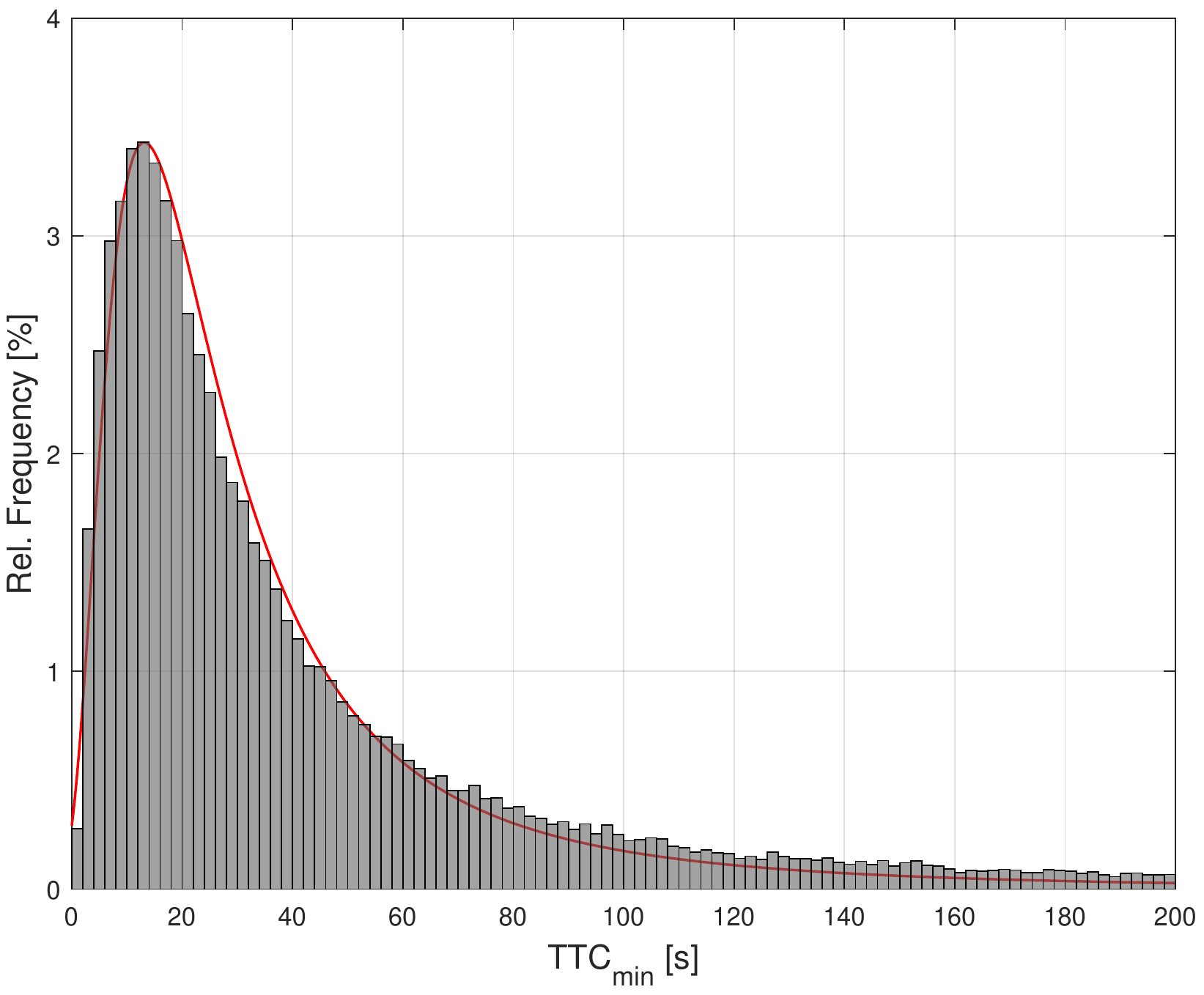}
	\vspace{-0.5cm}
	\caption{The distribution of the $\mathrm{TTC}_{\mathrm{min}}$ over all tracks. Values of $\mathrm{TTC} \geq \unit[200]{s}$, which account for 5.9\% of all tracks, are neglected. Negative TTC values are also not depicted (26\% of all tracks).}
	\label{fig:ttc_hist}
	\vspace{-0.2cm}
\end{figure}

In Fig. \ref{fig:ttc_hist} the distribution of all $\mathrm{TTC}_{\mathrm{min}}$ with a positive magnitude is given. The depicted pdf of the Generalized Extreme Value distribtion (GEV) is $\mathrm{TTC}_{\mathrm{min}} \sim GEV(19, 16, 0.5)$. The peak is around $\mathrm{TTC} \approx \unit[12]{s}$. 

The TTC histogram in \cite{Chen.2016}, derived from the 100-cars-study data, has a peak at around $\mathrm{TTC} = \unit[8]{s}$ throughout the recorded data. Contrary to the highD data set, the 100-cars-study was performed on different road classes, located in the United States. Publication \cite{Chen.2016} states, that the TTC values increase with higher velocities.

According to \cite{Gietelink.2002}, drivers typically start to apply braking around $\mathrm{TTC} = \unit[6]{s}$. That is analyzed for the highD data set by setting a threshold of at least $\unit[4]{s}$ tailgate driving after $\mathrm{TTC} = \unit[6.0\pm0.5]{s}$. In total 36.4\% of 1785 vehicles in this scenario had a acceleration of $a_x < \unitfrac[0]{m}{s^2}$, while 3.8\% had a acceleration of $a_x < \unitfrac[-1.5]{m}{s^2}$. The mean value of active braking scenarios is $a_x = \unitfrac[-2.2]{m}{s^2}$, the overall mean value of scenarios with negative is $a_x = \unitfrac[-0.9]{m}{s^2}$.

\begin{figure}[h]
\centering
	\includegraphics[width=\linewidth]{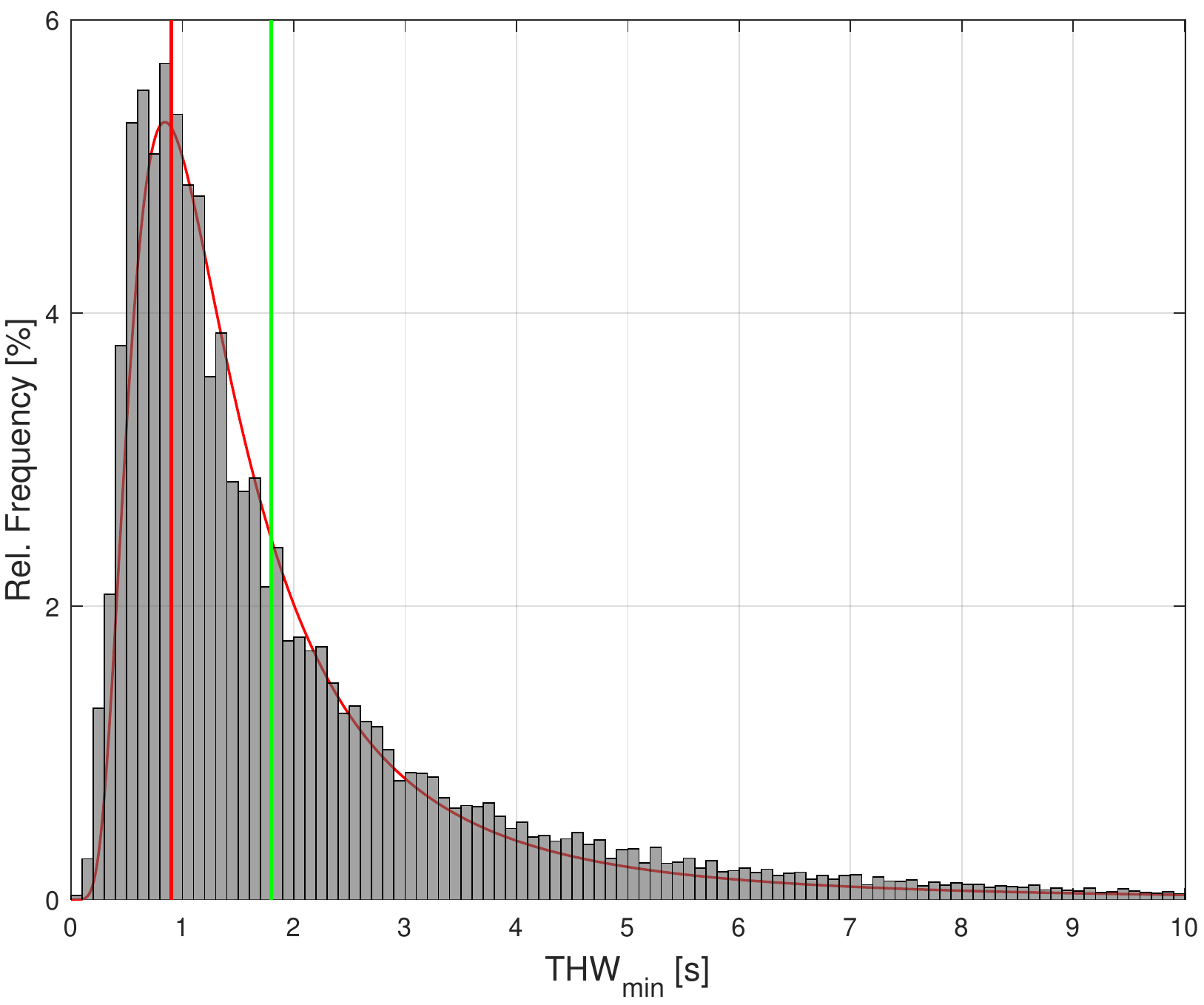}
	\vspace{-0.5cm}
	\caption{The distribution of the $\mathrm{THW}_{\mathrm{min}}$ over all tracks. Values of $\mathrm{THW} \geq \unit[10]{s}$, which account for 0.8\% of all tracks, are neglected. The green line indicates the juristic definition of safe time headway, the red line the endangering distance according to \cite{Filzek.2002, Allianz.2019}.}
	\label{fig:thw_hist}
	\vspace{-0.2cm}
\end{figure}

Fig. \ref{fig:thw_hist} depicts the $\mathrm{THW}_{\mathrm{min}}$ distribution. The depicted pdf is $\mathrm{THW}_{\mathrm{min}} \sim GEV(1.1, 0.7, 0.5)$. The green line indicates the juristic safe distance ($\mathrm{THW} \leq \unit[1.5]{s}$). The red line ($\mathrm{THW} \leq \unit[0.9]{s}$) indicates the threshold for an endangering distance in the juristic definition according to \cite{Filzek.2002, Allianz.2019}. If the THW value falls below these thresholds due to a cut-in maneuver of a target vehicle, the affected follower vehicle has to reach a safe time gap within few seconds again \footnote{Trucks and busses with a permissible total mass of above $\unit[3.5]{t}$ have to keep a distance of at least $\mathrm{DHW}=\unit[50]{m}$ for velocities above $\unitfrac[50]{km}{h}$ on German motorways. In inner city traffic the THW must be generally $\mathrm{THW} \geq \unit[1]{s}$ for all vehicles. In case of a cut-in maneuver from another vehicle in urban areas, the distance has to be reached again after $\unit[3]{s}$. \cite{Allianz.2019,STVO.2019}}. 
The proportion of vehicles, which undercut the green line for at least one time step is 56\%. The proportion of vehicles, which undercut the red line for at least one time step is 31\%.

\begin{figure}[h]
\centering
	\includegraphics[width=\linewidth]{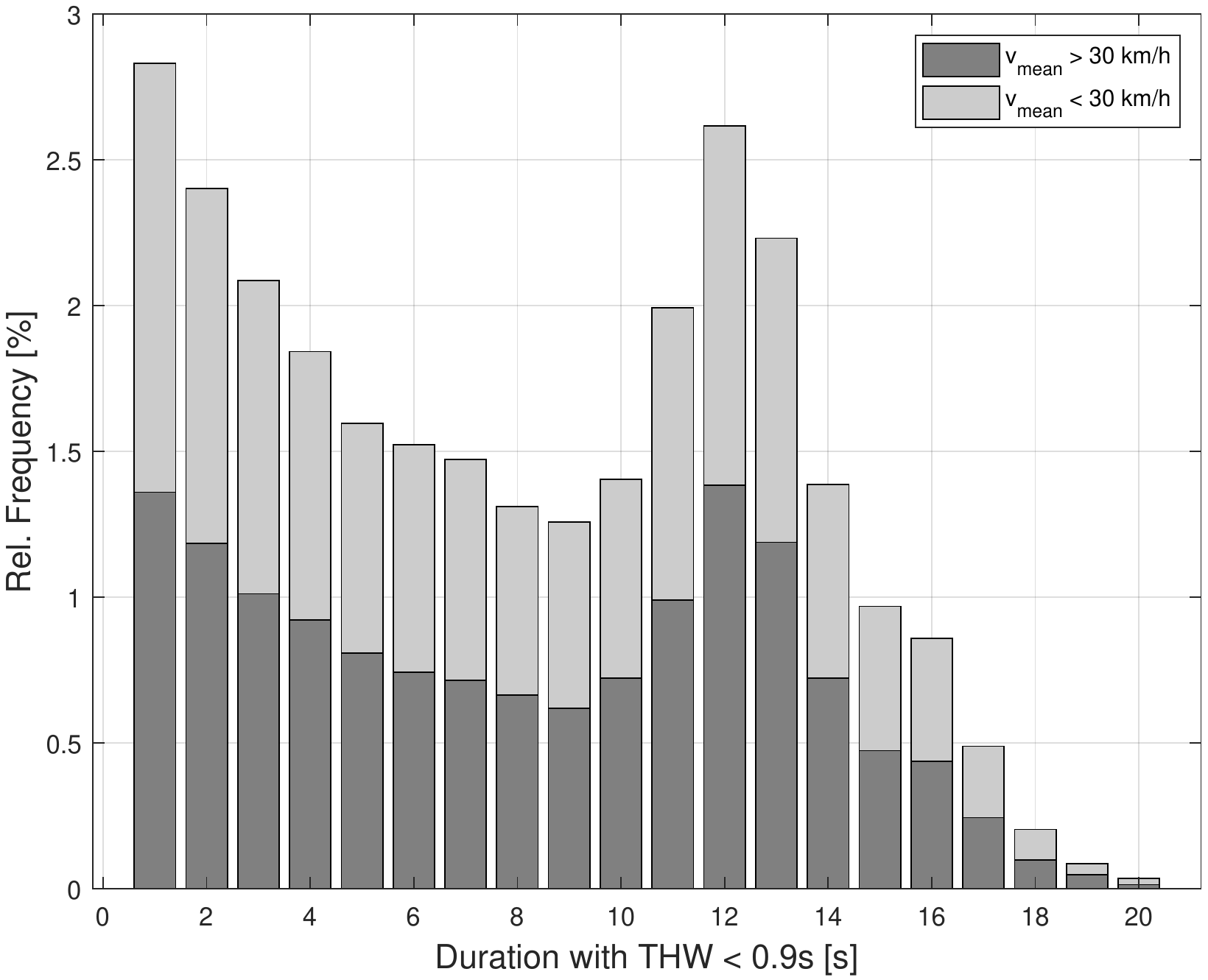}
	\vspace{-0.5cm}
	\caption{Duration when vehicles undercut $\mathrm{THW} \leq \unit[0.9]{s}$. Values of $\mathrm{THW} \geq \unit[5]{s}$ are excluded, as they are not considered to be car-following situations.}
	\label{fig:thwpunish_hist}
	\vspace{-0.2cm}
\end{figure}

Since the relative amount of vehicles exceeding the legal threshold for at least one time frame is surprisingly high, additionally all time frames over all tracks are evaluated. For this purpose, only car-following situations are observed. The transition from car-following to free-flow driving is assumed for $THW \geq \unit[5]{s}$. In \cite{Bosch.1999}, the distance for a free-flow driving is estimated with $\mathrm{DHW} \geq \unit[150]{m}$ for motorways in Germany. Still, the proportion of vehicles, which undercut $\mathrm{THW} \leq \unit[0.9]{s}$ with an average velocity $v \geq \unitfrac[30]{km}{h}$ for at least one second is 12.6\%. Around 8.6\% undercut the threshold for more than $\unit[5]{s}$ within the visible range. Taking all velocities into consideration, approximately doubles the rate. One recognizes accordingly the relatively higher portion of low THW values at low velocities. The relative occurrences for both velocity ranges are depicted in Fig. \ref{fig:thwpunish_hist}.

In a field test carried out on German motorways \cite{Filzek.2002}, the endangering legal threshold was undercut in 41\% of the time during tailgate driving. The test was carried out with 24 drivers and a total driving time of approximately 9 hours. 
As shown in Section \ref{microscopic_section}, the THW values correlate with the flow rate. Therefore, these numbers have to be interpreted with regard to the macroscopic basis of the data.

\begin{figure}[h]
\centering
	\includegraphics[width=\linewidth]{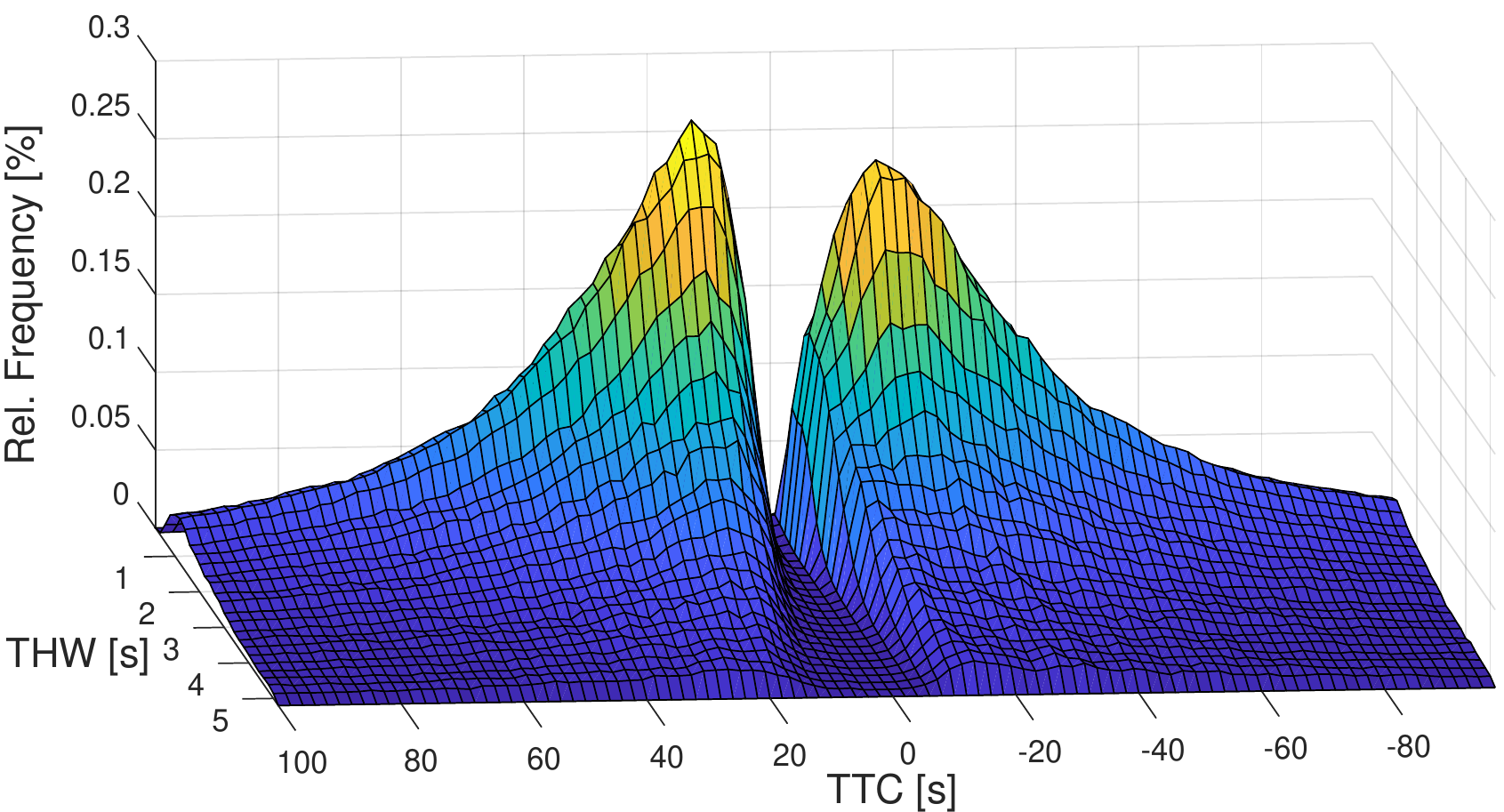}
	\vspace{-0.5cm}
	\caption{Combined THW and TTC distribution for each recorded time frame, which fulfills the boundary restrictions of $\mathrm{THW} \leq \unit[5]{s}$ on one axis and $ \mathrm{TTC} \geq \unit[-100]{s} \land \mathrm{TTC} \leq \unit[100]{s}$ on the second axis for better legibility.}
	\label{fig:TTCandTHW_all}
	\vspace{-0.2cm}
\end{figure}

Fig. \ref{fig:TTCandTHW_all} depicts the combined THW and TTC distribution for each time frame within defined boundaries. The boundaries are defined with $\mathrm{THW} \leq \unit[5]{s}$ on the first axis and $ \mathrm{TTC} \leq \unit[|100|]{s}$ on the second axis. The total proportion of the considered time frames in this figure corresponds to 55\% of the data, or 246 hours of driving, respectively. Obviously, the successive time frames within a vehicle track are correlated. A valley for small TTC values $\mathrm{TTC} \leq \unit[|4|]{s}$, as well as two prominent peaks at $\mathrm{TTC} \approx \unit[|15|]{s} \land \mathrm{THW} \approx \unit[1]{s}$ can be recognized.  

\begin{figure}[h]
\centering
	\includegraphics[width=\linewidth]{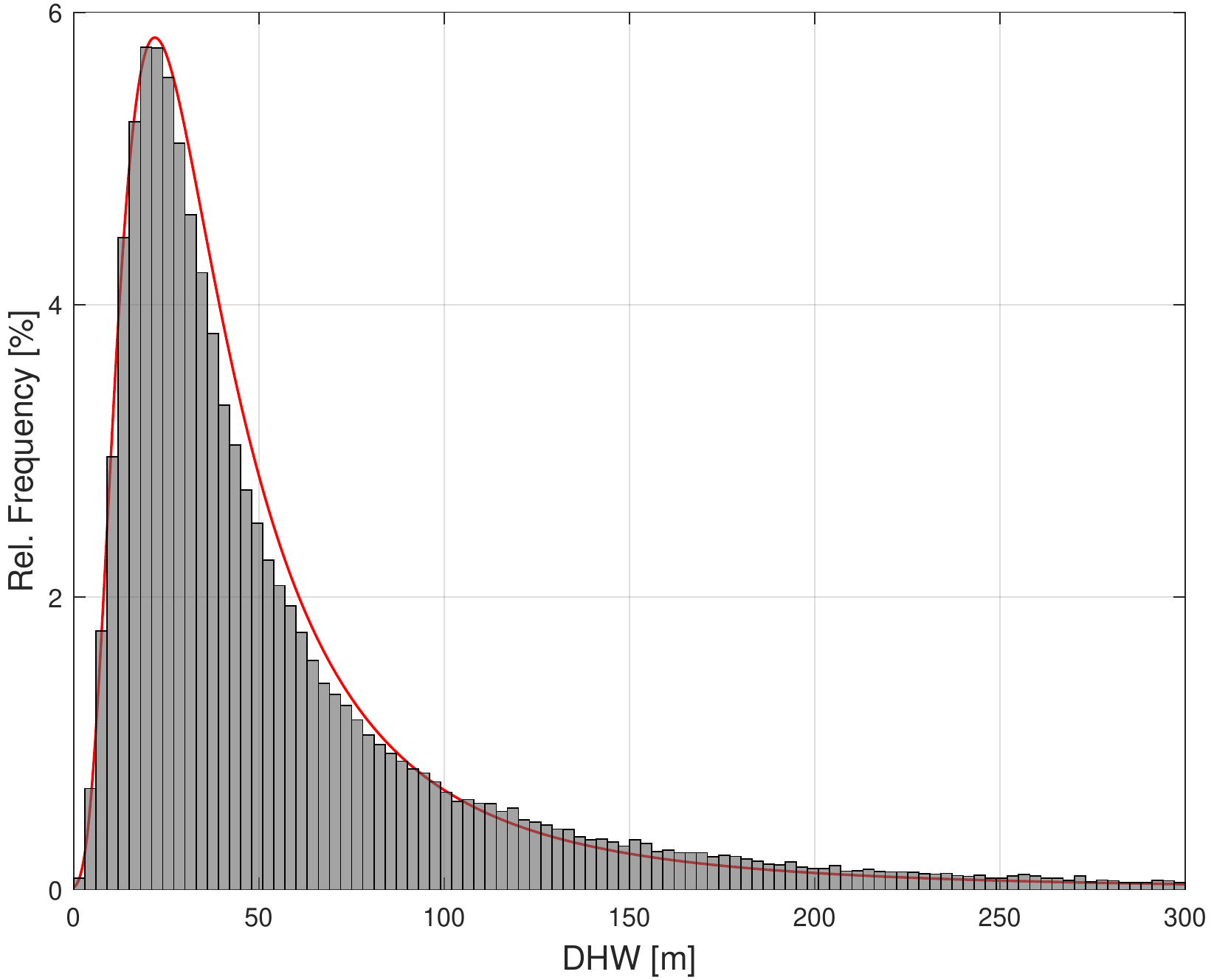}
	\vspace{-0.5cm}
	\caption{The distribution of the $\mathrm{DHW}_{\mathrm{min}}$ per track. Values of $\mathrm{DHW}_{\mathrm{min}} > \unit[300]{m}$, which account for 1.1\% of all tracks, are neglected.}
	\label{fig:dhw_hist}
	\vspace{-0.2cm}
\end{figure}

The distribution of $\mathrm{DHW}_{\mathrm{min}}$ in Fig. \ref{fig:dhw_hist} is related to the results of the $\mathrm{THW}_{\mathrm{min}}$ distribution. The depicted pdf is $\mathrm{DHW}_{\mathrm{min}} \sim GEV(29.8, 21.6, 0.5)$. If the drivers kept at least the legal distance, considering the given velocity distribution, a peak for $\mathrm{DHW}_{\mathrm{min}}$ should be located at around $\mathrm{DHW}_{\mathrm{min}} \approx \unit[40]{m}$.
The influence of the traffic density and the velocity on the DHW is driver specific \cite{Filzek.2002}. 

\section{Analysis on criticality measures}
\label{results_section}
This section is divided into two parts: first an analysis of the TTC and THW at critical low values. Second, the results regarding the RP indicator is compared to the results of \cite{Kondoh.2008}. 
The number of scenarios considered here is limited to the number of vehicles recorded ($M$). Only the most critical scenario for each vehicle within its recorded range is considered. 

\subsection{Results with TTC and THW}
\label{ttcthw_section}
This subsection discusses the findings regarding low TTC and THW values. Note, that only $\mathrm{THW}_{\mathrm{min}} \leq \unit[0]{s}$ and $\mathrm{TTC}_{\mathrm{min}} \leq \unit[0.8]{s}$ are manually validated in this work as described in Section \ref{data_section}.

Table \ref{table:thw_ttc_appearances} depicts the appearance of $\mathrm{THW}_{\mathrm{min}}$ and $\mathrm{TTC}_{\mathrm{min}}$ values within defined boundaries.
For $\mathrm{THW}_{\mathrm{min}} > \unit[0.5]{s}$, the occurrence of $\mathrm{THW}_{\mathrm{min}}$ falls approximately linearly with the $\mathrm{THW}_{\mathrm{min}}$ value.
A drop of occurrences at $\mathrm{THW}_{\mathrm{min}} \leq \unit[0.5]{s}$ can be seen. 
The $\mathrm{TTC}_{\mathrm{min}} \leq \unit[8]{s}$ occurrences drop exponentially with decreasing $\mathrm{TTC}_{\mathrm{min}}$ values, which is graphically depicted in Fig. \ref{fig:TTCandTHW_all}.

\begin{table}[h]
\begin{center}
\begin{tabular}{|c|c|c||c|c|c|}
\hline 
\rule[0ex]{0pt}{2.5ex} $\mathrm{THW} \lbrack\unit[]{s}\rbrack$ & $M_{x_{\mathrm{thw}}}$ & $\lbrack\unit[]{\%}\rbrack$ & $\mathrm{TTC} \lbrack\unit[]{s}\rbrack$ & $M_{x_{\mathrm{ttc}}}$ & $\lbrack\unit[]{\%}\rbrack$ \\ 
\hline
\hline 
\rule[0ex]{0pt}{2.5ex} 2 & 73452 & 66.5 & 8 & 8164 & 7.40\\ 
\hline 
\rule[0ex]{0pt}{2.5ex} 1 & 38607 & 35.0 & 4 & 2145 & 1.94 \\ 
\hline 
\rule[0ex]{0pt}{2.5ex} 0.$\overline{6}$ & 17674 & 16.0 & 2 & 311 & 0.28\\ 
\hline 
\rule[0ex]{0pt}{2.5ex} 0.5 & 8697 & 7.88 & 1 & 47 & 0.04\\ 
\hline 
\rule[0ex]{0pt}{2.5ex} 0.4 & 4432 & 4.01 & 0.8 & 29 & 0.03\\ 
\hline 
\rule[0ex]{0pt}{2.5ex} 0.25 & 1000 & 0.91 & 0.4 & 7 & 0.01\\ 
\hline 
\rule[0ex]{0pt}{2.5ex} 0.2 & 419 & 0.38 & 0.2 & 2 & 0.00\\ 
\hline 
\end{tabular} 
\end{center}
\caption{Appearance of THW and TTC values within defined boundaries}
\label{table:thw_ttc_appearances}
\end{table}

Now, the occurrences of critical scenarios according to the definition of \cite{nhtsa.2006} and \cite{Benmimoun.2012} are analyzed.
The Tables \ref{table:ATZ_risk_profiles} and \ref{table:100_cars_study_triggers} depict the number of occurrences in the highD data set in the ``highD'' column. The left sided number shows the overall occurrences, while the right hand sided number shows the occurrences, where no lane change was performed up to a time span of \unit[2]{s} after that critical time step. The actual time span considered depends on how long the vehicle was visible after the most critical time frame. Vehicles, which performed the lane change beyond the visible range, are treated here as if no lane change was performed. 
 
Given the risk level definition from \cite{Benmimoun.2012}, the occurrences of critical scenarios are depicted in Table \ref{table:ATZ_risk_profiles}, alongside with the rule set. Because brake light signals are not available, the threshold for braking is set to $a_x \leq \unitfrac[-1.5]{m}{s^2}$. All 24 samples of the TTC risk level 1, which did not perform a lane change
have a $\mathrm{TTC}_{\mathrm{min}} \approx \unit[1.7]{s}$ value. None of the vehicles reached the TTC risk level 2 or level 3.

\begin{table}[h]
\centering
	\includegraphics[width=\linewidth]{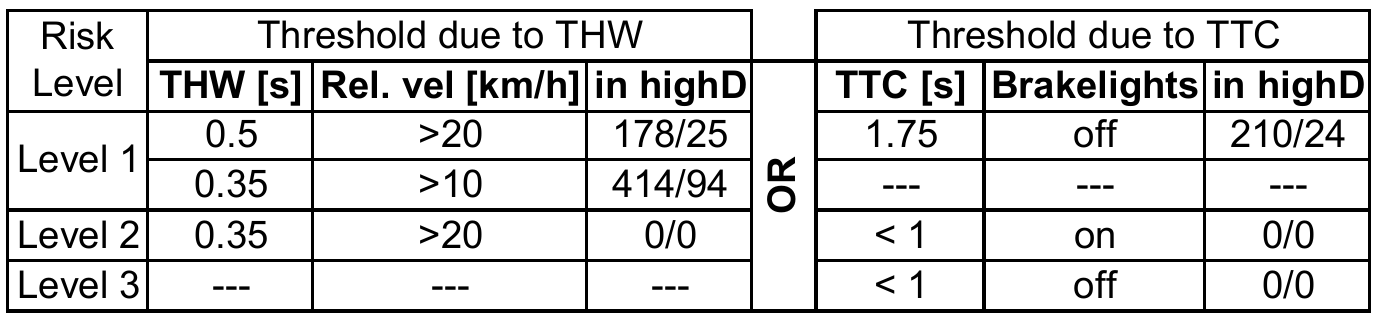}
	\vspace{-0.5cm}
	\caption{Risk Level definition for THW in combination with the relative velocity and TTC combined with the brakelight indication from \cite{Benmimoun.2012}. This table is enriched with the occurrences in the highD data set. The threshold for brake lights is set to $a_x \leq \unitfrac[-1.5]{m}{s^2}$, as bus data is not available. The highD column depicts on the left side the total number of occurrences, on the right side if no lane change was performed up to \unit[2]{s} after the critical time step.}
	\label{table:ATZ_risk_profiles}
	\vspace{-0.2cm}
\end{table}

The situation at the $\mathrm{THW}_{\mathrm{min}}$ risk level classification is similar. None of the vehicles reached level 2. The occurrence of $\mathrm{THW}_{\mathrm{min}} \leq \unit[0.35]{s}$ is relatively high, while relative velocities of $v \geq \unitfrac[20]{km}{h}$ appear less often. This is related to the frequent occurrence of low $\mathrm{THW}_{\mathrm{min}}$ values in congested traffic situations. 

Table \ref{table:100_cars_study_triggers} depicts a selection of the trigger points defined in the 100-cars-study to capture relevant events \cite{nhtsa.2006}. Relevant events are divided into three classes: incidents (crash-relevant conflicts and proximity conflicts), near crash and crash. The thresholds for $a_x$ and $a_y$ are high with regard to the velocities on German motorways. The TTC triggers are coupled with accelerations and in two cases additionally with distances. The other trigger points from \cite{nhtsa.2006}, namely TTC rear, the yaw rate and a button for drivers are not considered here. 

\begin{table}[h]
\centering
\begin{tabular}{|c|c|c|}
\hline 
\rule[0ex]{0pt}{2.5ex} \textbf{Trigger} & \textbf{Description} & \textbf{in highD}\\ 
\hline
\hline  
\rule[0ex]{0pt}{2.5ex} $a_y$ & $a_y \geq \unit[0.7]{g} \lor a_y \leq \unit[-0.7]{g}$  & 0/- \\ 
\hline 
\rule[0ex]{0pt}{2.5ex} $a_x$ & $a_x \geq \unit[0.6]{g} \lor a_x \leq \unit[-0.6]{g}$  & 2/- \\ 
\hline 
\rule[0ex]{0pt}{2.5ex} TTC & $\mathrm{TTC} \leq \unit[4]{s} \land a_x \geq \unit[0.5]{g}$ & 2/2 \\ 
\hline 
\rule[0ex]{0pt}{2.5ex} TTC & $\mathrm{TTC} \leq \unit[4]{s} \land a_x \leq \unit[-0.5]{g}$  & 0/0 \\ 
\hline 
\rule[0ex]{0pt}{2.5ex} TTC & \makecell{$\mathrm{TTC} \leq \unit[4]{s} \land a_x \geq \unit[0.4]{g}\, \land$ \\ $ a_x \leq \unit[0.5]{g} \land \mathrm{DHW} \leq \unit[100]{ft}$} & 8/8 \\ 
\hline 
\rule[0ex]{0pt}{2.5ex} TTC & \makecell{$\mathrm{TTC} \leq \unit[4]{s} \land a_x \leq \unit[-0.4]{g}\, \land$ \\ $a_x \geq \unit[-0.5]{g} \land \mathrm{DHW} \leq \unit[100]{ft}$}&5/5\\ 
\hline 
\end{tabular}
	\caption{Event triggers defined in the 100-cars-study \cite{nhtsa.2006}. The highD column depicts on the left side the total number of occurrences, on the right side if no lane change was performed up to $\unit[2]{s}$ after the critical time step. The textual description of \cite{nhtsa.2006} is converted to a logical expression here.}
	\label{table:100_cars_study_triggers}
	\vspace{-0.2cm} 
\end{table}

The approach of the 100-cars-study to capture critical scenarios is different compared to the risk level definition above. The event triggers are shifted towards less criticality in terms of the TTC value, while increasing the threshold for accelerations. This approach tends towards capturing critical scenarios, where the driver is actually aware of the situation. A threshold of $\mathrm{TTC}_{\mathrm{min}} \leq \unit[4]{s}$ increases the chance for drivers to react and initialize longitudinal or lateral actions compared to $\mathrm{TTC}_{\mathrm{min}} \leq \unit[1.75]{s}$, as discussed in Section \ref{relatedwork_section}. The approach of \cite{Benmimoun.2012} treats high accelerations separately. They can shift the risk level to upwards or downwards or not at all.

Next, the relative distribution of the $\mathrm{THW}_{\mathrm{min}}$ and $\mathrm{TTC}_{\mathrm{min}}$ values are analyzed with respect to velocities and accelerations. The velocities and accelerations are the mean values of a \unit[1.0]{s} period, \unit[0.5]{s} before and after the most critical time frame. The range for $\mathrm{THW}_{\mathrm{min}}$ is limited to $\mathrm{THW}_{\mathrm{min}} \leq \unit[1]{s}$. The range for $\mathrm{TTC}_{\mathrm{min}}$ is limited to $\mathrm{TTC}_{\mathrm{min}} \leq \unit[8]{s}$. It should be noted, that the samples for $\mathrm{TTC}_{\mathrm{min}} \leq \unit[1.0]{s}$ are very limited and are therefore only of limited meaning (see Table \ref{table:thw_ttc_appearances}). 

The results are depicted in three dimensional bar plots. The sum of all bar values across one THW or TTC threshold row equals 100\%.
 
\begin{figure}[h]
\centering
	\includegraphics[width=\linewidth]{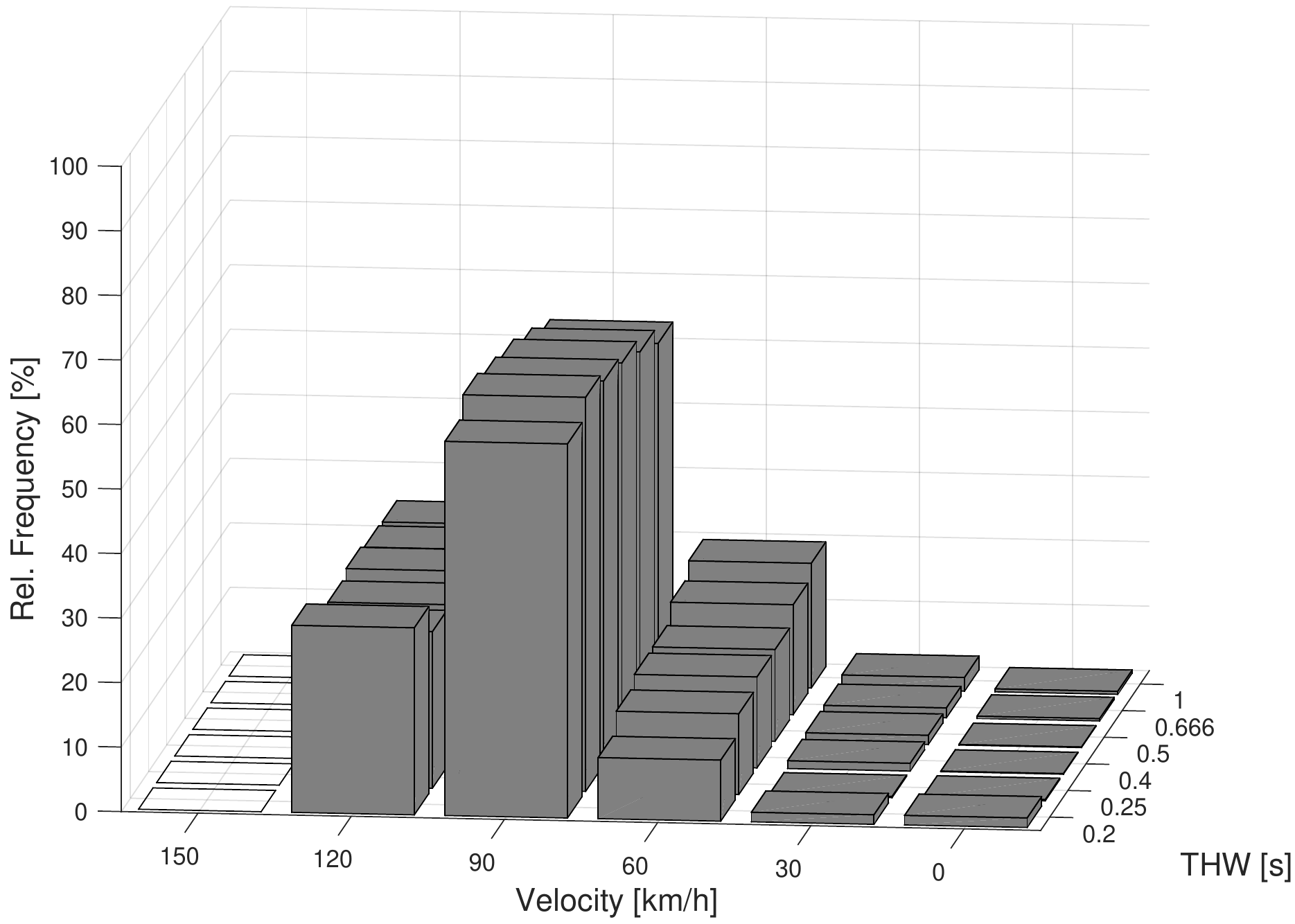}
	\vspace{-0.5cm}
	\caption{The relative distribution of $\mathrm{TTC}_{\mathrm{min}}$ across several velocity ranges. The relative proportion of $\mathrm{THW}_{\mathrm{min}}$ values is independent of the velocity.}
	\label{fig:thw_vs_velocity_bar3}
	\vspace{-0.2cm}
\end{figure}

Fig. \ref{fig:thw_vs_velocity_bar3} depicts the $\mathrm{THW}_{\mathrm{min}}$ proportions with respect to the velocity. Note, that the bars at $v = \unitfrac[0]{km}{h}$ include velocities from $v = \unitfrac[0]{km}{h}$ up to $v < \unitfrac[30]{km}{h}$. This principle holds for all following figures in this Section.

The $\mathrm{THW}_{\mathrm{min}}$ proportions are generally independent of the velocity. Low $\mathrm{THW}_{\mathrm{min}}$ values appear more often in the lower velocity range of $v \leq \unitfrac[30]{km}{h}$, compared to the proportion of low velocities in the overall velocity distribution. This coincides with the results regarding undercutting the legal threshold of $\mathrm{THW}_{\mathrm{min}} \leq \unit[0.9]{s}$ depicted in Fig. \ref{fig:thwpunish_hist}.
The proportion in the velocity range of $v \geq \unitfrac[90]{km}{h}$ is higher than the proportion of vehicles compared to the trucks. This coincides with the graph in Fig. \ref{fig:density_vs_flowrate_vs_THW}, which shows that trucks usually keep in average a higher THW to their leader vehicle.    

The following Fig. \ref{fig:ttc_vs_velocity_bar3} depicts the relative $\mathrm{TTC}_{\mathrm{min}}$ distribution with regard to the velocity. Again, the proportion of the velocity ranges of $v \leq \unitfrac[30]{km}{h}$ and $v \geq \unitfrac[90]{km}{h}$ with small $\mathrm{TTC}_{\mathrm{min}}$ are overproportionally.  

\begin{figure}[h]
\centering
	\includegraphics[width=\linewidth]{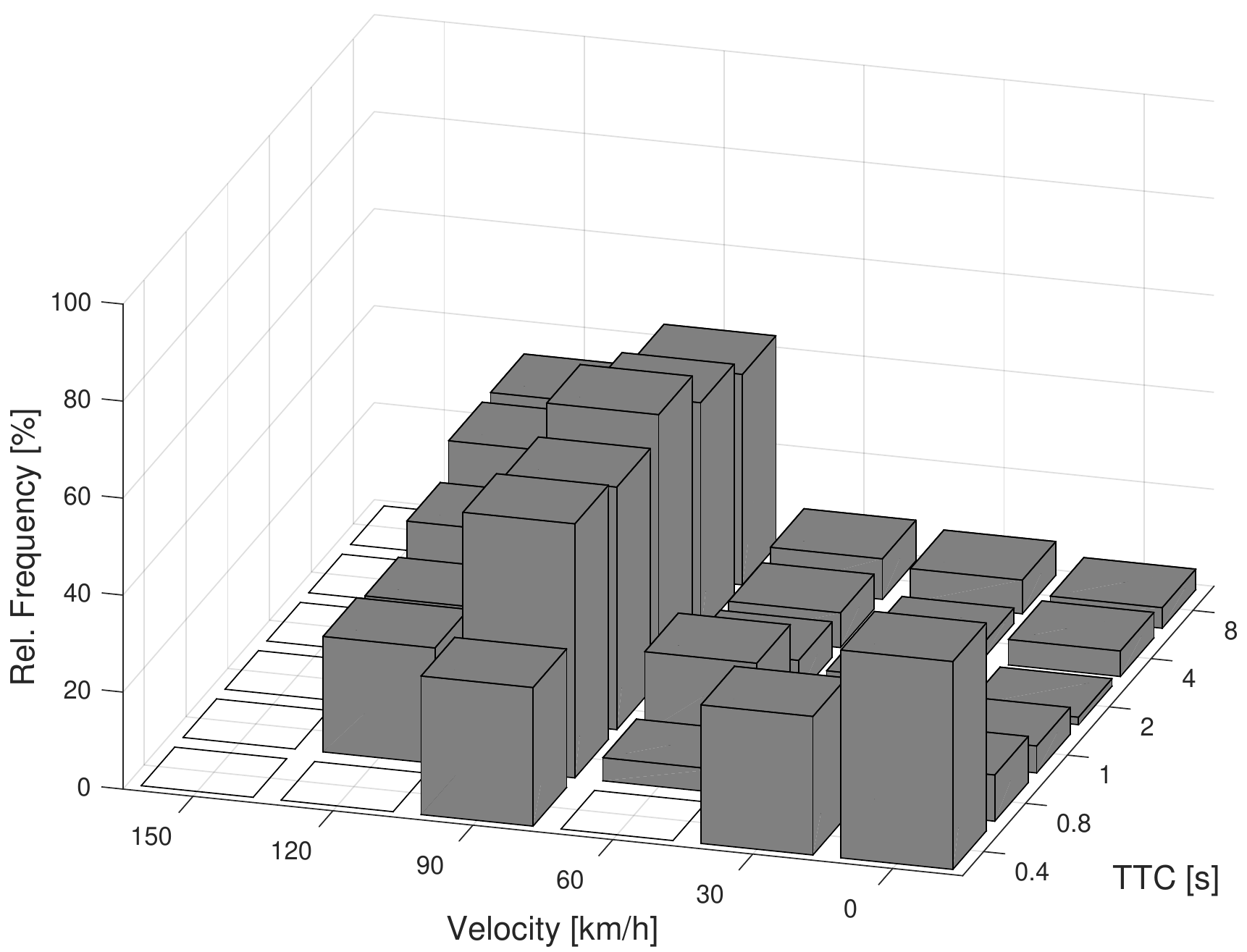}
	\vspace{-0.5cm}
	\caption{The relative distribution of $\mathrm{TTC}_{\mathrm{min}}$ across several velocity ranges. The proportion of the velocity ranges of $v \leq \unitfrac[30]{km}{h}$ and $v \geq \unitfrac[90]{km}{h}$ are overproportionally.}
	\label{fig:ttc_vs_velocity_bar3}
	\vspace{-0.2cm}
\end{figure}

\begin{figure}[h]
\centering
	\includegraphics[width=\linewidth]{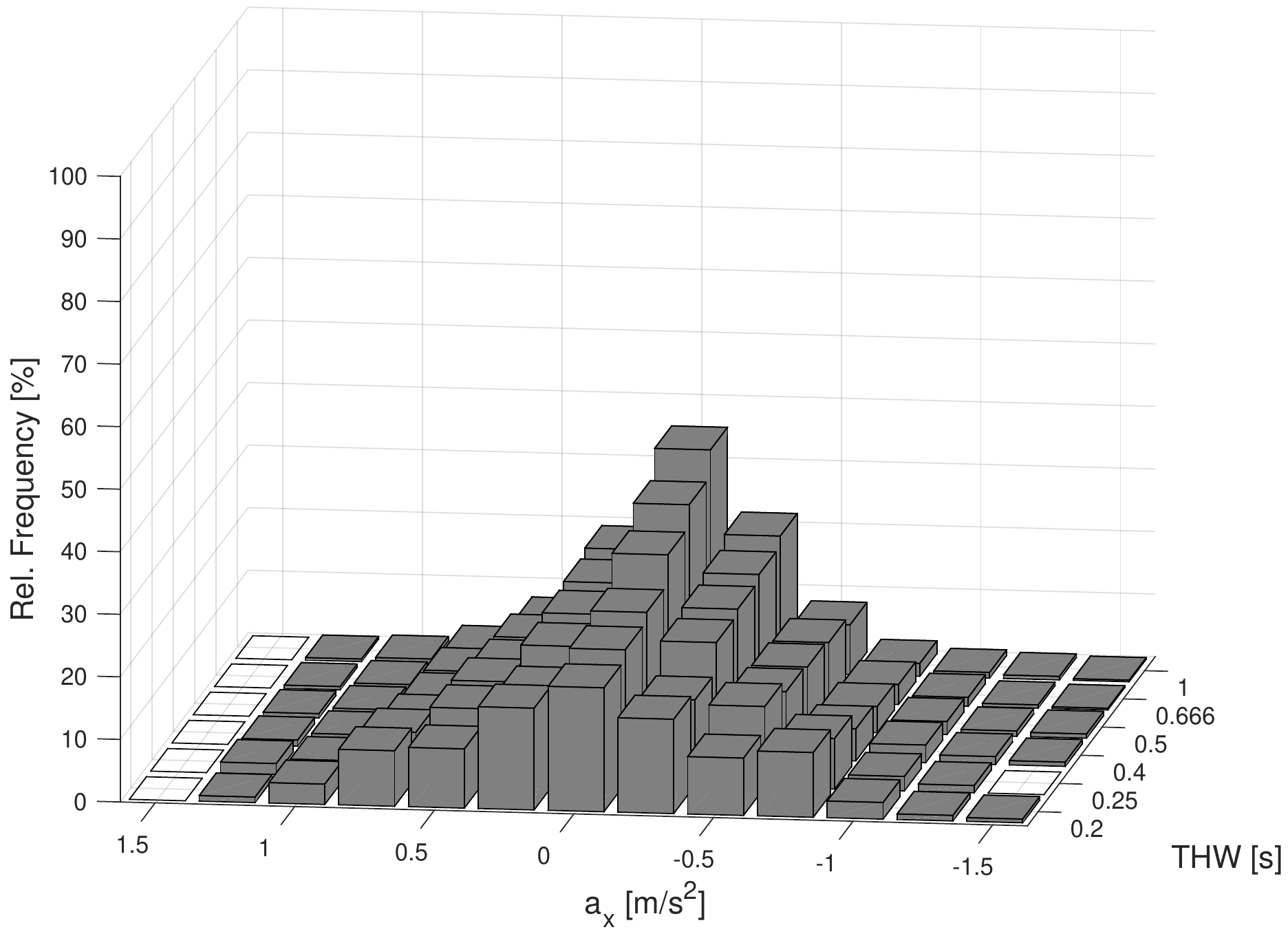}
	\vspace{-0.5cm}
	\caption{The relative distribution of $\mathrm{THW}_{\mathrm{min}}$ across several longitudinal acceleration ranges}
	\label{fig:thw_vs_accx_bar3}
	\vspace{-0.2cm}
\end{figure}

Fig. \ref{fig:thw_vs_accx_bar3} depicts the $\mathrm{THW}_{\mathrm{min}}$ values with respect to the longitudinal acceleration. With decreasing $\mathrm{TTC}_{\mathrm{min}}$ values, the percentage of vehicles with acceleration in both directions increases. In the region of $\mathrm{THW}_{\mathrm{min}} \leq \unit[0.4]{s}$ vehicles tend rather to accelerate. This indicates a high likelihood of an overtaking and lane change maneuver as discussed before.

\begin{figure}[h]
\centering
	\includegraphics[width=\linewidth]{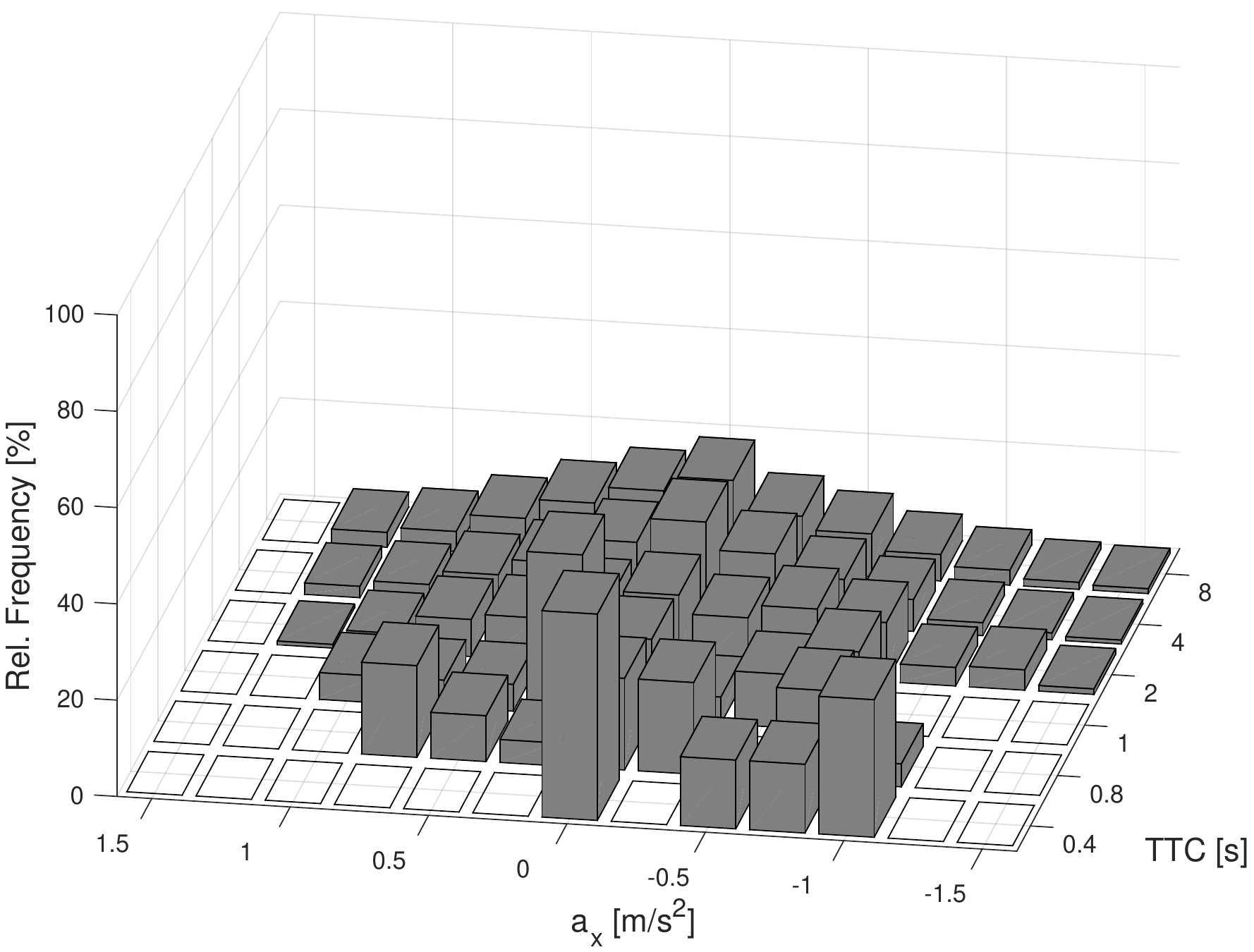}
	\vspace{-0.5cm}
	\caption{The relative distribution of $\mathrm{TTC}_{\mathrm{min}}$ across several longitudinal acceleration ranges}
	\label{fig:ttc_vs_accx_bar3}
	\vspace{-0.2cm}
\end{figure}

Fig. \ref{fig:ttc_vs_accx_bar3} depicts the $\mathrm{TTC}_{\mathrm{min}}$ values with respect to the longitudinal acceleration. In the case of $\mathrm{TTC}_{\mathrm{min}} \geq \unit[4]{s}$ the distribution between positive and negative acceleration is balanced. The prominent peak at $\mathrm{TTC}_{\mathrm{min}} \geq \unit[0.4]{s}$ and zero acceleration depicts those examples, where the vehicle almost finished the lane change or a neighboring vehicle crossed the ego vehicles lane slightly (see Section \ref{data_section}).   

\begin{figure}[h]
\centering
	\includegraphics[width=\linewidth]{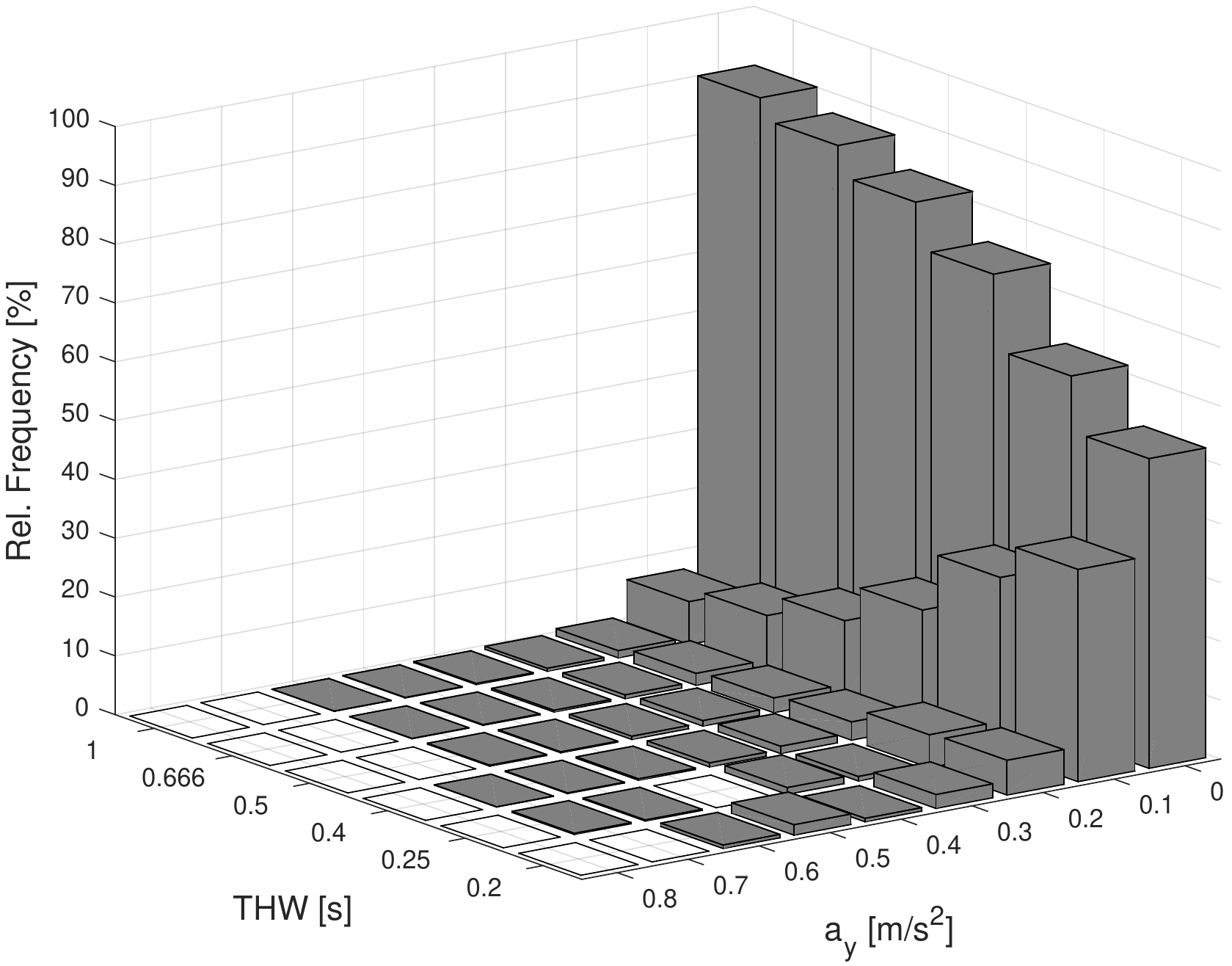}
	\vspace{-0.5cm}
	\caption{The relative distribution of $\mathrm{THW}_{\mathrm{min}}$ across several lateral acceleration ranges}
	\label{fig:thw_vs_accy_bar3}
	\vspace{-0.2cm}
\end{figure}

For larger values of lateral acceleration the proportion of vehicles with low $\mathrm{THW}_{\mathrm{min}}$ increases as depicted in Fig. \ref{fig:thw_vs_accy_bar3}, indicating lane changes. 

\begin{figure}[h]
\centering
	\includegraphics[width=\linewidth]{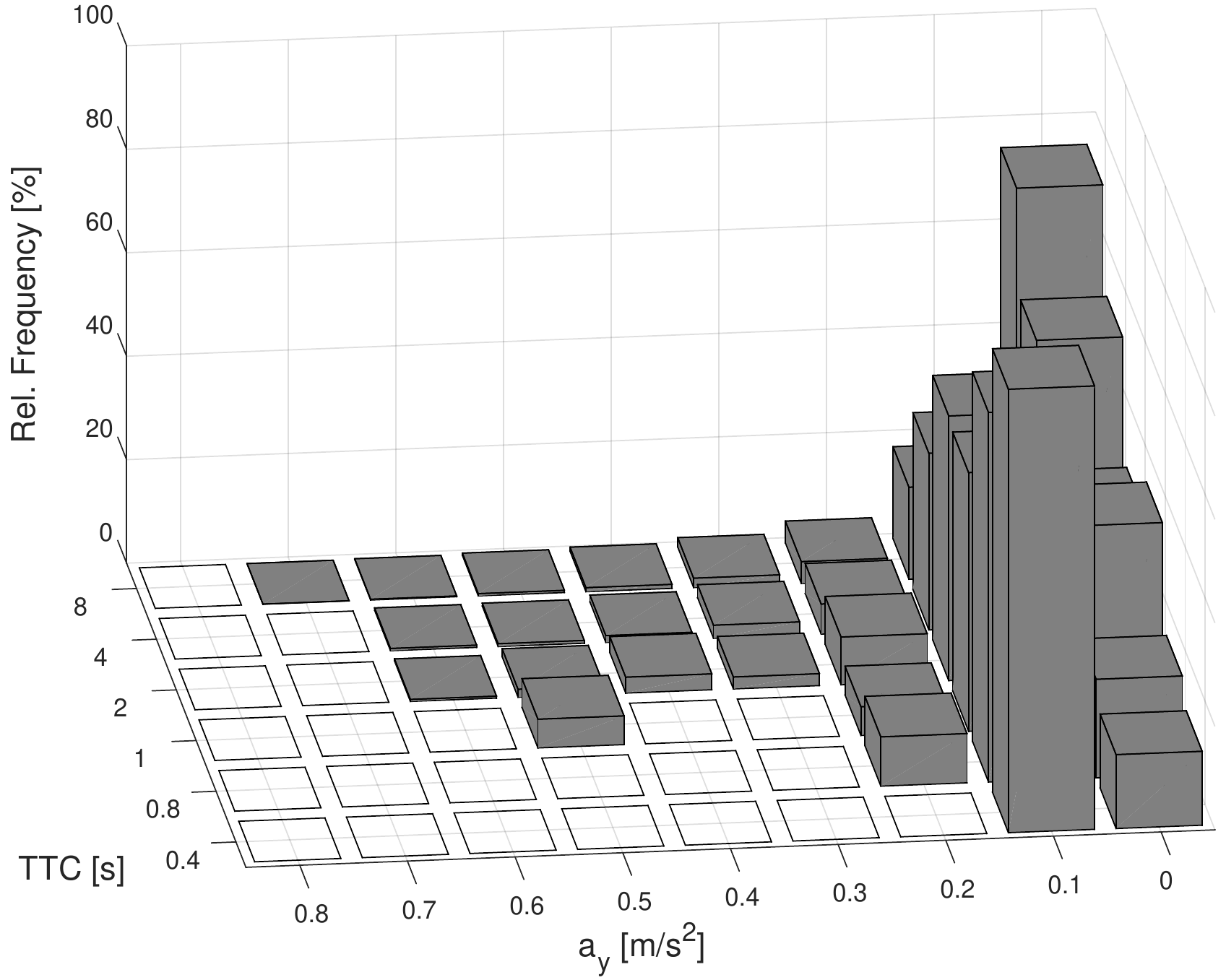}
	\vspace{-0.5cm}
	\caption{The relative distribution of $\mathrm{TTC}_{\mathrm{min}}$ across several lateral acceleration ranges}
	\label{fig:ttc_vs_accy_bar3}
	\vspace{-0.2cm}
\end{figure}

Fig. \ref{fig:ttc_vs_accy_bar3} depicts the relative distribution of $\mathrm{TTC}_{\mathrm{min}}$ across several lateral acceleration ranges. The number of samples for $a_y \geq \unitfrac[0.2]{m}{s^2}$ and small $\mathrm{TTC}_{\mathrm{min}}$ values is limited, so that no conclusion can be drawn about the tendency.

Finally, the statements from this subsection are summarized. Scenarios which appear to be critical, if only considering the TTC or THW minimum values, turn out to be rather normal driving scenarios on highways in most cases. The longitudinal and lateral accelerations are close to zero or in the typical range appearing during lane change maneuvers. The proportion of low velocity ranges, which can be found in congested traffic on highways, are higher than their actual proportion over the complete data set. These critical values often occur because of the oscillating relative distances and represent normal driving scenarios.

\subsection{Results with RP}
\label{rp_section}
\begin{figure}[h]
\centering
	\includegraphics[width=\linewidth]{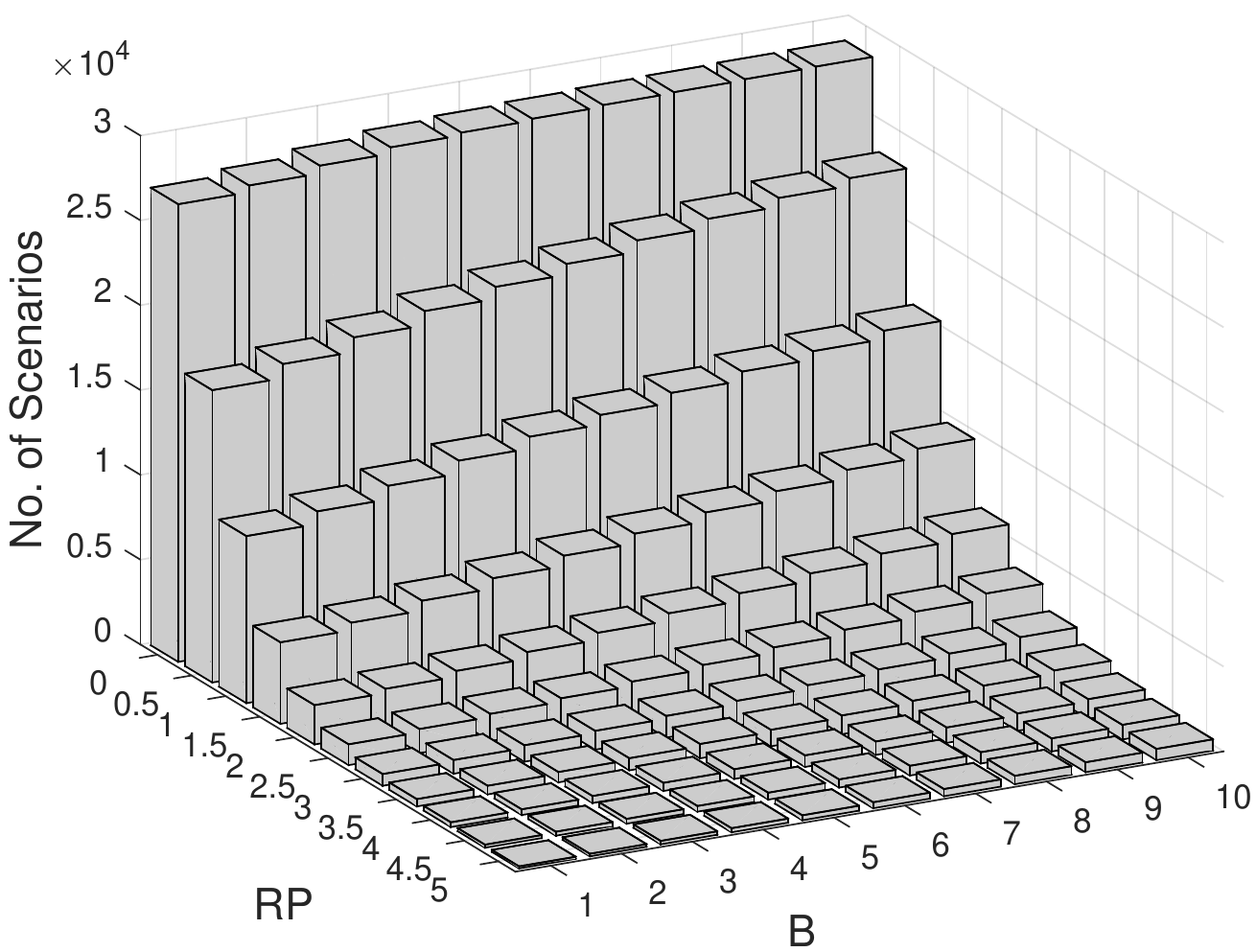}
	\vspace{-0.5cm}
	\caption{Distribution of RP values by varying the variable $B$ and a fixed $A=1$. Bars with $RP = 0$ represent all tracks of the data set, which fulfilled the tailgating restriction of $\pm \unit[4]{s}$.}
	\label{fig:RP}
	\vspace{-0.2cm}
\end{figure}

This section analyses distributions of the RP indicator. The variable $A$ is set to $A=1$, $B$ is varied over $\mathrm{B} \in [1,2\dots,10]$ and $\mathrm{RP} \in [0.5,1.0\dots,5.0]$. Fig. \ref{fig:RP} depicts the results. The authors define $\mathrm{S}_{\mathrm{RP}_{214}}$ to be the variable set, where drivers latest start to brake. Additionally, the following vehicle must be located behind its leader vehicle four seconds before and four seconds after the critical point in time. In the highD data set in total $M_{\mathrm{RP}_{214}} = 3323$ of such scenarios appear, which corresponds to 11.8\% of all tracks which fulfilled the time span requirement.  

\begin{figure}[h]
\centering
	\includegraphics[width=\linewidth]{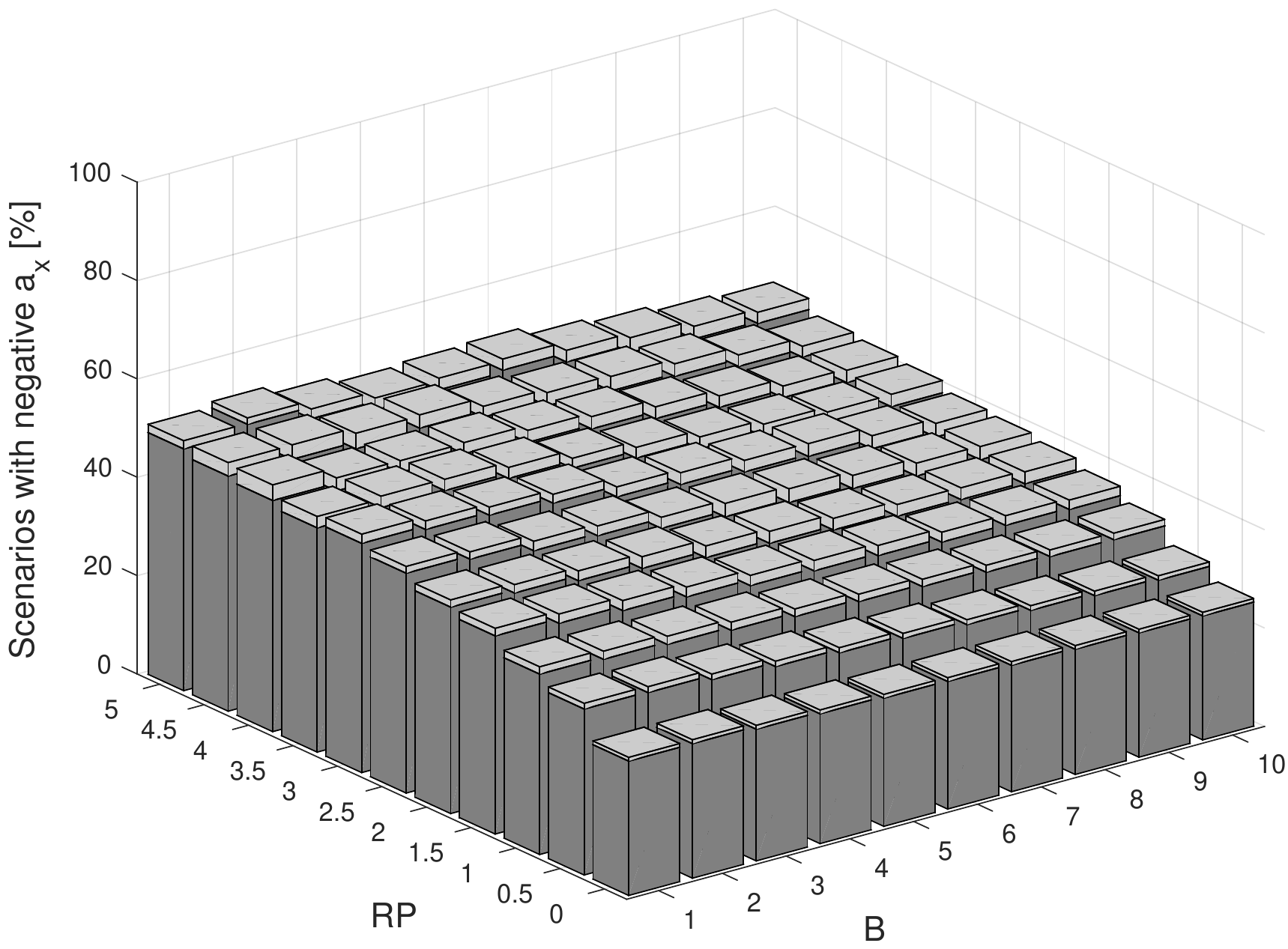}
	\vspace{-0.5cm}
	\caption{Relative share of scenarios, where the follower had a maximum negative acceleration of $a_x \leq \unitfrac[0]{m}{s^2}$ (dark gray) or $a_x \leq \unitfrac[-1.5]{m}{s^2}$ (light gray) at the critical time step or within $\unit[0.20]{s}$ after the critical time step.}
	\label{fig:relBrake}
	\vspace{-0.2cm}
\end{figure}

Next, an analysis is carried out, if the vehicles actually braked as proposed in \cite{Kondoh.2008}. In Fig. \ref{fig:relBrake} the proportion of scenarios are depicted, where the follower vehicle has a minimum acceleration of $a_x \leq \unitfrac[0]{m}{s^2}$ (dark gray) or $a_x \leq \unitfrac[-1.5]{m}{s^2}$ (light gray) within a period of $\unit[0.2]{s}$ after the most critical time frame. 

The total proportion of vehicles with a negative longitudinal acceleration and a variable set $\mathrm{S}_{\mathrm{RP}_{\mathrm{214}}}$ is 41.7\% of $M_{\mathrm{RP}_{214}}$ vehicles. Only 2.4\% of the vehicles performed active braking. In none of the depicted variable sets the rate exceeds 52\%. These results are contrary to the conclusion from \cite{Kondoh.2008}. In \cite{Kondoh.2008}, the real world data was collected with 15 experienced drivers driving on freeways and surface roads around Minneapolis/St. Paul. 

\begin{figure}[h]
\centering
	\includegraphics[width=\linewidth]{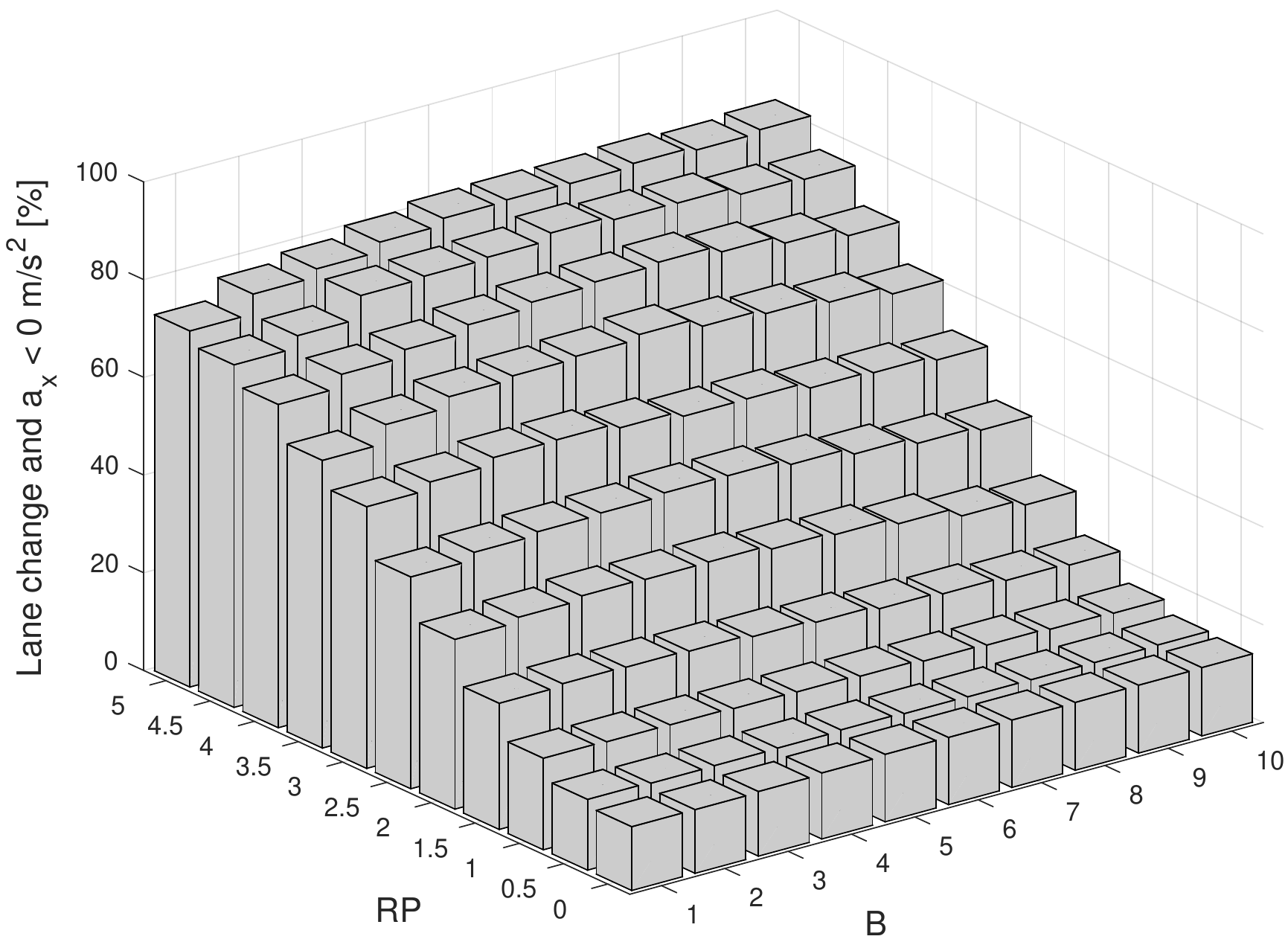}
	\vspace{-0.5cm}
	\caption{Relative share of scenarios, where the following vehicle performed a lane change while having a negative acceleration with $a_x \leq \unitfrac[0]{m}{s^2}$ after the most critical time frame.}
	\label{fig:lnChange_relBrakeAccNeg}
	\vspace{-0.2cm}
\end{figure}

\begin{figure}[h]
\centering
	\includegraphics[width=\linewidth]{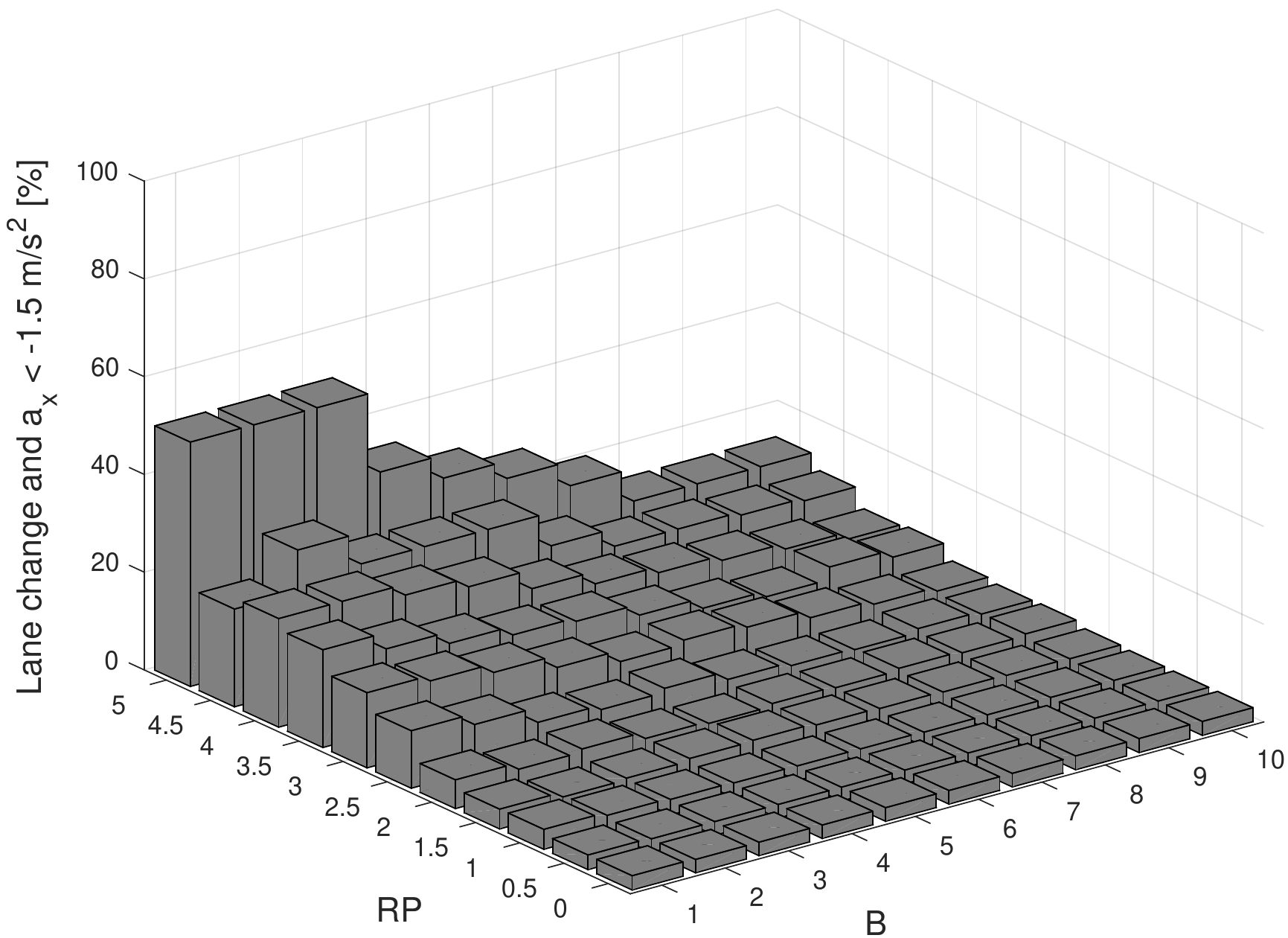}
	\vspace{-0.5cm}
	\caption{Relative share of scenarios, where the following vehicle performed a lane change while performing active braking with $a_x \geq \unitfrac[-1.5]{m}{s^2}$ after the most critical time frame.}
	\label{fig:lnChange_relBrakebrOn}
	\vspace{-0.2cm}
\end{figure}

One reason for not performing a brake application at $RP \geq\ 2.0$ can be found in scenarios, when the follower vehicle performs a sequential combination of a closing, lane change and overtake maneuver. Therefore the appearance of lane changes is analyzed as well. The results are depicted in Fig. \ref{fig:lnChange_relBrakeAccNeg} and \ref{fig:lnChange_relBrakebrOn}. Fig. \ref{fig:lnChange_relBrakeAccNeg} depicts the proportion of follower vehicles, which had a negative acceleration and performed a lane change within the time span of $\pm \unit[4]{s}$. For the variable set $\mathrm{S}_{\mathrm{RP}_{214}}$ the total proportion of lane changing vehicles with negative acceleration is 36.4\%, compared to $M_{\mathrm{RP214}}$. With $a_x \leq \unitfrac[-1.5]{m}{s^2}$ only 3.8\% (3 vehicles out of 79) performed a lane change. The quantity of available samples with active braking is small though.

Summarizing the above, around 41.7\% of all vehicles of $\mathrm{S}_{\mathrm{RP}_{\mathrm{214}}}$ had a negative acceleration and 2.4\% performed active braking. From those, approximately 36.4\% of 1306 vehicles with $a_x \leq \unitfrac[0]{m}{s^2}$ performed a lane change. At $a_x \leq \unitfrac[-1.5]{m}{s^2}$, 3.8\% of 79 vehicles performed a lane change. The majority of vehicles do not brake at $\mathrm{S}_{\mathrm{RP}_{\mathrm{214}}}$ in this data set. If they do have to brake, then mostly a lane change within \unit[4]{s} was not performed.

\section{CONCLUSIONS}
\label{conclusions_section}
The presented work analyses the highD data set. The macroscopic analysis includes the fundamental traffic stream variables and depicts the relationship to other variables like the THW or the proportion of trucks. Additionally, the influence of the traffic flow rate on the lane load and lane change rate is shown.   

The microscopic analysis includes velocities, accelerations, distances, TTC, THW and RP. Histograms and pdf are provided and outliers are discussed separately.
Critical values of the TTC, THW, which are provided in the data set, mostly appear during lane change maneuvers. Combining these criticality indicators with additional variables like accelerations and furthermore applying a motion model is necessary to get more reliable estimations of the risk potential.

\addtolength{\textheight}{-12cm}   



\section*{ACKNOWLEDGMENT}
The authors acknowledge the financial support by the Federal Ministry of Education and Research of
Germany (BMBF) in the framework of FH-Impuls (project number 03FH7I02IA).

\bibliographystyle{IEEEtran}
\bibliography{ref}
\end{document}